\documentclass[12pt]{article}
\usepackage[margin=1.2in]{geometry}
\usepackage[T1]{fontenc}
\usepackage[utf8]{inputenc}
\usepackage{lmodern}
\usepackage{amsmath,amssymb,amsthm,mathtools}
\usepackage{booktabs,array}
\usepackage[round,authoryear]{natbib}
\usepackage{microtype}
\usepackage{enumitem}
\usepackage{tabularx,needspace,capt-of}

\usepackage{hyperref}

\numberwithin{equation}{section}
\newtheorem{theorem}{Theorem}
\newtheorem{proposition}{Proposition}
\newtheorem{lemma}{Lemma}
\newtheorem{example}{Example}
\newtheorem{remark}{Remark}
\newcommand{\R}{\mathbb R}
\newcommand{\E}{\mathbb E}
\newcommand{\Prb}{\mathbb P}
\newtheorem{corollary}{Corollary}
\newcolumntype{Y}{>{\raggedright\arraybackslash}X}
\newcolumntype{P}[1]{>{\raggedright\arraybackslash}p{#1}}
\providecommand{\1}{\mathbf 1}

\providecommand{\dd}{\,\mathrm d}

\newcommand{\pstep}[1]{\par\Needspace{4\baselineskip}Step #1.\ }

\hypersetup{pdftitle={Universal Nonexistence of Monotone Equilibrium in Multiunit Auctions with Risk Aversion},pdfauthor={Wei He, Yeneng Sun, and Hanping Xu}}

\title{Universal Nonexistence of Monotone Equilibrium in\\Multiunit Auctions with Risk Aversion}
\author{Wei He\thanks{Department of Economics, The Chinese University of Hong Kong, Hong Kong.}
\and Yeneng Sun\thanks{Departments of Economics and Mathematics, National University of Singapore, Singapore.}
\and Hanping Xu\thanks{Institute for Advanced Study in Mathematics, Harbin Institute of Technology, Harbin.}}
\date{\today}

\begin{document}

\pagenumbering{gobble} 

\maketitle
\thispagestyle{plain}

\vspace{1.5em}
\begin{abstract}
We establish universal nonexistence results for monotone pure-strategy equilibrium in multiunit auctions with risk-averse bidders. Fix any number of bidders greater than one and any profile of strictly increasing concave utilities, with at least one strictly risk-averse bidder. Under uniform pricing, for every supply of at least three units, we construct independent, full-support type distributions under which neither lowest-winning-bid nor highest-losing-bid pricing admits an admissible monotone equilibrium. Under discriminatory pricing, we construct a finite supply and independent, full-support type distributions that preclude monotone pure-strategy equilibrium. We extend these results to all-pay auctions and finite bid grids. 
\end{abstract}


\newpage

\pagenumbering{arabic} 
\setcounter{page}{1}

\section{Introduction}\label{sec:introduction}

Multiunit auctions are central institutions in markets for government securities, electricity, and other standardized goods. The economic scale is substantial: in 2025, the U.S. Treasury alone issued nearly \$30 trillion in marketable securities through auctions. Uniform-price and discriminatory auctions are two principal formats for these transactions. The United States uses uniform pricing for Treasury securities, while France and Germany award successful competitive bids at their respective bid prices.\footnote{TreasuryDirect reports 444 public auctions and approximately \$29.7 trillion in marketable securities issued in 2025; the figure measures gross issuance. The institutional descriptions follow \citet{USTreasury}, \citet{AFT}, and \citet{GermanFinanceAgency}. The United Kingdom uses discriminatory pricing for competitive bids in conventional gilt auctions and uniform pricing for index-linked gilt auctions; see \citet{UKDMO}.} In both formats, bidders submit demand schedules and compete over quantities as well as prices. Understanding their behavior is essential to evaluating how these markets allocate goods and how much revenue they generate.

For risk-neutral bidders with independent private values, \citet{Reny1999}, \citet{McAdams2003,McAdams2006},\footnote{  \citet{McAdams2006} further shows that every mixed-strategy equilibrium in his model has a monotone pure-strategy counterpart with the same ex post allocation and interim expected payments. }  and \citet{JacksonSwinkels2005} establish the existence of monotone pure-strategy equilibrium in a range of multiunit auction models. In a monotone equilibrium, a bidding strategy is nondecreasing in the bidder's vector of marginal values. This property gives a tractable way to establish equilibrium existence when demand schedules are difficult to characterize explicitly, and it facilitates equilibrium analysis for studying allocation and revenue.

Risk aversion changes the trade-offs bidders face when submitting bids. A bidder chooses her demand schedule before knowing how many units she will receive or how much she will pay. She may be willing to sacrifice expected surplus to reduce her exposure to low-surplus outcomes. Empirical works have well documented that risk aversion plays an important role in the auction markets. For example, \citet{ArmantierSbai2006} find evidence of heterogeneous risk aversion in French Treasury auctions, and \citet{Hafner2023} shows that accounting for risk aversion changes estimates of rents in Swiss import-quota auctions.

In single-unit first-price auctions, the existence of monotone equilibrium with risk-averse bidders has been extensively studied in various settings; see,  for example, \cite{MaskinRiley1984,MaskinRiley2000}, \cite{LizzeriPersico2000}, and \cite{RenyZamir2004}. This raises a basic question: can we guarantee the existence of monotone equilibrium in multiunit auctions with risk-averse bidders? In particular, can even mild risk aversion preclude monotone equilibrium in the standard framework with strictly increasing payoffs and  independent private values under full-support distributions? \cite{McAdams2006} raises a related open question about monotone best replies.\footnote{See also \cite{Reny2011} for detailed discussions.} He explains why monotone best replies can fail under discriminatory pricing with two decreasing absolute risk averse (DARA) bidders and two units, while suggesting that they may exist under highest-losing-bid uniform pricing with DARA bidders.\footnote{In a two-bidder, two-unit discriminatory auction, \cite{McAdams2006} writes: ``Consequently, if I have decreasing absolute risk aversion (DARA), then I will tend to be `less risk averse' and hence bid less in the auction for the second unit. Of course, this logic depends crucially on my having DARA utility and applies only to the discriminatory auction. It remains an open question whether bidders always have monotone best replies in uniform-price auctions given risk-averse bidders. It seems plausible to me that bidders may always have monotone best replies in the uniform $(S+1)$-st price auction given DARA bidders.''}

We establish two universal nonexistence theorems. Fix any number of bidders greater than one and any profile of strictly increasing, concave utilities, with at least one strictly risk-averse bidder. Under uniform pricing, with at least three units, there are independent type distributions under which neither highest-losing-bid nor lowest-winning-bid pricing\footnote{Under highest-losing-bid pricing, all winners pay the highest rejected bid for each unit; under lowest-winning-bid pricing, they pay the lowest accepted bid. These rules generate different incentives for the bidders. With a single unit, they reduce to the second-price and first-price auctions, respectively. Our highest-losing-bid construction addresses the questions in  \cite{McAdams2006}: it shows that strictly risk-averse bidders, including those with DARA preferences, need not have monotone best replies when rivals follow monotone strategies.} admits an admissible monotone pure-strategy equilibrium.\footnote{Admissibility excludes weakly dominated bids. We prove the stronger statement that neither rule admits a no-overbidding pure-strategy equilibrium in which even the designated strictly risk-averse bidder's last bid is nondecreasing in her marginal values. Her other bids and her rivals' strategies need not be monotone.} The same distributions work for both pricing rules. Under discriminatory pricing, we establish nonexistence of monotone pure-strategy equilibrium without any admissibility or no-overbidding restriction. Whenever there is one strictly risk-averse bidder, we identify an environment with finitely many units and independent type distributions under which no monotone pure-strategy equilibrium exists.

In both theorems, the constructed densities are bounded above and away from zero, the utility functions are taken as given, no particular functional form or minimum degree of strict risk aversion is required, and all but one bidder may be risk neutral. The two nonexistence results highlight a fundamental conflict between risk aversion and monotonicity.  Even one strictly risk-averse bidder, however mild her risk aversion is, can rule out admissible monotone equilibrium in both uniform-price and discriminatory auctions.


To illustrate the intuition, we present Example~\ref{exam:uniform} in which there are two bidders and three units under highest-losing-bid pricing. Each winner pays the highest rejected bid for every unit she receives. Bidder 1 is strictly risk averse. She submits her bids before knowing whether her rival has a modest value for one unit and negligible values for the others (weak), or high values for two (strong). Against the weak rival, she can leave one unit to the rival and buy the other two almost for free. She can secure all three units by raising her third bid above the rival’s first bid. Since bids are ordered, her second bid must be at least as high. Against the strong rival, the rival’s values are sufficiently high that the gain from a second unit exceeds the total payment needed to secure both. The rival therefore wins two units, leaving bidder 1 with one unit. When bidder 1’s second bid loses, it can set the price she pays for her remaining unit. Keeping her second and third bids low would reduce that payment; raising them to compete for all three units makes the one-unit outcome more expensive. She must choose a single bidding vector that balances these consequences: obtaining an additional unit when competition is weak and paying more even when she obtains only one unit when competition is strong.

Now raise bidder 1’s second marginal value, holding her first and third values fixed. Against a weak rival, she earns more surplus whether she buys two units or three. The additional surplus from securing the third unit is unchanged, but strict concavity makes this increment yield a smaller utility gain. When she wins only one unit, neither her surplus nor the utility cost of a higher payment changes. We can choose the probability of a strong rival so that the gain from competing for the third unit outweighs the payment cost for the lower type, but not for the higher type. Thus, sufficiently small increases in the first and third values preserve the reversal: a bidder can value every unit more and still optimally bid less for the third.

Under discriminatory pricing, bidder 1 pays her own winning bids. Thus, a high third bid commits her to paying at least that much for any earlier unit she wins. In our three-unit example, she would sacrifice too much surplus against a weak rival if she raised her second bid enough to prevent a strong rival from winning two units. Then the strong rival can profitably secure two units and win them sufficiently often that some higher types prefer low third bids to avoid an expensive one-unit outcome. Monotonicity would require lower types to submit equally low or lower third bids. The weak rival could earn a positive payoff by outbidding them for one unit. Since higher types are not easier to outbid, earning that payoff in equilibrium requires her to win against lower types sufficiently often. Then some lower types would rather raise their bids and secure the third unit than keep losing it to the weak rival.\footnote{With two units and CARA utilities, we show in  Appendix~\ref{app:main-results} that a monotone no-overbidding equilibrium exists under uniform pricing (with both highest-losing-bid and lowest-winning-bid pricing) and discriminatory pricing. With three units, an intermediate marginal value can change the bidder's willingness to compete for a later unit.  We provide examples in Sections~\ref{sec:uniform} and~\ref{sec:discriminatory} to demonstrate that nonexistence can arise even with CARA preferences.}

Our results are related to the nonexistence example of \citet{McAdams2007b}, who shows that risk aversion can rule out monotone equilibrium in weakly undominated pure strategies in a two-unit highest-losing-bid uniform pricing auction. His construction permits satiation and has lower-dimensional support for marginal-value vectors. Working with strictly increasing utilities and full-dimensional type densities, \citet{Reny2011} explains the failure of single crossing under the coordinatewise order and establishes existence under a novel order on types, including in discriminatory auctions with CARA bidders. We take the utility functions as given and establish nonexistence for every profile of strictly increasing, concave payoffs (with one strictly concave), with independent type distributions whose densities are bounded above and away from zero on the entire space of ordered marginal values.


Further work on single-unit auctions accommodates private information beyond the value of the object. \citet{McAdams2007a} proves uniqueness within the class of monotone pure-strategy equilibria for symmetric risk-averse bidders with affiliated signals and interdependent values.  \citet{GentryLiLu2015} obtain monotone equilibrium under CARA or CRRA utility when values, wealth, and risk preferences are all private information.  \citet{PY2019} establish monotone approximate interim equilibria in asymmetric first-price auctions with affiliated signals and interdependent values, and give conditions for obtaining exact equilibria as limits.  \citet{OlszewskiRenySiegel2026} address equilibrium existence more broadly, allowing mixed strategies, non-quasi-linear utilities, and correlated, possibly atomic private values.\footnote{They also show that modifying tie-breaking at a single bid can restore existence when the sufficient conditions fail.}  Our results concern monotone pure-strategy equilibrium in multiunit auctions.

Our paper also relates to the literature on bidding incentives and auction design in  markets with divisible goods. \citet{Wilson1979} and \citet{BackZender1993} show how strategic bidding can sustain low-price equilibria in uniform-price auctions. \citet{AusubelEtAl2014} study differential bid shading across units and its consequences for efficiency and revenue in both uniform-price and discriminatory auctions. \citet{PyciaWoodward2026} characterize pay-as-bid equilibrium with symmetrically informed, risk-neutral bidders and derive implications for supply and information disclosure. Our focus is on whether equilibrium bids can be nondecreasing as risk-averse bidders' marginal values increase in multiunit auctions.

The remainder of the paper is organized as follows.  Section~\ref{sec:model} presents the model. Sections~\ref{sec:uniform} and~\ref{sec:discriminatory} establish the uniform-price and discriminatory results, respectively, and Section~\ref{sec:conclusion} concludes. Appendix~\ref{app:main-results} contains the general proofs and the two-unit CARA existence result. Appendix~\ref{app:further-results} gives the extensions to all-pay auctions and finite bid grids.

\section{The model}\label{sec:model}

There are $n\geq2$ bidders and $m\geq2$ identical units. Bidder $i$ privately observes her marginal values $t_i=(t_{i1},\ldots,t_{im})$, ordered so that she values earlier units at least as much as later ones. Her type space is
\[
T_m=\{t\in[0,1]^m:t_1\geq\cdots\geq t_m\}.
\]
The value of $q$ units is $V_q(t_i)=\sum_{k=1}^q t_{ik}$, with $V_0=0$. Types are independent across bidders. Each bidder's distribution has a density with respect to $m$-dimensional Lebesgue measure on $T_m$, and the coordinates of her own type may be dependent. We write $U_A$ for the uniform distribution on a set $A$ with positive volume.

The general nonexistence results will hold for distributions satisfying
\[
0<\underline f_i\leq f_i(t)\leq\overline f_i<\infty
\quad\text{for almost every }t\in T_m,\qquad i=1,\ldots,n.
\]
In particular, the distributions have full support.\footnote{These bounds describe the distributions we construct. They are not imposed on every example or on the existence result in the setting with two units.}

Bidder $i$ evaluates surplus $s$ using a utility function $u_i(s)$. Throughout the analysis, we fix utilities satisfying
\begin{equation}\label{eq:utilities}
\begin{gathered}
u_i:\R\longrightarrow\R\text{ is strictly increasing and concave for every }i,\\
u_1\text{ is strictly concave}.
\end{gathered}
\end{equation}
Here, we assume that one bidder is strictly risk averse, while other bidders may be risk neutral. We require no differentiability condition.

Each bidder submits one bid per unit, in nonincreasing order. The feasible bid vectors are
\[
B_m=\{b\in\R_+^m:b_1\geq\cdots\geq b_m\}.
\]
There is no upper bound on bids. The $m$ highest unit bids win. To resolve ties, draw a uniform random ordering of bidders, independently of their types, and use the same ordering at every tied bid. Within a bidder's vector, earlier units receive priority. A bidder who wins $q$ units therefore wins her first $q$ bids. Write $Q_i$ for her quantity. Let $b_{(k)}$ be the $k$th highest bid among all bidders.

The auction formats differ only in what a winner pays.
\begin{itemize}
\item Under \emph{uniform pricing}, every accepted unit is purchased at a common cutoff price: either the highest losing bid, $b_{(m+1)}$, or the lowest winning bid, $b_{(m)}$. We consider both conventions.
\item Under \emph{discriminatory pricing}, the bidder pays each of her accepted bids.
\end{itemize}
Her surplus is
\[
S_i^{\mathrm{UP}}=V_{Q_i}(t_i)-Q_i p
\quad\text{and}\quad
S_i^{\mathrm D}=\sum_{k=1}^{Q_i}(t_{ik}-b_{ik}),
\]
respectively. Bidder $i$'s objective is to maximize expected utility $\E[u_i(S_i)]$.

A \emph{pure-strategy Bayesian equilibrium} consists of measurable bidding strategies $\sigma_i:T_m\to B_m$ that maximize interim expected utility for almost every type $t_i\in T_m$. A \emph{no-overbidding equilibrium} satisfies $\sigma_{ik}(t_i)\leq t_{ik}$ almost everywhere. This restricts equilibrium bids, not the deviations used to test optimality.

We use the conventional coordinatewise order on both values and bids.  A strategy $\sigma_i$ is \emph{monotone} if there is a measurable
set $T_i^*\subseteq T_m$ of probability one such that for $t,t'\in T_i^*$,
\[ t\leq t'
\quad\Longrightarrow\quad
\sigma_i(t)\leq\sigma_i(t').
\]
Thus, a bidder whose marginal values all rise never lowers any unit bid. A \emph{monotone equilibrium} is a pure-strategy Bayesian equilibrium in which every bidder's strategy is monotone.\footnote{Changes on null sets do not affect the definition or our conclusions.} Finally, a bid at a given type is \emph{admissible} if no lottery over feasible bids gives at least as much expected utility against every profile of opponents’ bids and strictly more against some such profile. An equilibrium is admissible if the bids are admissible for almost every type.

\section{Uniform-price auctions}\label{sec:uniform}

We first establish universal nonexistence of admissible monotone pure-strategy equilibrium under uniform pricing.

\begin{theorem}[Universal nonexistence under uniform pricing]\label{thm:uniform}
Fix $n\geq2$ and $m\geq3$. There are independent type distributions under which neither highest-losing-bid nor lowest-winning-bid uniform pricing admits an admissible monotone equilibrium.
\end{theorem}

\begin{remark}\label{rem:uniform_prices}
We indeed prove a stronger result. We construct one collection of densities under which the universal nonexistence result holds for any price
\[
p_\alpha=(1-\alpha)b_{(m)}+\alpha b_{(m+1)}
\]
and every $\alpha\in[0,1]$. The price $p_\alpha$ corresponds to the highest losing bid when $\alpha=1$, and lowest winning bid when $\alpha=0$.\footnote{As pointed out by \citet{McAdams2006}, ``The Paris and Amsterdam exchanges also open trading with batch auctions, which may be thought of, at least approximately, as uniform $S+1/2$-th price auctions'', which corresponds to the case $\alpha=1/2$.}
\end{remark}

Fix a type $t$ and a feasible bid vector $b$. Cap each bid at its corresponding marginal value: $\widehat b_k=\min\{b_k,t_k\}$, for $k=1,\ldots,m$. The capped bids remain nonnegative and weakly decreasing. The following lemma will be useful.

\begin{lemma}\label{lem:caps}
For any profile of rival bids, bidding $\widehat b$ yields nonnegative surplus, which is at least as much surplus as bidding $b$.

Thus, every admissible bid vector satisfies $b_k\leq t_k$ for all $k$. Under highest-losing-bid pricing, $b_1=t_1$ by admissibility.
\end{lemma}

When a bidder caps each bid at its marginal value, she weakly reduces the common price. Any units she loses as a consequence could not have contributed positive surplus at the original price. Under highest-losing-bid pricing, bidding the first marginal value also weakly dominates a lower first bid: it can add a profitable first unit without raising the payment when that unit was already won.

To illustrate Theorem~\ref{thm:uniform}, we present an example with two bidders and three units under highest-losing-bid pricing.

\begin{example}\label{exam:uniform}
We call bidder 1's rival weak when she has one small positive marginal value and strong when she has two high marginal values. A weak rival leaves bidder 1 a choice between two units at a negligible price and all three units at a small positive price. A strong rival seeks two units, so bidder 1's bid for the third unit can raise the price of the one unit she retains.

We start with two benchmark types for bidder 1,
\[
t^L=(x,y_L,z),\qquad t^H=(x,y_H,z),
\qquad 0<z<y_L<y_H<x<\tfrac12.
\]
Only the second marginal value changes. Write $v_L=x+y_L$ and $v_H=x+y_H$ for the values of the first two units. Since bidder 1 is strictly risk averse,
\[
u_1(v_L+z)-u_1(v_L)>u_1(v_H+z)-u_1(v_H).
\]
Both types value the third unit at $z$, but the higher type attaches a smaller utility gain to that unit.

We choose the weak rival's first marginal value to be a small number $c>0$, with her other values close to zero. By Lemma~\ref{lem:caps}, her first bid is close to $c$ and her other bids are close to zero. Bidder 1 can leave one unit to this rival and obtain two units at a negligible price. Alternatively, to secure all three, she must bid above the rival's first bid for her third unit. The rival's first bid then sets the price, so bidder 1 pays approximately $c$ on each of the three units.

We use this observation to bound the incentives for bidder 1's two types. For the low type, we allow a total payment of $3c$ for three units and compare this outcome with two units obtained for free. For the high type, we ignore payments altogether, giving an upper bound on the gain from acquiring the third unit. Choose $c>0$ with $3c<z$, sufficiently small that
\[
A_L=u_1(v_L+z-3c)-u_1(v_L)>A_H:=u_1(v_H+z)-u_1(v_H)>0.
\]
Such a $c$ exists by continuity. Thus, the low type's gain remains larger even after allowing for the payment on all three units.

We choose the strong rival's values so that she must win two units in any candidate admissible equilibrium, as verified below. Bidder 1 then receives one unit, valued at $x$ by both benchmark types. If her third bid is at least $c$, her second bid must also be at least $c$. That second bid loses and makes the price of her winning unit at least $c$. The utility cost of increasing the payment on this unit from zero to $c$ is
\[
C=u_1(x)-u_1(x-c)>0.
\]
This comparison is the same for both types because their first marginal values are identical.

In the limiting comparison, for a low type whose third bid is below $c$, submitting bids just above $c$ for both later units yields a gain of at least $A_L$ against the weak rival and a loss of at most $C$ against the strong rival. For a high type whose third bid is at least $c$, lowering both later bids toward zero saves at least $C$ against the strong rival and sacrifices at most $A_H$ against the weak rival. Choose the probability $\delta$ of a strong rival so that
\[
(1-\delta)A_H<\delta C<(1-\delta)A_L.
\]
Both deviations are then strictly profitable whenever the corresponding bid restriction is violated. Optimality requires low types to bid at least $c$ for the third unit and high types to bid below $c$. The strict inequalities allow these opposing incentives to survive in small neighborhoods of the benchmark types, yielding the contradiction to monotonicity.

We next place this comparison inside an auction in which both bidders choose their bids optimally. Choose two rival values satisfying
\[
\max\{x,2y_H\}<d_2<d_1<1.
\]
For a sufficiently small $\varepsilon>0$, let
\[
\begin{aligned}
L&=[x,x+\varepsilon]\times[y_L,y_L+\varepsilon]\times[z,z+\varepsilon],\\
H&=[x+2\varepsilon,x+3\varepsilon]\times[y_H,y_H+\varepsilon]\times[z+2\varepsilon,z+3\varepsilon],\\
C_0&=\{t:c\leq t_1\leq c+\varepsilon,\ 0\leq t_3\leq t_2\leq\varepsilon\},\\
D&=[d_1,d_1+\varepsilon]\times[d_2,d_2+\varepsilon]\times[0,\varepsilon].
\end{aligned}
\]
All four regions have positive three-dimensional volume. Every type in $H$ strictly exceeds every type in $L$ in every coordinate. Independently across bidders, set
\[
\mu_1=\tfrac12 U_L+\tfrac12 U_H,\qquad
\mu_2=(1-\delta)U_{C_0}+\delta U_D.
\]
The weak rival lies in $C_0$ and the strong rival in $D$. We choose $\varepsilon$ small enough to preserve all value separations and the strict utility comparisons above. In particular,
\[
c+2\varepsilon<z,\quad x+3\varepsilon<d_2,\quad\text{and}\quad 2(y_H+\varepsilon)<d_2.
\]
\end{example}

\begin{proposition}\label{prop:uniform_example}
For all sufficiently small $\varepsilon>0$, the auction in Example~\ref{exam:uniform} has no admissible pure equilibrium in which bidder 1's third bid is coordinatewise nondecreasing.
\end{proposition}

\begin{proof}
Suppose that such an equilibrium exists. By Lemma~\ref{lem:caps}, the first bid equals the first value and no bid exceeds its marginal value. Bidder 1 always wins at least one unit: her first bid is at least $x$, while the rival's third bid is at most $\varepsilon$.

We first show why the strong rival must win two units. She can raise her first two bids to at least $R=x+3\varepsilon$, leaving her third bid unchanged. She then wins exactly two units and pays at most $s=y_H+\varepsilon$ per unit. If she already won two, this increase in winning bids leaves the price unchanged. If she previously won at most one, the increase in surplus is at least $d_2-2s>0$. Since her utility is strictly increasing, she wins two units almost surely. Since types are independent among bidders, almost every type of bidder 1 wins exactly one unit against almost every strong rival.

Consider a low type whose third bid is below $c$. Against the weak rival she wins at most two units, because the rival's truthful first bid is at least $c$. Against the strong rival she wins exactly one. Now let her deviate to
\[
d^L=(1,c+2\varepsilon,c+2\varepsilon).
\]
She wins all three units against the weak rival at a price no greater than $c+\varepsilon$. Against the strong rival she wins at least her first unit at a price no greater than $c+2\varepsilon$. Any additional units she obtains have values above that price.

As the type regions shrink to their benchmark values, the resulting lower bound on her expected utility gain converges to
\[
(1-\delta)A_L-\delta C>0.
\]
By continuity, the deviation is strictly profitable for all sufficiently small $\varepsilon$. The convergence is uniform over low types and candidate equilibrium bids, since the payoff bounds depend only on the type regions and the price bounds. Hence, almost every low type must bid at least $c$ for the third unit.

Next, consider a high type whose third bid is at least $c$. Against the weak rival, her surplus cannot exceed the total value of all three units. Against the strong rival she wins exactly one unit. Her losing second bid is no lower than her third bid, so she pays at least $c$ for that unit. Let her instead submit $d^H=(1,2\varepsilon,2\varepsilon)$. She obtains at least two units against the weak rival and at least one unit against the strong rival, in both cases at a price no greater than $2\varepsilon$. Any additional units contribute positive surplus. As the type regions shrink to their benchmark values, the lower bound on her expected utility gain converges to
\[
\delta C-(1-\delta)A_H>0.
\]
The convergence is uniform over high types and candidate equilibrium bids, since the payoff bounds depend only on the type regions and the stated price bounds.

We chose $\delta$ so that
\[
(1-\delta)A_H<\delta C<(1-\delta)A_L.
\]
Both limiting gains are strictly positive. By continuity, we can choose a single $\varepsilon>0$ small enough that the two deviations remain profitable for all types under consideration. By the optimality requirement, we have that
\[
\sigma_{13}(t)\geq c\quad U_L\text{-almost everywhere},\qquad
\sigma_{13}(t)<c\quad U_H\text{-almost everywhere}.
\]
Every type in $H$ strictly exceeds every type in $L$ in each coordinate. These inequalities contradict monotonicity.
\end{proof}

In the example above, the bidder reduces her willingness to compete because her second value rises, not because her third value falls. Indeed, in the perturbed regions the high type also has strictly higher first and third values. These small increases cannot offset the strict incentive reversal.

\begin{remark}\label{rem:uniform_extension}
The three-unit argument separates three roles: a first unit that bidder 1 retains against a strong rival, an intermediate unit whose higher value raises her surplus against a weak rival, and a final unit for which she may compete less aggressively. With more units, the same comparison can be arranged. Additional bidders can be assigned values sufficiently small that they do not overturn the strict incentives. In the proof of Theorem~\ref{thm:uniform}, we provide this construction for both highest-losing-bid and lowest-winning-bid uniform pricing in the general case.
\end{remark}

\section{Discriminatory auctions}\label{sec:discriminatory}

Under discriminatory pricing, a bidder pays her own winning bids. Therefore, a high bid for a later unit commits her to paying at least as much for each earlier unit she obtains. This payment effect can prevent monotone equilibrium even without an admissibility restriction.

\begin{theorem}[Universal nonexistence under discriminatory pricing]\label{thm:discriminatory}
Fix $n\geq2$. There is an integer $m\geq3$ and independent type distributions under which the discriminatory auction has no monotone pure-strategy equilibrium.\footnote{In this theorem, we fix the utilities before choosing the supply and type distributions.}
\end{theorem}

The following example illustrates the nonexistence result in Theorem~\ref{thm:discriminatory} in the setting with two CARA bidders and three units.

\begin{example}[A three-unit CARA example]\label{exam:discriminatory}
There are two bidders, both with utility $u(s)=1-e^{-10s}$. Bidder 1's values are concentrated near two ordered vectors. Her high types value the second unit much more than her low types do, while their third values are nearly the same. Her rival is usually weak, with one modest positive marginal value. With a small probability, the rival is strong and values two units close to one.

To secure all three units against the weak rival, bidder 1 must outbid her for the third unit. Her first two bids must be at least as high as her third. If the strong rival wins two units, bidder 1 receives only one unit and pays her first bid. Thus, a high third bid also places a lower bound on her payment when she wins only one unit. A rise in the second marginal value adds surplus only when at least two units are won. Even under CARA utility, it can reduce the utility gain from competing for a third unit relative to the cost of a higher payment when only one unit is won.

Bidder 1 can buy all three units cheaply against the weak rival. A high second bid would sacrifice too much surplus in this (frequent) event. Hence, the strong rival can secure two units at low cost. To earn at least the payoff available from this deviation, the strong rival must win at least two units sufficiently often. At this frequency, some high types of bidder 1 prefer to lower their bids and save on the payment when they win only one unit, even at the cost of giving up the third unit against the weak rival. Monotonicity would require low types to submit equally low or lower third bids.

The weak rival can bid above all these low third bids while still paying less than her first marginal value. This gives her a positive payoff against low types, which her equilibrium strategy must match in expectation. For any given bid, the weak rival is at least as likely to win against a low type as against a high type, because the low type's third bid cannot be higher. Hence, earning the payoff available from the deviation requires the weak rival to win against low types sufficiently often. At that frequency, some low types of bidder 1 would prefer to raise their bids and secure all three units against the weak rival. This contradicts the low third bids required by monotonicity.

To be concrete, set $\eta=10^{-6}$, $\alpha=1/1000$, and $\delta=1/5000$. Types are independent across bidders and uniform within the regions in Table~\ref{tab:discriminatory}, with the indicated probabilities. The small low-value component $E$ allows us to derive the required bid bounds from equilibrium optimality.

\begin{table}[htbp]
\centering
\caption{Type regions in the discriminatory example}\label{tab:discriminatory}
\small
\setlength{\tabcolsep}{6pt}
\begin{tabular}{@{}llcccc@{}}
\toprule
Bidder & Region & First value & Second value & Third value & Probability\\
\midrule
1 & $L$ & $[.800,.801]$ & $[.520,.521]$ & $[.470,.471]$ & $(1-\alpha)/2$\\
1 & $H$ & $[.995,.996]$ & $[.990,.991]$ & $[.472,.473]$ & $(1-\alpha)/2$\\
1 & $E$ & $[0,\eta]$ & $[0,\eta]$ & $[0,\eta]$ & $\alpha$\\
\addlinespace
2 & $C$ & $[.069,.070]$ & $[.0001,.0002]$ & $[0,\eta]$ & $1-\delta$\\
2 & $D$ & $[.999,1]$ & $[.999,1]$ & $[0,\eta]$ & $\delta$\\
\bottomrule
\end{tabular}
\end{table}

Each region is intersected with $T_3$. Every type in $H$ strictly exceeds every type in $L$ in each coordinate. All regions have positive three-dimensional volume, but the distributions do not have full support on $T_3$.
\end{example}

\begin{proposition}\label{prop:discriminatory_example}
The auction in Example~\ref{exam:discriminatory} has no pure-strategy Bayesian equilibrium in which bidder 1's third bid is coordinatewise nondecreasing.
\end{proposition}

\begin{remark}
Proposition~\ref{prop:discriminatory_example} holds with unrestricted nonnegative bids or with bids bounded by one. No admissibility or no-overbidding restriction is imposed.
\end{remark}

\begin{proof}
Suppose that such an equilibrium exists.

We first derive the bid bounds. Capping discriminatory bids at their marginal values weakly increases surplus in every outcome and strictly increases it whenever an overbid wins. Thus, winning overbids have probability zero in equilibrium. The low-value component $E$ rules out every overbid by bidder 2. Indeed, any positive overbid faces a positive probability that all of bidder 1's values lie below it. In that event, either the overbid wins or bidder 1 wins with a bid that exceeds her value. Both possibilities contradict the observation above.

Thus, bidder 2's third bid is at most $\eta$. An $E$ type of bidder 1 cannot bid above $\eta$, since her first bid would win at a loss. An overbid by a type in $L\cup H$ would exceed $0.470$ and win against the weak rival, whose bids are at most $0.070$. Hence, bidder 1 does not overbid on $L\cup H$.

For a type $t=(x,y,z)$ of bidder 1, define
\[
J_t(b)=\E\!\left[e^{-10(S_1(t,b,\sigma_2(t_2))-x)}\right].
\]
Maximizing expected CARA utility is equivalent to minimizing $J_t$. This normalization lets us express the following bounds in terms of surplus relative to the first value $x$.

The bid $(0.07,0.07,0.07)$ wins all three units against $C$ almost surely and at least one unit against $D$. It yields surplus relative to $x$ of at least $0.78$ in the former event and at least $-0.07$ in the latter. Therefore, every equilibrium bid for a type in $L\cup H$ must satisfy
\[
J_t(\sigma_1(t))\leq K:=e^{-7.8}+\delta e^{0.7}.
\]
If her second bid were at least $0.38$, she would win at least two units against $C$ and pay at least $0.76$ for the first two. Since $y+z\leq1.464$, her surplus relative to $x$ would be at most $0.704$, whether she won two units or three. This would imply
\[
J_t(\sigma_1(t))\geq(1-\delta)e^{-7.04}>K.
\]
Therefore, her second bid must be below $0.38$ almost everywhere on $L\cup H$.

A strong rival can bid $(0.381,0.381,0)$ and win at least two units almost surely. This guarantees surplus of at least $2(0.999-0.381)=1.236$. Let $p_D$ be this rival type's equilibrium probability of winning at least two units. Her surplus is at most one when she wins at most one unit, and never exceeds $2+\eta$. Her expected utility in equilibrium cannot exceed $(1-p_D)u(1)+p_Du(2+\eta)$. She would prefer the bid $(0.381,0.381,0)$ if this upper bound were below $u(1.236)$. Hence,
\[
p_D\geq\frac{u(1.236)-u(1)}{u(2+\eta)-u(1)}.
\]
Even if she won two units against every low type and every $E$ type, this bound would still require
\[
\Prb(Q_2\geq2\mid t_1\in H,\ t_2\in D)>\frac45.
\]

Suppose bidder 1's third bid were at least $0.05$ almost everywhere on $H$. When the strong rival wins at least two units, bidder 1 either wins one unit and pays at least $0.05$, or wins nothing. In either case, her surplus relative to $x$ is at most $-0.05$. Averaging over high types gives
\[
\E_{t\sim U_H}J_t(\sigma_1(t))>\delta\frac45 e^{0.5}.
\]
A high type can instead bid $(0.001,0.001,0)$. She wins at least two units against $C$ and at least one against $D$, with surplus relative to $x$ of at least $0.988$ and $-0.001$, respectively. Thus,
\[
J_t((0.001,0.001,0))\leq e^{-9.88}+\delta e^{0.01}<\delta\frac45 e^{0.5}.
\]
Comparing this uniform bound with the preceding average shows that a positive-measure set of high types would strictly gain by submitting this bid. Hence, a positive-measure subset of $H$ must have third bids below $0.05$.

Every high type values each unit more than every low type. Under the assumed strategy, bidder 1 cannot submit a higher third bid at a low type than at a high type. Hence, almost every low type also bids below $0.05$ for the third unit. A weak rival can bid $(0.051,0,0)$ and win a unit against almost every low type, obtaining surplus of at least $0.018$. Her surplus is nonnegative in every other event. Put $r=(1-\alpha)/2$, and let $p_L$ and $p_H$ be this rival type's equilibrium probabilities of winning at least one unit conditional on $L$ and $H$. For any given first bid, the weak rival is at least as likely to win a unit against a low type as against a high type, because the low type submits no higher third bid. Hence, $p_H\leq p_L$. By bidding $(0.051,0,0)$, the weak rival can obtain expected utility of at least $r u(0.018)$. In equilibrium, she earns zero surplus if she wins nothing and at most $0.070201$ if she wins. She would prefer the deviation unless
\[
r u(0.018)\leq(rp_L+rp_H+\alpha)u(0.070201)\leq\big((1-\alpha)p_L+\alpha\big)u(0.070201).
\]
It follows that
\[
p_L\geq\frac{r u(0.018)/u(0.070201)-\alpha}{1-\alpha}>\frac4{25}.
\]
Whenever the weak rival wins a unit, bidder 1 obtains at most two. For a low type, her surplus relative to $x$ is then at most $y\leq0.521$. Therefore, averaging over low types gives
\[
\E_{t\sim U_L}J_t(\sigma_1(t))\geq(1-\delta)\frac4{25}e^{-5.21}>K.
\]
This contradicts the upper bound given by the bid $(0.07,0.07,0.07)$.
\end{proof}

\begin{remark}
The proof for Theorem~\ref{thm:discriminatory} uses two constructions. If bidder 1's utility on nonnegative surplus is not exponential, we construct a choice between a certain surplus and a lottery over two surpluses. She reverses her preference when the total value of the units she wins under both choices increases. Another bidder's optimal behavior generates the opposing bids needed for this comparison. For exponential utility, we use blocks of units to separate the surplus earned against a weak rival from the payments incurred against a strong one. The block sizes depend on the given utility.

By assigning small probabilities to values near zero, we rule out overbidding in equilibrium. We also show that adding a sufficiently small uniform component to each type distribution preserves nonexistence.
\end{remark}

\section{Conclusion}\label{sec:conclusion}

We establish two universal nonexistence results for monotone pure-strategy equilibrium in multiunit auctions. The results apply to any number of bidders greater than one and any given profile of strictly increasing, concave utilities, provided at least one bidder is strictly risk averse. Under uniform pricing, for every supply of at least three units, we construct independent type distributions under which neither standard pricing rule admits an admissible monotone equilibrium. Under discriminatory pricing, no admissibility restriction is needed. In both cases, the constructed densities are bounded above and away from zero on the entire type space.

The two-unit case helps explain why the third unit matters. With CARA utilities, both uniform-price rules and discriminatory pricing with two units admit a monotone no-overbidding pure-strategy equilibrium for any number of bidders and any independent type densities. The CARA coefficients may differ across bidders and may be zero. Under highest-losing-bid pricing, equilibrium strategies can be chosen in the form
\[
\sigma_i(x,y)=(x,\beta_i(y)),\qquad 0\leq\beta_i(y)\leq y,
\]
with $\beta_i$ nondecreasing. When the first bid equals the first value, CARA preferences make the ranking of second bids independent of that value. With three units, a higher second value can instead increase surplus when two or three units are won without changing surplus when only one is won. The additional surplus from a third unit then yields less utility, while the utility cost of paying more for the first unit remains unchanged. Our examples show how this difference can reverse bidding incentives even under CARA preferences.  Appendix~\ref{app:main-results} proves the two-unit existence result. We also establish in Appendix~\ref{app:further-results} an analogous nonexistence result for all-pay auctions, and provide results on finite bid grids, showing that discrete bidding does not by itself restore monotone equilibrium.

Our nonexistence results show that even independent private values and regular full-support densities do not ensure admissible monotone equilibrium under uniform pricing or monotone equilibrium under discriminatory pricing. These results suggest that to study bidding or compare auction formats when bidders are risk averse, one may have to allow for bidding behavior that is nonmonotone in marginal values.

\clearpage
\appendix
\raggedbottom
\displaywidowpenalty=10000
\makeatletter
\let\appendix@originalseccntformat\@seccntformat
\newcommand{\appendix@format@section}{}
\renewcommand{\@seccntformat}[1]{%
  \ifcsname appendix@format@#1\endcsname
    \csname appendix@format@#1\endcsname
  \else
    \appendix@originalseccntformat{#1}%
  \fi}
\makeatother
\section{Appendix A}\label{app:main-results}

\subsection{Proof of Theorem~\ref*{thm:uniform}}\label{app:uniform-proof}
Under highest-losing-bid pricing with quasilinear utility, the no-overbidding and first-bid implications appear in \citet[Lemmas 1 and 2]{MarkakisTelelis2012}. We give the proof for the utilities in \textup{(\ref{eq:utilities})} and the two uniform pricing rules for completeness.

\begin{proof}[Proof of Lemma~\ref{lem:caps}]
Fix a bidder of type \(t_i\). Cap each bid at its marginal value, so that \(\widehat b_{ij}=\min\{b_{ij},t_{ij}\}\). This vector remains ordered. Fix the opposing bids and the priority. Lowering own bids weakly reduces both the quantity and the price. If own unit \(j\) is lost, its bid must have fallen to \(t_{ij}\). Indeed, acceptance of an unchanged bid depends only on its rank among opposing bids and its own index. The \((m-j+1)\)st highest opposing bid now precedes unit \(j\), so it is at least \(t_{ij}\). This opposing bid was originally rejected because own unit \(j\) originally won. Hence \(t_{ij}\le b_{(m+1)}\le p\), where \(p\) is the original uniform price. The lost unit therefore had nonpositive surplus at the original price. Each retained unit costs weakly less, so total surplus weakly increases. The capped surplus is nonnegative because every accepted capped bid is at most its marginal value.

If \(b_{ij}>t_{ij}\) for some \(j\), choose \(v\in(t_{ij},b_{ij})\). Let one opponent bid \(v\) on every unit and all other opponents bid zero. Unit \(j\) originally wins at a price of at least \(v\), since there are at least \(m+1\) bids at or above \(v\). It loses after capping. Its original marginal surplus was strictly negative, so total surplus strictly increases for every priority. Strictly increasing utility preserves the comparisons, so admissibility rules out overbidding.

For the first-bid implication in Section~\ref{sec:uniform}, suppose the price is the highest losing bid and \(b_{i1}<t_{i1}\). Raise only the first bid to \(t_{i1}\). If it was already accepted, quantity and price are unchanged. Otherwise, only the first bid can become accepted and any resulting surplus is nonnegative. To obtain strict improvement, let one rival bid \(v\in(b_{i1},t_{i1})\) on all units and let all other rival bids be zero. The bidder goes from zero units to one at price \(v\), so her surplus increases from zero to \(t_{i1}-v>0\) for every priority.
\end{proof}

\begin{proof}[Proof of Theorem~\ref{thm:uniform}]
Use the auction in Section~\ref{sec:model} with the two uniform prices in Section~\ref{sec:uniform}, \(p_0=b_{(m)}\) and \(p_1=b_{(m+1)}\). For fixed rival bids and priority, a bidder's \(j\)th bid wins if at most \(m-j\) rival bids precede it. Thus changing other own bids within the ordered bid space leaves the acceptance of an unchanged bid unaffected.

By Lemma~\ref{lem:caps}, it suffices to exclude pure equilibria with no overbidding in which bidder~1's last bid is coordinatewise nondecreasing. We refer to these as candidate equilibria.

\pstep{1} We construct the type distributions. Put \(x=0.40\), \(y_L=0.60/m\), \(y_H=0.90/m\), and \(z=0.39/m\). Consider parameters satisfying
\begin{equation}\label{eq:parameters}
0<c<\min\{0.01/m,z/(4m)\},\qquad
0<\delta<1,\qquad
0<\varepsilon<\min\{c/20,0.001/m^2\}.
\end{equation}
For integers \(r\ge2\), let \(\rho_r=\varepsilon/r\) and \(\tau_r=1/r\). Write \(\mathcal T_j[A,B]=\{w\in\R^j\mid B\ge w_1\ge\cdots\ge w_j\ge A\}\). The type regions are
\begin{equation}\label{eq:regions}
\begin{aligned}
&L_m= [x,x+\varepsilon]\times[y_L,y_L+\varepsilon]\times \mathcal T_{m-2}[z,z+\varepsilon],\\
&H_m= [x+2\varepsilon,x+3\varepsilon]\times[y_H,y_H+\varepsilon]\times \mathcal T_{m-2}[z+2\varepsilon,z+3\varepsilon],\\
&C_m(r)= [c,c+\rho_r]\times \mathcal T_{m-1}[0,\rho_r],\\
&D_m(r)= [0.95,0.95+\rho_r]\times \mathcal T_{m-2}[0.90,0.90+\rho_r]\times[0,\rho_r],\\
&E_m(r)= \mathcal T_m[0,\rho_r].
\end{aligned}
\end{equation}
All regions have positive volume and lie in \(T_m\). Every high type strictly exceeds every low type coordinatewise. Every marginal value in these two regions is at least \(z\).

Let \(v_L=x+y_L+(m-3)z\) and \(v_H=x+y_H+(m-3)z\). We will choose \(c,\delta\) to satisfy
\begin{equation}\label{eq:delta_requirement}
\begin{aligned}
(1-\delta)[u_1(v_H+z)-u_1(v_H)]
&<\delta[u_1(x)-u_1(x-c)]\\
&<(1-\delta)[u_1(v_L+z-mc)-u_1(v_L)].
\end{aligned}
\end{equation}
Strict concavity gives
\(u_1(v_L+z)-u_1(v_L)>u_1(v_H+z)-u_1(v_H)\), since a fixed positive surplus increment is worth strictly less at a higher initial surplus.\footnote{For \(a<b\) and \(d>0\), the points \(a+d\) and \(b\) are strict convex combinations of \(a\) and \(b+d\) with complementary weights. Adding the two strict concavity inequalities gives \(u(a+d)+u(b)>u(a)+u(b+d)\).}
Choose \(c>0\) within the range in \eqref{eq:parameters} small enough that \(A_L>A_H\), where
\[
A_L=u_1(v_L+z-mc)-u_1(v_L),
\qquad A_H=u_1(v_H+z)-u_1(v_H)>0.
\]
Such a choice exists by continuity. Set \(C_u=u_1(x)-u_1(x-c)>0\) and \(\delta=(A_L+A_H)/(2C_u+A_L+A_H)\). Then \(0<\delta<1\) and
\begin{equation*}
(1-\delta)A_L-\delta C_u
=\delta C_u-(1-\delta)A_H
=\frac{1-\delta}{2}(A_L-A_H)=:\gamma>0.
\end{equation*}

This verifies \eqref{eq:delta_requirement}.
The basic type distributions are \(\mu_1^r=\tfrac12 U_{L_m}+\tfrac12 U_{H_m}\) for bidder~1, \(\mu_2^r=(1-\delta)U_{C_m(r)}+\delta U_{D_m(r)}\) for bidder~2, and \(\mu_i^r=U_{E_m(r)}\) for every bidder \(i\ge3\).
In the auction indexed by \(r\), types are drawn independently from
\(\widehat\mu_i^r=(1-\tau_r)\mu_i^r+\tau_r U_{T_m}\). Since \(T_m\) has volume \(1/m!\), if \(g_i^r\) denotes the basic density, the actual density satisfies
\[
\tau_rm!\le\widehat f_i^r=(1-\tau_r)g_i^r+\tau_rm!
\le(1-\tau_r)\lVert g_i^r\rVert_\infty+\tau_rm!
\quad\text{almost everywhere}.
\]
These independent distributions have full support and satisfy the density bounds in Section~\ref{sec:model} at every finite \(r\). They are the same for both pricing rules.

\pstep{2}
Steps~2--5 apply to every fixed \(\varepsilon\) satisfying \eqref{eq:parameters}, with \(c,\delta\) chosen in Step~1. We choose \(\varepsilon\) sufficiently small in Step~6. Throughout Steps~2--5, \(\varepsilon\) is fixed as \(r\) increases. Consider a pricing rule and an index \(r\) at which a candidate equilibrium \(\sigma^r=(\sigma_i^r)_{i=1}^n\) exists. We may impose no overbidding everywhere by modifying strategies on own-type null sets. Bidder~1's last bid remains nondecreasing on a full-measure set of her types. At the realized type \(t_i^r\), write \(b_i^r=\sigma_i^r(t_i^r)\) for the equilibrium bid vector, with coordinates \(b_{ij}^r=\sigma_{ij}^r(t_i^r)\), and put \(\beta_r=b_{1m}^r\). Let \(Q_i^r\) be bidder \(i\)'s equilibrium quantity and \(p^r\) the common price. Write \(S_i^r=V_{Q_i^r}(t_i^r)-Q_i^rp^r\) for her uniform-price surplus \(S_i^{\mathrm{UP}}\) at these outcomes. A superscript \(r,d\) gives the corresponding outcome when bidder \(i\) uses \(d_i^r(t_i^r)\) and her rivals keep their equilibrium bids, with the same tie priority. Write \(\Delta S_i^{r,d}=S_i^{r,d}-S_i^r\) and \(H_i^{r,d}=u_i(S_i^{r,d})-u_i(S_i^r)\).

All deviations below satisfy the no-overbidding bounds. Since the price is nonnegative and no higher than any winning bid, both original and deviating surpluses lie in \([0,m]\). Put \(W_i=u_i(m)-u_i(0)\), \(\ell_i=u'_{i,+}(m)\), and \(L_i=u'_{i,-}(0)\). Then
\begin{equation}\label{eq:derivatives}
0<\ell_i\le L_i<\infty,\qquad
\ell_i(y-x)\le u_i(y)-u_i(x)\le L_i(y-x)
\quad(0\le x\le y\le m).
\end{equation}
If \(\Delta S_i^{r,d}\ge0\), \eqref{eq:derivatives} gives \(H_i^{r,d}\ge\ell_i\Delta S_i^{r,d}\). If \(\Delta S_i^{r,d}<0\), applying the upper slope bound to the utility loss gives \(H_i^{r,d}\ge L_i\Delta S_i^{r,d}\). Consequently,
\[
\Delta S_i^{r,d}\ge g-h,\quad g,h\ge0
\quad\Longrightarrow\quad
H_i^{r,d}\ge\ell_i g-L_i h.
\]
Indeed, for a nonnegative surplus change,
\(H_i^{r,d}\ge\ell_i(g-h)\ge\ell_i g-L_i h\).
For a negative surplus change,
\(H_i^{r,d}\ge L_i(g-h)\ge\ell_i g-L_i h\).\footnote{Finite concavity on \(\R\) gives finite one-sided derivatives and bounds secant slopes between the right derivative at \(m\) and the left derivative at \(0\). Strict increase gives \(u'_{i,+}(m)\ge u_i(m+1)-u_i(m)>0\).}

Write \(\E^r,\Prb^r\) for expectation and probability when each \(t_j^r\) is drawn independently from \(\mu_j^r\), with the independent priority permutation \(\pi\) from Section~\ref{sec:model}. A subscript \(L\) or \(H\) replaces bidder~1's mixture by \(U_{L_m}\) or \(U_{H_m}\). A subscript \(C\) or \(D\) replaces bidder~2's mixture by \(U_{C_m(r)}\) or \(U_{D_m(r)}\). For example, \(\E_D^r\) averages over bidder~2's uniform draw from \(D_m(r)\), the other bidders' basic draws, and priority. Under the basic distributions we call bidder~2's two components states \(C\) and \(D\).

Fix a type \(t_i\) at which interim optimality holds. With probability \(w_r=(1-\tau_r)^{n-1}\), all opponents' types are drawn from their basic components. Denote the deviation's expected utility gain against the basic opponents by \(G_i^r(t_i)\). On the remaining mixture component the expected gain is at least \(-W_i\). Interim optimality therefore gives \(w_rG_i^r(t_i)-(1-w_r)W_i\le0\). Integrating the resulting bound against any normalized own type distribution \(\nu_i\) absolutely continuous with respect to \(\widehat\mu_i^r\) yields
\begin{equation}\label{eq:obedience}
\E_{\nu_i,\mu_{-i}^r,\pi}H_i^{r,d}\le e_{ir},
\qquad e_{ir}=W_i[(1-\tau_r)^{-(n-1)}-1]=O(r^{-1}).
\end{equation}
The expectation in \eqref{eq:obedience} first averages over rival types drawn from \(\mu_{-i}^r=\bigotimes_{j\ne i}\mu_j^r\) and over priority, conditional on bidder \(i\)'s type. It then averages over that type using \(\nu_i\). For example, \(\nu_2=U_{D_m(r)}\) gives \(\E_D^r\) for bidder~2. With \(\nu_1=U_{L_m}\) it gives \(\E_L^r\) for bidder~1, so the probabilities of states \(C\) and \(D\) are still \(1-\delta\) and \(\delta\).

Let \(K_r^D\) be the probability that bidder~1 wins at least two units in state \(D\). Let \(K_r^C(a)\) be the probability that she wins all \(m\) units with a last bid at most \(a\) in state \(C\). Thus
\begin{equation*}
K_r^D=\Prb_D^r(Q_1^r\ge2),\qquad
K_r^C(a)=\Prb_C^r(Q_1^r=m,\ \beta_r\le a),\qquad 0<a<c.
\end{equation*}
\pstep{3}
Suppose \(\alpha=1\). We use two deviations by bidder 2. Her deviation on \(D_m(r)\) bounds \(K_r^D\), and her deviation on \(C_m(r)\) bounds \(K_r^C(a)\). Let \(R_*=\tfrac12(0.90/m+0.90/(m-1))\) and \(\kappa=0.90-(m-1)R_*=0.45/m\). For bidder 2 in \(D_m(r)=[0.95,0.95+\rho_r]\times \mathcal T_{m-2}[0.90,0.90+\rho_r]\times[0,\rho_r]\), raise the first \(m-1\) bids to at least \(R_*\). Thus
\begin{equation}\label{eq:expensivestrategy}
d_{2j}^{r,D}(t_2)=
\begin{cases}
\max\{\sigma_{2j}^r(t_2),R_*\},&t_2\in D_m(r),\ j\le m-1,\\
\sigma_{2j}^r(t_2),&\text{otherwise}.
\end{cases}
\end{equation}
The deviation preserves order and the no-overbidding bounds. Under the basic opponent distributions, bidder 1 has \(b_{12}^r\le y_H+\varepsilon<R_*\) and each extra bidder has all bids below \(R_*\). Thus bidder 2 wins at least \(m-1\) units. Acceptance of her unchanged last coordinate is preserved by the rank observation above.

If \(Q_2^r\le m-2\), her original \((m-1)\)st bid lost to at least two opposing bids, so it is below \(R_*\). This bid now equals \(R_*\) and is accepted. Hence \(Q_2^{r,D}=m-1\) and \(p^{r,D}\le R_*\). Writing \(q=Q_2^r\), her original surplus is \(V_q(t_2^r)-qp^r\le V_{m-2}(t_2^r)\). Her new surplus is \(V_{m-1}(t_2^r)-(m-1)p^{r,D}\). Subtracting gives
\begin{equation}\label{eq:forcinggain}
\begin{aligned}
\Delta S_2^{r,D}
&=V_{m-1}(t_2^r)-(m-1)p^{r,D}-[V_q(t_2^r)-qp^r]\\
&\ge V_{m-1}(t_2^r)-(m-1)R_*-V_{m-2}(t_2^r)\\
&=t_{2,m-1}^r-(m-1)R_*\ge\kappa.
\end{aligned}
\end{equation}
This comparison holds under both pricing rules. When \(Q_2^r\ge m-1\), the raised bids were already accepted, so the quantity and the highest losing bid are unchanged. The surplus gain is zero. By \eqref{eq:derivatives}, \(H_2^{r,D}\ge\ell_2\kappa\) on \(\{Q_2^r\le m-2\}\), with zero gain elsewhere in state \(D\). Since \(Q_1^r\ge2\) implies \(Q_2^r\le m-2\), applying \eqref{eq:obedience} with bidder~2's own type drawn from \(U_{D_m(r)}\) gives
\[
e_{2r}\ge \E_D^r[H_2^{r,D}]
\ge \ell_2\kappa\Prb_D^r(Q_2^r\le m-2)
\ge \ell_2\kappa K_r^D.
\]
Dividing by \(\ell_2\kappa>0\) gives \(K_r^D\le e_{2r}/(\ell_2\kappa)\).

For bidder 2 in \(C_m(r)=[c,c+\rho_r]\times \mathcal T_{m-1}[0,\rho_r]\), replace the first bid with her first value,
\begin{equation}\label{eq:cheapstrategy}
d_2^{r,C}(t_2)=
\begin{cases}
(t_{21},\sigma_{22}^r(t_2),\ldots,\sigma_{2m}^r(t_2)),&t_2\in C_m(r),\\
\sigma_2^r(t_2),&\text{otherwise}.
\end{cases}
\end{equation}
If the first bid already wins, quantity and price are unchanged, so the surplus gain is zero. Otherwise bidder 2 starts with zero surplus and wins at most one unit after the deviation, with nonnegative surplus by the no-overbidding bounds. On \(A_r^C(a)=\{Q_1^r=m,\ \beta_r\le a\}\), the new first bid is at least \(c>a\). It displaces bidder 1's last bid, which becomes the highest losing bid. The quantity increases from zero to one, and the price is \(\beta_r\). Hence \(S_2^r=0\) and \(S_2^{r,C}=V_1(t_2^r)-\beta_r=t_{21}^r-\beta_r\ge c-a\). By \eqref{eq:derivatives}, \(H_2^{r,C}\ge\ell_2(c-a)\) on \(A_r^C(a)\), with a nonnegative gain elsewhere in state \(C\). Applying \eqref{eq:obedience} with bidder~2's own type drawn from \(U_{C_m(r)}\) gives
\[
e_{2r}\ge \E_C^r[H_2^{r,C}]
\ge \ell_2(c-a)\Prb_C^r(A_r^C(a))
=\ell_2(c-a)K_r^C(a).
\]
Dividing by \(\ell_2(c-a)>0\) gives \(K_r^C(a)\le e_{2r}/[\ell_2(c-a)]\).

\pstep{4}
Consider the lowest-winning-bid price, \(\alpha=0\). We first bound the gap between bidder 1's first two bids on \(F_r=\{Q_1^r=1,\ Q_2^r=m-1\}\) under \(\Prb_D^r\). Each deviation in this step changes bidder~1's bids on \(L_m\cup H_m\) and leaves her strategy unchanged elsewhere.

For \(h>0\), raise bidder 1's bids to \(d_{1j}^{r,+h}(t_1)=\min\{\sigma_{1j}^r(t_1)+h,t_{1j}\}\). This vector is ordered and satisfies the no-overbidding bounds. Write \(q,p\) for her original quantity and price, and \(q',p'\) for the corresponding outcomes after this deviation. Then \(q'\ge q\). Each submitted bid increases by at most \(h\), so the \(m\)th highest bid also increases by at most \(h\). Indeed, any bid originally at most \(p\) remains at most \(p+h\), leaving at most \(m-1\) bids above \(p+h\). Hence \(p'\le p+h\). Each extra unit has nonnegative marginal surplus, so
\begin{equation}\label{eq:raise_surplus_identity}
\Delta S_1^{r,+h}
=q(p-p')+\sum_{j=q+1}^{q'}(t_{1j}^r-p')\ge-mh.
\end{equation}
Take \(h=2\rho_r\). In \(D_m(r)\), if \(Q_1^r=0\), all \(m\) winning bids are opposing bids and at most \(m-1\) of them can exceed \(\rho_r\). Hence \(p\le\rho_r\). Every new own bid is at least \(2\rho_r\) for large \(r\), so bidder 1 wins at least one unit at \(p'\le3\rho_r\). Her surplus rises from zero to at least \(z-3\rho_r\ge z/2\). On every other basic profile the loss is at most \(2m\rho_r\). Let \(J_r\) be the indicator of state \(D\) and \(Q_1^r=0\). The two surplus bounds give
\(\Delta S_1^{r,+2\rho_r}\ge (z/2)J_r-2m\rho_r\). By \eqref{eq:derivatives},
\(H_1^{r,+2\rho_r}\ge\ell_1(z/2)J_r-2L_1m\rho_r\).
With bidder~1's own type drawn from \(\mu_1^r\), the expectation of \(J_r\) is \(\delta P_{1r}^D\), where \(P_{1r}^D=\Prb_D^r(Q_1^r=0)\). Thus \eqref{eq:obedience} gives
\[
e_{1r}\ge\E^r[H_1^{r,+2\rho_r}]
\ge\delta\ell_1(z/2)P_{1r}^D-2L_1m\rho_r.
\]
Consequently \(P_{1r}^D\le(e_{1r}+2L_1m\rho_r)/(\delta\ell_1 z/2)=O(r^{-1})\).

Next replace only bidder 1's first bid by
\(\min\{\sigma_{11}^r(t_1),\max\{\sigma_{12}^r(t_1),2\rho_r\}\}\).
This bid is at least her second bid and at most her original first bid. Whenever it is reduced, the new bid is at least \(2\rho_r\), above the opposing \(m\)th bid in either \(C_m(r)\) or \(D_m(r)\). All own bids therefore keep their acceptance status by the rank observation above. The price weakly falls. On \(F_r\), the original price is \(p^r=\min\{b_{11}^r,b_{2,m-1}^r\}\ge b_{12}^r\), while the new price is at most \(b_{12}^r+2\rho_r\). Her surplus changes from \(t_{11}^r-p^r\) to \(t_{11}^r-p^{r,d}\), so the gain is at least \((p^r-b_{12}^r-2\rho_r)_+\), where \(x_+=\max\{x,0\}\). On all other basic profiles the gain is nonnegative. By \eqref{eq:derivatives} and \eqref{eq:obedience},
\[
e_{1r}\ge\E^r[H_1^{r,d}]
\ge\delta\ell_1\E_D^r[(p^r-b_{12}^r-2\rho_r)_+\1_{F_r}].
\]
On \(F_r\), \(p^r-b_{12}^r\le(p^r-b_{12}^r-2\rho_r)_++2\rho_r\). Multiplying by \(\1_{F_r}\) and taking the expectation in state \(D\) gives
\[
X_r:=\E_D^r[(p^r-b_{12}^r)\1_{F_r}]
\le\frac{e_{1r}}{\delta\ell_1}+2\rho_r\Prb_D^r(F_r)
\le\frac{e_{1r}}{\delta\ell_1}+2\rho_r=O(r^{-1}).
\]

Let \(Y_r=\E_D^r[(b_{11}^r-b_{12}^r)\1_{F_r}]\), \(h_r=r^{-2/3}\), and \(\theta_r=r^{-1/3}\). Partition \(F_r\) into four disjoint events. The first event is \(F_r\cap\{b_{11}^r-b_{12}^r\le h_r\}\), whose contribution to \(Y_r\) is at most \(h_r\). The second event is the part of \(F_r\) with \(b_{11}^r-b_{12}^r>h_r\) and \(p^r-b_{12}^r>h_r/2\). Since the bid gap is at most one, its contribution is bounded by
\[
\Prb_D^r(F_r\cap\{p^r-b_{12}^r>h_r/2\})
\le \frac{2}{h_r}\E_D^r[(p^r-b_{12}^r)\1_{F_r}]
=\frac{2X_r}{h_r}.
\]
On the remaining profiles in \(F_r\), the bid gap exceeds \(h_r\) and \(p^r-b_{12}^r\le h_r/2\). Hence \(p^r<b_{11}^r\), so \(p^r=b_{2,m-1}^r\). The third event consists of these profiles with \(t_{12}^r-p^r\le\theta_r\). No overbidding gives
\[
b_{2,m-1}^r-h_r/2=p^r-h_r/2
\le b_{12}^r\le t_{12}^r\le p^r+\theta_r=b_{2,m-1}^r+\theta_r.
\]
Under \(\Prb_D^r\), \(t_{12}^r\) has density at most \(1/\varepsilon\) and is independent of \(b_{2,m-1}^r\). Thus
\[
\Prb_D^r\!\left(
b_{2,m-1}^r-h_r/2\le t_{12}^r\le b_{2,m-1}^r+\theta_r
\,\middle|\,b_{2,m-1}^r\right)
\le\frac{\theta_r+h_r/2}{\varepsilon}
\quad\text{almost surely}.
\]
Taking expectations gives the same bound for the interval event. The third event is contained in this interval event, so its contribution to \(Y_r\) is at most \((\theta_r+h_r/2)/\varepsilon\).

Let \(E_r\) be the fourth event, consisting of profiles in \(F_r\) for which \(b_{11}^r-b_{12}^r>h_r\), \(p^r-b_{12}^r\le h_r/2\), and \(t_{12}^r-p^r>\theta_r\). Apply the capped increase with \(h=h_r\) to every own type in \(L_m\cup H_m\). On \(E_r\), \(b_{12}^r+h_r\ge p^r+h_r/2>p^r\) and \(t_{12}^r>p^r+\theta_r>p^r\). Thus the new second bid \(\min\{b_{12}^r+h_r,t_{12}^r\}\) strictly exceeds \(p^r=b_{2,m-1}^r\). Since \(t_{12}^r\ge z\), the new bid is also at least \(h_r>\rho_r\) for large \(r\). The opposing \((m-1)\)st bid is at most \(\max\{b_{2,m-1}^r,\rho_r\}\), so bidder 1 wins a second unit. Her original surplus is \(t_{11}^r-p^r\). At the new price \(p^{r,d}\le p^r+h_r\), her new surplus is at least \(t_{11}^r+t_{12}^r-2p^{r,d}\), since further units have nonnegative marginal surplus. The gain is at least \(t_{12}^r-p^r-2h_r>\theta_r-mh_r\) on \(E_r\), while \eqref{eq:raise_surplus_identity} bounds the loss elsewhere by \(mh_r\). For large \(r\), \(\theta_r>mh_r\). Thus \eqref{eq:derivatives} gives utility gain at least \(\ell_1(\theta_r-mh_r)\) on \(E_r\) in state \(D\) and utility loss at most \(L_1mh_r\) elsewhere. Averaging and using \eqref{eq:obedience} yields
\[
e_{1r}\ge\E^r[H_1^{r,+h_r}]
\ge\delta\ell_1(\theta_r-mh_r)\Prb_D^r(E_r)-L_1mh_r.
\]
Hence \(\Prb_D^r(E_r)\le(e_{1r}+L_1mh_r)/[\delta\ell_1(\theta_r-mh_r)]\).
Combining the four bounds yields
\begin{equation}\label{eq:spread}
Y_r\le h_r+\frac{2X_r}{h_r}
+\frac{\theta_r+h_r/2}{\varepsilon}
+\frac{e_{1r}+L_1mh_r}{\delta\ell_1(\theta_r-mh_r)}
=O(r^{-1/3}).
\end{equation}

\pstep{5}
Continue with \(\alpha=0\). We use the same two deviations by bidder 2, now allowing for their effects on the lowest-winning-bid price. In \(D_m(r)\), apply bidder 2's deviation \eqref{eq:expensivestrategy}. It guarantees at least \(m-1\) units, and the last bid has unchanged acceptance. If she originally wins at most \(m-2\) units, her original surplus is at most \(V_{m-2}(t_2^r)\). Her new quantity is \(m-1\), the new price is at most \(R_*\), and her surplus is at least \(V_{m-1}(t_2^r)-(m-1)R_*\). The gain is therefore at least \(\kappa\), as in \eqref{eq:forcinggain}. On \(F_r\), bidder~2 wins \(m-1\) units both before and after the deviation. Only her already accepted bids are raised, so the identities of the winning bids are unchanged. In particular, bidder~1's first bid still wins. The old price is at least \(b_{12}^r\), and the new price is at most \(b_{11}^r\). Bidder~2's surplus changes from \(V_{m-1}(t_2^r)-(m-1)p^r\) to \(V_{m-1}(t_2^r)-(m-1)p^{r,D}\). Thus the loss is at most \((m-1)(b_{11}^r-b_{12}^r)\). If bidder 2 originally wins \(m\) units, her last bid remains the price, giving zero gain. In the remaining case \(Q_2^r=m-1,\ Q_1^r=0\), the utility loss is at most \(W_2\).

By \eqref{eq:derivatives}, the gain on \(\{Q_2^r\le m-2\}\) is at least \(\ell_2\kappa\), and the utility loss on \(F_r\) is at most \(L_2(m-1)(b_{11}^r-b_{12}^r)\). Set \(P_{2r}^D=\Prb_D^r(Q_2^r\le m-2)\). Averaging the four cases over bidder~2's \(D_m(r)\) component and applying \eqref{eq:obedience} gives
\begin{align*}
e_{2r}\ge\E_D^r[H_2^{r,D}]
&\ge\ell_2\kappa P_{2r}^D-L_2(m-1)Y_r
-W_2\Prb_D^r(Q_2^r=m-1,\ Q_1^r=0)\\
&\ge\ell_2\kappa P_{2r}^D-L_2(m-1)Y_r-W_2P_{1r}^D.
\end{align*}
Since \(Q_1^r\ge2\) implies \(Q_2^r\le m-2\), this yields
\[
K_r^D\le P_{2r}^D\le\frac{e_{2r}+L_2(m-1)Y_r+W_2P_{1r}^D}{\ell_2\kappa}=O(r^{-1/3}).
\]
Also \(p^r\in[b_{12}^r,b_{11}^r]\) on \(F_r\), and \(\Prb_D^r(F_r^c)\le P_{2r}^D+P_{1r}^D\). Since all original bids and prices lie in \([0,1]\), it follows that
\begin{equation}\label{eq:firstpricebound}
\E_D^r|p^r-b_{11}^r|\le Y_r+P_{2r}^D+P_{1r}^D.
\end{equation}

To control the gap between the first and last bids, let bidder 1 replace each own bid on \(L_m\cup H_m\) by \(\min\{\sigma_{1m}^r(t_1)+2\rho_r,t_{1j}\}\). The vector is ordered and satisfies the no-overbidding bounds. For large \(r\) each new bid is at least \(2\rho_r\). The deviation then wins at least \(m-1\) units in \(C_m(r)\) and at least one in \(D_m(r)\). In both cases the price is at most \(\beta_r+2\rho_r\).

In \(C_m(r)\), the deviation preserves all units when bidder 1 originally wins \(m\). Indeed, if all her bids are first set to \(\beta_r\), any opposing bid tied at \(\beta_r\) still loses by the common bidder priority. Increasing these bids subject to their no-overbidding bounds preserves their acceptance. Thus the new quantity is at least the old quantity on every \(C_m(r)\) profile. The old price is at least \(\beta_r\), and every additional unit gives nonnegative marginal surplus. The surplus loss is at most \(2m\rho_r\).

In \(D_m(r)\), if the original quantity is zero, the deviation gives nonnegative surplus. If the original quantity is one, the original surplus is \(t_{11}^r-p^r\) and the new surplus is at least \(t_{11}^r-\beta_r-2\rho_r\), even if additional units are won. Since the original last bid is rejected, \(p^r\ge\beta_r\). The utility gain is therefore at least \(\ell_1(p^r-\beta_r)-2L_1\rho_r\). If the original quantity is at least two, the utility loss is bounded by \(W_1\). Let \(d\) denote this deviation. In state \(C\), \eqref{eq:derivatives} gives \(\E_C^r[H_1^{r,d}]\ge-2L_1m\rho_r\). In state \(D\), averaging over the three quantity cases gives
\[
\E_D^r[H_1^{r,d}]
\ge\ell_1\E_D^r[(p^r-\beta_r)\1_{\{Q_1^r=1\}}]
-2L_1\rho_r\Prb_D^r(Q_1^r=1)-W_1K_r^D.
\]
Bidder~1 faces both states. Thus \eqref{eq:obedience} gives
\begin{align*}
e_{1r}&\ge\E^r[H_1^{r,d}]
=(1-\delta)\E_C^r[H_1^{r,d}]+\delta\E_D^r[H_1^{r,d}]\\
&\ge\delta\ell_1\E_D^r[(p^r-\beta_r)\1_{\{Q_1^r=1\}}]
-2L_1m\rho_r-\delta W_1P_{2r}^D.
\end{align*}
The last inequality uses \(K_r^D\le P_{2r}^D\) and \((1-\delta)m+\delta\Prb_D^r(Q_1^r=1)\le m\).

Write \(Z_r=\E^r(b_{11}^r-\beta_r)=\E_D^r(b_{11}^r-\beta_r)\), by independence of types. On \(\{Q_1^r=1\}\), split the gap at \(p^r\). On its complement, use the upper bound one on the gap. Then \eqref{eq:firstpricebound} and \(\Prb_D^r(Q_1^r\ne1)\le P_{1r}^D+P_{2r}^D\) give
\begin{align*}
Z_r
&\le\E_D^r|b_{11}^r-p^r|+
\E_D^r[(p^r-\beta_r)\1_{\{Q_1^r=1\}}]+P_{1r}^D+P_{2r}^D\\
&\le Y_r+2(P_{2r}^D+P_{1r}^D)
+\frac{e_{1r}+2L_1m\rho_r+\delta W_1P_{2r}^D}{\delta\ell_1}.
\end{align*}
Thus \(Z_r\to0\).

Finally, apply bidder 2's deviation \eqref{eq:cheapstrategy} in \(C_m(r)\). Let \(A_r^C(a)=\{Q_1^r=m,\ \beta_r\le a\}\). If bidder 2 originally wins at least one unit, her quantity is unchanged. The old price is at least \(\beta_r\). The new price is at most \(\max\{b_{11}^r,\rho_r\}\), since all accepted bids other than her first bid satisfy that bound. Writing \(q=Q_2^r\), her surplus changes from \(V_q(t_2^r)-qp^r\) to \(V_q(t_2^r)-qp^{r,C}\). The loss is therefore at most \(m(b_{11}^r-\beta_r+\rho_r)\). If her first bid originally loses, her old surplus is zero, and the new surplus is nonnegative. On \(A_r^C(a)\), she wins exactly one unit after the deviation. The price is at most \(b_{11}^r\), so her new surplus \(t_{21}^r-p^{r,C}\) is at least \(c-b_{11}^r\ge c-a-(b_{11}^r-\beta_r)\). Together these cases give
\[
\Delta S_2^{r,C}\ge(c-a)\1_{A_r^C(a)}-m(b_{11}^r-\beta_r+\rho_r).
\]
Applying \eqref{eq:derivatives} to the gain and loss terms separately gives
\[
H_2^{r,C}\ge\ell_2(c-a)\1_{A_r^C(a)}-L_2m(b_{11}^r-\beta_r+\rho_r).
\]
The distribution of bidder~1's bid gap is the same under \(C\) as under \(D\). Taking the expectation with bidder~2's own type drawn from \(U_{C_m(r)}\) and applying \eqref{eq:obedience} yields
\[
e_{2r}\ge\E_C^r[H_2^{r,C}]
\ge\ell_2(c-a)K_r^C(a)-L_2m(Z_r+\rho_r).
\]
Therefore \(K_r^C(a)\le[e_{2r}+L_2m(Z_r+\rho_r)]/[\ell_2(c-a)]\).
Table~\ref{tab:bidder2surplus} summarizes the comparisons for bidder 2 in Steps 3 and 5. The deviation raises the first \(m-1\) bids to at least \(R_*\) in \(D_m(r)\) and raises only the first bid to \(t_{21}^r\) in \(C_m(r)\). Write \(q=Q_2^r\) for the original quantity and \(g_r=b_{11}^r-\beta_r\). The surplus change is the deviating surplus minus the original surplus. Whenever quantity is unchanged, this change is \(q(p^r-p^{r,d})\).

\begin{center}
\begin{minipage}{\linewidth}
\small
\captionof{table}{Quantity and surplus changes under bidder 2's deviations}\label{tab:bidder2surplus}
\renewcommand{\arraystretch}{1.2}
\begin{tabularx}{\linewidth}{@{}P{0.07\linewidth}Y P{0.18\linewidth} P{0.18\linewidth} P{0.23\linewidth}@{}}
\toprule
State & Original event & Quantity change & \(\Delta S_2\) at \(\alpha=1\) & \(\Delta S_2\) at \(\alpha=0\)\\
\midrule
\(D\) & \(q\le m-2\) & \(q\to m-1\) & \(\ge\kappa\) & \(\ge\kappa\)\\
\(D\) & \(q=m-1,\ Q_1^r=1\) & \(m-1\to m-1\) & \(0\) & \(\ge-(m-1)(b_{11}^r-b_{12}^r)\)\\
\(D\) & \(q=m-1,\ Q_1^r=0\) & \(m-1\to m-1\) & \(0\) & \(\ge-m\)\\
\(D\) & \(q=m\) & \(m\to m\) & \(0\) & \(0\)\\
\midrule
\(C\) & \(q\ge1\) & \(q\to q\) & \(0\) & \(\ge-m(g_r+\rho_r)\)\\
\(C\) & \(A_r^C(a)\) & \(0\to1\) & \(t_{21}^r-\beta_r\ge c-a\) & \(\ge\max\{0,c-a-g_r\}\)\\
\(C\) & \(q=0\) outside \(A_r^C(a)\) & \(0\to0\) or \(1\) & \(\ge0\) & \(\ge0\)\\
\bottomrule
\end{tabularx}
\end{minipage}
\end{center}

We have therefore shown under both pricing rules that, for each fixed \(a\in(0,c)\),
\begin{equation}\label{eq:statebounds}
K_r^D\longrightarrow0,\qquad K_r^C(a)\longrightarrow0.
\end{equation}
These convergence statements refer to the bounds derived above. For every fixed admissible choice of the parameters and \(a\), their right-hand sides are bounded by deterministic quantities that tend to zero with \(r\). The bounds apply to every candidate equilibrium at any index where one exists. Their constants depend only on the fixed parameters, the utilities, and \(a\).

\pstep{6}
We use the choice of \(c\) and \(\delta\) in Step~1 to derive a contradiction under each pricing rule. 
For a type \(t=(t_1,\ldots,t_m)\) of bidder 1, define
\begin{align*}
F_0(t)&=(1-\delta)u_1(V_{m-1}(t))+\delta u_1(t_1),\\
F_+(t)&=(1-\delta)u_1(V_m(t)-mc)+\delta u_1(t_1-c),\\
G_b(t)&=(1-\delta)u_1(V_m(t))+\delta u_1(t_1-b)
\qquad (0<b\le c).
\end{align*}
At the low reference type \((x,y_L,z,\ldots,z)\), \(F_+-F_0=\gamma\). At the high reference type \((x,y_H,z,\ldots,z)\), \(F_0-G_c=\gamma\). Set \(\lambda=\gamma/2\). By continuity we can fix \(\varepsilon>0\) within the range in \eqref{eq:parameters} so that
\begin{equation}\label{eq:margins}
F_+(t)-F_0(t)\ge\lambda\quad(t\in L_m),
\qquad
F_0(t)-G_c(t)\ge\lambda\quad(t\in H_m).
\end{equation}
Here \(L_m=[x,x+\varepsilon]\times[y_L,y_L+\varepsilon]\times \mathcal T_{m-2}[z,z+\varepsilon]\) and \(H_m=[x+2\varepsilon,x+3\varepsilon]\times[y_H,y_H+\varepsilon]\times \mathcal T_{m-2}[z+2\varepsilon,z+3\varepsilon]\). Put \(M=mL_1\). Fix \(a\in(0,c)\) close enough to \(c\) that \(\delta L_1(c-a)<\lambda/4\). Fix \(\eta>0\) small enough that \(c+\eta<z\) and \(M\eta<\lambda/4\). Fix \(\varepsilon,a,\eta\) at these choices. Apply Steps~2--5 with this fixed \(\varepsilon\). All parameters are now fixed before choosing \(r\).

Take an index \(r\) with \(\rho_r<\eta\) at which a candidate equilibrium exists. We test two deviations for bidder 1, denoted \(L\) and \(H\). Deviation \(L\) replaces all \(m\) bids by \(c+\eta\) on \(A_r^L=\{t\in L_m\mid \sigma_{1m}^r(t)\le a\}\) and leaves the bid unchanged elsewhere. Deviation \(H\) replaces all bids by \(\eta\) on \(H_m\) and leaves them unchanged elsewhere. Both are feasible and respect the no-overbidding bounds.

Under the basic distributions, the opposing region \(C_m(r)=[c,c+\rho_r]\times \mathcal T_{m-1}[0,\rho_r]\) has probability \(1-\delta\). At most one opposing bid can exceed \(\rho_r\) there, and it is at most \(c+\rho_r\). Thus the low type deviation wins all \(m\) units and the high type deviation wins at least \(m-1\). In \(D_m(r)=[0.95,0.95+\rho_r]\times \mathcal T_{m-2}[0.90,0.90+\rho_r]\times[0,\rho_r]\), at most \(m-1\) opposing bids exceed \(\rho_r\), so either deviation wins at least one unit. At either pricing rule the deviating price is at most the flat bid, since at least one own bid wins. Additional units contribute nonnegative surplus. Table~\ref{tab:finalsurplus} collects the bounds. The high rows concern types whose original last bid exceeds \(a\).

\begin{center}
\begin{minipage}{\linewidth}
\small
\captionof{table}{Surplus bounds for bidder 1 under the basic distributions}\label{tab:finalsurplus}
\renewcommand{\arraystretch}{1.25}
\begin{tabularx}{\linewidth}{@{}P{0.17\linewidth}Y Y P{0.22\linewidth}@{}}
\toprule
Type and state & Original surplus at most & Deviating surplus at least & Exception to original bound\\
\midrule
\(t\in A_r^L,\ C\) & \(V_{m-1}(t)\) & \(V_m(t)-m(c+\eta)\) & \(Q_1^r=m\)\\
\(t\in A_r^L,\ D\) & \(t_1\) & \(t_1-c-\eta\) & \(Q_1^r\ge2\)\\
\(t\in H_m,\ C\) & \(V_m(t)\) & \(V_{m-1}(t)-(m-1)\eta\) & None\\
\(t\in H_m,\ D\) & \(t_1-a\) & \(t_1-\eta\) & \(Q_1^r\ge2\)\\
\bottomrule
\end{tabularx}
\end{minipage}
\end{center}

For the first three rows, the original surplus bounds use the indicated quantity limits and nonnegative prices. For the last row, zero units give zero surplus. If the original quantity is one, the last bid \(\beta_r>a\) is rejected, so either original price is at least \(\beta_r\). This gives \(S_1^r\le t_1-a\).

First apply the low type deviation. Let \(P_r^L=U_{L_m}(A_r^L)\). Under \(U_{L_m}\), the probability in state \(C\) of the joint event \(\{t_1^r\in A_r^L,\ Q_1^r=m\}\) is at most \(2K_r^C(a)\). In state \(D\), the probability of \(\{t_1^r\in A_r^L,\ Q_1^r\ge2\}\) is at most \(2K_r^D\). The factor of two comes from the weight \(1/2\) of \(U_{L_m}\) in bidder~1's basic distribution. On each exception the excess utility above utility at the stated surplus bound is at most \(W_1\). Thus its contribution to the weighted utility comparison is at most \(B_{Lr}=2W_1[(1-\delta)K_r^C(a)+\delta K_r^D]\). Apply \(u_1\) to the surplus bounds before averaging. Replacing \(c+\eta\) by \(c\) in the deviating bounds increases their weighted utility by at most \(M\eta\) per selected type, by \eqref{eq:derivatives}. Since the deviation makes no change outside \(A_r^L\), \eqref{eq:obedience} and \eqref{eq:margins} yield
\begin{align}
e_{1r}\ge \E_L^r H_1^{r,L}
&\ge \int_{A_r^L}(F_+(t)-F_0(t))\,\dd U_{L_m}(t)-M\eta P_r^L-B_{Lr}\notag\\
&\ge(\lambda-M\eta)P_r^L-B_{Lr}.
\label{eq:lowfinal}
\end{align}
For sufficiently large \(r\), \(e_{1r}+B_{Lr}<\lambda/2\), so \(P_r^L<2/3\). A positive measure of low types therefore bid above \(a\) on the last unit. Choose one such type from the full measure set on which the last bid is nondecreasing. Every high type exceeds it coordinatewise. It follows that almost every high type has \(\beta_r>a\) in this same equilibrium. This is the only use of monotonicity.

We can now apply the high type deviation. Under the normalized high component \(U_{H_m}\), the conditional probability of the exception in \(D\) is at most \(2K_r^D\). Its contribution to the weighted utility comparison is at most \(B_{Hr}=2\delta W_1K_r^D\). Removing the payments from the deviating surplus bounds increases their weighted utility by at most \(M\eta\). Also \(G_a(t)-G_c(t)\le\delta L_1(c-a)\), by \eqref{eq:derivatives}. Applying \eqref{eq:obedience} and \eqref{eq:margins} to the high rows gives
\begin{align}
e_{1r}\ge \E_H^r H_1^{r,H}
&\ge\int_{H_m}(F_0(t)-G_a(t))\,\dd U_{H_m}(t)-M\eta-B_{Hr}\notag\\
&\ge\lambda-\delta L_1(c-a)-M\eta-B_{Hr}
>\lambda/2-B_{Hr}.
\label{eq:highfinal}
\end{align}
By \eqref{eq:statebounds}, for every sufficiently large finite \(r\) we have \(\rho_r<\eta\), \(e_{1r}+B_{Lr}<\lambda/2\), and \(e_{1r}+B_{Hr}<\lambda/2\). The bounds are uniform over the candidate equilibria at each \(r\). The second inequality establishes the low type conclusion. The third contradicts \eqref{eq:highfinal} in the same game.

For each pricing rule, candidate equilibria are therefore excluded at every sufficiently large finite \(r\). Choose one \(r\) above both thresholds. Its actual type distributions have the required density bounds and work for both pricing rules. Lemma~\ref{lem:caps} then excludes admissible monotone pure equilibria.
\end{proof}


\subsection{Proof of Theorem~\ref*{thm:discriminatory}}\label{app:discriminatory-proof}
Use the model in Section~\ref{sec:model} and the discriminatory payment rule in Section~\ref{sec:discriminatory}. Subtracting \(u_i(0)\) from each utility does not change preferences, so take \(u_i(0)=0\) throughout this subsection. Write \(u=u_1\), which is strictly concave by \textup{(\ref{eq:utilities})}. We retain \(\sigma_i\) for bidding strategies and \(S_i^{\mathrm D}\) for discriminatory surplus. We use the rank observation, the utility bounds \eqref{eq:derivatives}, and the conditioning argument leading to \eqref{eq:obedience} in the proof of Theorem~\ref{thm:uniform}. The following no-overbidding result is specific to discriminatory pricing.

\begin{lemma}\label{lem:disc-caps}
Suppose the types are independent and \(\Prb(t_{i1}<a)>0\) for every \(a>0\) and every bidder \(i\). Then every pure-strategy Bayesian equilibrium satisfies \(\sigma_{ij}(t_i)\le t_{ij}\) for all \(j\) at almost every \(t_i\).
\end{lemma}
\begin{proof}
Fix an equilibrium \(\sigma=(\sigma_i)\), and write \(b_i=\sigma_i(t_i)\) at realized types. Fix the opponents' bids and the priority. Cap each own bid at its marginal value. The resulting vector remains ordered. By the rank observation in the proof of Theorem~\ref{thm:uniform}, an unchanged bid has unchanged acceptance. A reduced bid that still wins costs less. A reduced bid that loses after winning had negative marginal surplus. Thus the cap weakly raises surplus everywhere and strictly raises it whenever an overbid wins. Interim optimality rules out a winning overbid with positive probability. Expected equilibrium utility is finite because bidding zero guarantees at least \(u_i(0)\) and surplus is bounded above by \(m\).

If a bidder overbids her first value with positive probability, some rational \(a>0\) satisfies \(\Prb(t_{i1}<a<b_{i1})>0\). Independence gives positive probability to this event together with \(t_{j1}<a\) for every opponent \(j\). Every marginal value is then below \(a\), but a highest bid exceeds \(a\) and wins. Hence some bidder has a winning overbid with positive probability, a contradiction. First bids are therefore capped almost surely for all bidders. Fix an own type at which interim optimality holds. If \(b_{ij}>t_{ij}\), the opponents' first values are all strictly below \(b_{ij}\) with positive probability. Their first bids are then also strictly below it. The overbid wins on this event, again a contradiction.
\end{proof}

The next two lemmas apply when \(u\) is not exponential on nonnegative surplus. The first gives two ordered comparison types with reversed preferred bids. The second identifies the limiting bid distributions used in the construction.

\begin{lemma}\label{lem:th2-probes}
Suppose \(u\) is strictly concave and its restriction to \([0,\infty)\) is not of the form \(A-Be^{-rs}\) with \(B,r>0\). There exist an integer \(k\ge1\), numbers \(W,y>0\) satisfying \(W/k<1\) and \(y<\min\{1/2,W/k\}\), and ordered comparison parameters \(B<Y<X_L/k<X_H/k<1\), where \(B=y/(k+1)\). For the distribution in \eqref{eq:th2-f2} and the reduced payoff in \eqref{eq:th2-reduced}, there are thresholds \(0<q_-<q_+<B\) such that every maximizing bid at \((X_L,Y)\) exceeds \(q_+\) and every maximizing bid at \((X_H,Y)\) is below \(q_-\).
\end{lemma}
\begin{proof}
Let \(E\) be the differentiability set of \(u\) in \((0,\infty)\). Finite concavity makes \(u\) locally absolutely continuous and \(E\) has full measure. If the graph of \((u(x),u'(x))\) for \(x\in E\) lies on a line, then \(u'=\gamma_0+\gamma_1u\) almost everywhere. The line is nonvertical since \(u\) is strictly increasing. Concavity implies \(\gamma_1\le0\), and \(\gamma_1=0\) would make \(u\) affine. Solving the equation with \(\gamma_1<0\) and using continuity at zero gives \(u(x)=C-De^{-rx}\) on \([0,\infty)\), with \(r,D>0\). In the present case the graph therefore fails to be collinear on some bounded interval \(I=(a_0,a_1)\subset(0,\infty)\).

There are noncollinear triples with arbitrarily small span in \(I\). Otherwise every sufficiently short subinterval would have a collinear graph. Overlapping such intervals share two graph points since \(E\) is dense, so the lines would agree throughout \(I\). Choose an integer \(k>a_1\), and choose differentiability points \(s<W<t\) giving such a triple with \(t-s<\min\{a_0/(3k),1/6\}\). Let \(\lambda=\tfrac{u(t)-u(W)}{u(t)-u(s)}\) and \(b_*=(W-s)/k\). Put \(y=t-W+(k+1)b_*\) and \(B=y/(k+1)\). Then \(0<b_*<B<y<W/k<1\) and \(y<1/2\). For \(0\le b\le B\) set
\begin{equation}\label{eq:th2-f2}
F_2(b)=\frac{u(W)-u(W-kb)}{u(W+y-(k+1)b)-u(W-kb)}.
\end{equation}
Extend \(F_2\) by one above \(B\). On \([0,B]\), the denominator is positive and the numerator lies between zero and the denominator. Both state utilities decrease strictly in \(b\), so holding their weighted average at \(u(W)\) requires the weight \(F_2(b)\) on the higher utility to rise strictly. Thus \(F_2\) is continuous and strictly increasing from zero to one.

For \(X,Y>0\) and \(b\in[0,1]\), consider the continuous function
\begin{equation}\label{eq:th2-reduced}
\Phi(X,Y;b)=(1-F_2(b))u(X-kb)+F_2(b)u(X+Y-(k+1)b).
\end{equation}
All maximizers lie in \([0,B]\), since \(\Phi\) is strictly decreasing above \(B\). At \((W,y)\) the function equals \(u(W)\) throughout \([0,B]\). At \(b_*\), the two surpluses are \(s,t\), with probabilities \(\lambda,1-\lambda\). The derivative at zero of
\(\lambda u(s+h)+(1-\lambda)u(t+h)-u(W+h)\) is nonzero by noncollinearity. Hence, for some arbitrarily small \(d>0\), \(b_*\) strictly beats both endpoints at \((W+d,y)\) if this derivative is positive, and at \((W-d,y)\) if it is negative. Both endpoint payoffs at \((W\pm d,y)\) equal \(u(W\pm d)\).

In the positive case, take \(X_L=W\), \(X_H=W+d\), and \(Y=y+e\). At the low type \(B\) is the unique maximizer for every sufficiently small \(e>0\). For \(b<B\), the winning surplus at \(Y=y\) is at least \(W\), so its utility increment from raising \(Y\) by \(e\) is at most \(u(W+e)-u(W)\) by concavity. Since \(F_2(b)<1\), the expected increment is strictly smaller than at \(B\). For small \(e>0\), the strict comparison of \(b_*\) with \(B\) at the high type survives by continuity, so every high maximizer lies strictly below \(B\).

In the negative case, take \(X_L=W-d\), \(X_H=W\), and \(Y=y-e\). At the high type, every \(b\in(0,B]\) has positive winning probability, so reducing \(Y\) lowers its payoff strictly below \(u(W)\). Zero keeps payoff \(u(W)\). For sufficiently small \(e>0\), \(b_*\) still strictly beats zero at the low type, so its maximizers are bounded away from zero. In either case choose \(d,e\) small enough that \(B<Y<X_L/k<X_H/k<1\). Compactness gives \(0<q_-<q_+<B\) with
\begin{equation}\label{eq:th2-separation}
\arg\max_b \Phi(X_L,Y;b)\subset(q_+,1],\qquad
\arg\max_b \Phi(X_H,Y;b)\subset[0,q_-).
\end{equation}
The complementary closed bid sets have strictly lower maximal payoffs.
\end{proof}

Set \(m=2k+1\). Use reference types \(a_1=((W/k)^k,y,0^k)\) and \(a_2=((1/2)^k,y,0^k)\), where powers denote repeated coordinates. Write \(w_i=\sum_{h=1}^k a_{ih}\), and let \(v_i^-(b)=u_i(w_i-kb)\) and \(v_i^+(b)=u_i(w_i+y-(k+1)b)\). These are the utilities from winning \(k\) or \(k+1\) units with all active bids equal to \(b\).

For the auxiliary contest in the next lemma, let \(C_i\in[0,y]\) be independent common bids with distribution functions \(F_i\). In this limiting contest, let \(Q_i\in\{0,\ldots,k+1\}\) denote the winning quantity capped at \(k+1\), and impose \(Q_1+Q_2\le m\). If the bids differ and are positive, the higher bidder wins \(k+1\) units and the lower bidder wins \(k\). If only one bid is positive, that bidder wins \(k+1\) units. At a positive tie the two quantities are \(k\) and \(k+1\). At a zero tie only feasibility is imposed. Write \(\varphi_i(b,q)=u_i(V_q(a_i)-qb)\) for the utility from bid \(b\) and quantity \(q\). Suppose there are constants \(v_i\) such that \(\E[\varphi_i(C_i,Q_i)\mid C_i]=v_i\) almost surely. Suppose also that, for \(j\ne i\) and \(0\le b\le B\),
\begin{equation}\label{th2:pooled-obedience}
v_i\ge U_i^+(b)\equiv v_i^-(b)+F_j(b)[v_i^+(b)-v_i^-(b)].
\end{equation}
This bound treats a tie as winning the last unit. Step~3 of Case~1 below obtains this bound by taking actual deviation bids down to \(b\) from above. It does not change the tie rule in the auction.

\begin{lemma}\label{lem:th2-pooled}
Under the independence, allocation, and conditional-payoff assumptions above and the deviation bound \eqref{th2:pooled-obedience}, both common bids have continuous distributions supported on \([0,B]\). Their distribution functions are given by \eqref{th2:pooled-cdf}, and \(v_i=u_i(w_i)\) for both bidders.
\end{lemma}
\begin{proof}
At zero, \eqref{th2:pooled-obedience} gives \(v_i\ge u_i(w_i)\). A positive bid above \(B\) gives strictly less utility than \(u_i(w_i)\) for either possible quantity. Thus the conditional-payoff equality restricts both distributions to \([0,B]\). On this interval \(v_i^+(b)>v_i^-(b)\). Independence and the ranking conditions bound the actual conditional payoff at \(b\) by \(U_i^+(b)\). At zero, a positive opposing bid receives \(k+1\) units, so the zero bidder receives at most \(k\). If both bids are zero, each bidder receives at most \(k+1\). Hence the same upper bound holds at zero. The conditional-payoff equality and \eqref{th2:pooled-obedience} therefore imply \(U_i^+(b)=v_i\) for \(F_i\)-almost every \(b\), in particular at each own atom. At a shared atom, equality would require both bidders to receive \(k+1\) units on the tie event. This contradicts feasibility, including at a shared zero atom.

Suppose bidder \(i\) alone has a positive atom at \(b\). Then \(U_j^+\) jumps up at \(b\) by the atom's mass times \(v_j^+(b)-v_j^-(b)>0\). Hence \(U_j^+(c)<v_j\) throughout a sufficiently small interval directly below \(b\). Thus bidder \(j\) has no mass there, since \(U_j^+=v_j\) \(F_j\)-almost surely. Bidder \(i\) can shade into this interval without changing her winning probability, since bidder \(j\) has no atom at \(b\). Both state payoffs strictly increase. This contradicts \eqref{th2:pooled-obedience}, so neither distribution has positive atoms.

Since positive atoms have been excluded, \(U_i^+\) is continuous at every positive bid. The equality \(U_i^+(b)=v_i\) therefore extends from almost every own bid to every positive point in bidder \(i\)'s support.

Let \(\beta_i=\sup\operatorname{supp}F_i\). If \(\beta_i>\beta_j\), choose a positive support bid \(b>\beta_j\) for bidder \(i\) and take \(c\in(\beta_j,b)\). Both bids win the last unit surely, so \(U_i^+(c)=v_i^+(c)>v_i^+(b)=U_i^+(b)=v_i\), contrary to \eqref{th2:pooled-obedience}. Thus the upper endpoints coincide at some \(\beta\). If \(\beta=0\), both distributions are concentrated at zero, which contradicts the exclusion of a shared zero atom. Hence \(\beta>0\). At this common endpoint, \(F_j(\beta)=1\) and the support equality gives \(v_i=v_i^+(\beta)=u_i(w_i+y-(k+1)\beta)\).

If bidder \(i\) alone has an atom at zero, her actual conditional payoff there is at most \(u_i(w_i)\). Together with the deviation bound at zero, this implies \(v_i=u_i(w_i)\). The endpoint equality and strict monotonicity of \(u_i\) then give \(\beta=B\), so \(v_j=u_j(w_j)\) as well. Yet \(U_j^+(0)=u_j(w_j)+F_i(0)[u_j(w_j+y)-u_j(w_j)]>v_j\), again contrary to \eqref{th2:pooled-obedience}. Thus neither distribution has an atom at zero.

The positive supports coincide because, on any open interval where \(F_j\) is constant, \(U_i^+\) strictly decreases and no interior bid can be optimal. A common support gap with positive right endpoint is impossible because either bidder can shade from that endpoint into the gap without affecting her winning probability. This also excludes a positive common lower endpoint. The supports are therefore \([0,\beta]\). Taking bids down to zero gives \(v_i=u_i(w_i)\), so \(\beta=B\). Indifference yields
\begin{equation}\label{th2:pooled-cdf}
F_j(b)=\frac{u_i(w_i)-v_i^-(b)}{v_i^+(b)-v_i^-(b)}
\quad (0\le b\le B,\ j\ne i).
\end{equation}
The quotient is continuous with endpoint values zero and one. It is strictly increasing because both state payoffs fall as the bid rises, so preserving the baseline payoff requires a strictly higher winning probability.
\end{proof}

\begin{proof}[Proof of Theorem~\ref{thm:discriminatory}]
It suffices to exclude pure equilibria in which bidder~1's strategy is coordinatewise nondecreasing on a full-measure set. We consider the two possible forms of \(u\) on nonnegative surplus.

\par
Case 1. Suppose \(u\) is not exponential on nonnegative surplus. Use the parameters in Lemma~\ref{lem:th2-probes}, with \(m=2k+1\), and the reference types above.

\pstep{1}
We construct the type distributions. For small \(\varepsilon>0\), set \(A_i^\varepsilon=T_m\cap\{t\mid\|t-a_i\|_\infty<\varepsilon\}\) for \(i=1,2\), and let \(E^\varepsilon=T_m\cap[0,\varepsilon]^m\). The low and high comparison regions \(L^\varepsilon,H^\varepsilon\) are open cubes of radius \(\varepsilon/(100k)\). Their centers \(\ell,h\) have coordinates
\[
\begin{array}{ll}
\ell_j=X_L/k+\varepsilon(k+1-2j)/(10k),&
h_j=X_H/k+\varepsilon(k+1-2j)/(10k),\quad 1\le j\le k,\\
\ell_{k+1}=Y-\varepsilon,&h_{k+1}=Y+\varepsilon,\\
\ell_{k+1+j}=\varepsilon(k+1-j)/(4k),&
h_{k+1+j}=\varepsilon(k+1-j)/(2k),\quad 1\le j\le k.
\end{array}
\]
For sufficiently small \(\varepsilon\), both cubes lie in the interior of \(T_m\). The last \(k\) coordinates are at most \(\varepsilon\), and every high type strictly exceeds every low type coordinatewise. All the regions have positive volume. Define the basic type distributions as follows.
\[
\begin{aligned}
\mu_1^\varepsilon&=(1-\varepsilon)U_{A_1^\varepsilon}+\tfrac{\varepsilon}{3}(U_{E^\varepsilon}+U_{L^\varepsilon}+U_{H^\varepsilon}),\\
\mu_2^\varepsilon&=(1-\varepsilon)U_{A_2^\varepsilon}+\varepsilon U_{E^\varepsilon},\qquad
\mu_i^\varepsilon=U_{E^\varepsilon}\quad(i\ge3).
\end{aligned}
\]
Draw types independently from \(\widehat\mu_i^\varepsilon=(1-\varepsilon^2)\mu_i^\varepsilon+\varepsilon^2U_{T_m}\). The density calculation in Step~1 of the proof of Theorem~\ref{thm:uniform} gives the bounds required in Section~\ref{sec:model} for each fixed \(\varepsilon>0\). In particular, Lemma~\ref{lem:disc-caps} applies.

Suppose these auctions admit pure equilibria \(\sigma^\varepsilon=(\sigma_1^\varepsilon,\ldots,\sigma_n^\varepsilon)\) along a sequence \(\varepsilon\downarrow0\). In realized-outcome expressions, write \(b_{ij}^\varepsilon=\sigma_{ij}^\varepsilon(t_i^\varepsilon)\), where \(t_i^\varepsilon\) is the realized type. Write \(Q_i^\varepsilon\) and \(S_i^{\mathrm D,\varepsilon}\) for the resulting quantity and surplus. All limits below use this sequence. Modify bids on own-type null sets so that the value caps hold everywhere. We first derive conclusions that do not require monotonicity.

\pstep{2}
Fix an active bidder \(i\in\{1,2\}\) and a type \(t_i\in A_i^\varepsilon\) at which interim optimality holds. Write \(b=\sigma_i^\varepsilon(t_i)\), \(c=b_{k+1}\), and \(M=\max\{c,2\varepsilon\}\). Replace her first \(k\) bids by \(M\), leaving the remaining bids unchanged. For small \(\varepsilon\), the resulting vector is ordered and satisfies the value caps. Write \(S_i^{\mathrm D,d}\) for surplus after this deviation, with opposing bids and priority fixed.

Let \(G^\varepsilon\) be the event that the other active bidder is drawn from her reference component and each extra bidder is drawn from \(U_{E^\varepsilon}\). Its probability is \(p_\varepsilon=(1-\varepsilon^2)^{n-1}(1-\varepsilon) \ge(1-2\varepsilon)^{n-1}\). On \(G^\varepsilon\), at most \(k+1\) opposing bids exceed \(\varepsilon\). Every lowered coordinate among the first \(k\) bids remains above \(\varepsilon\), so it wins both before and after the deviation. The total payment saving on these units is
\(A(b)=\sum_{j=1}^k(b_j-M)_+\), where \(x_+=\max\{x,0\}\).
A coordinate can be raised only when \(M=2\varepsilon\). The added payment on retained units is then at most \(2k\varepsilon\), and each newly won unit has nonnegative surplus. Unchanged coordinates keep their acceptance status. Therefore
\(S_i^{\mathrm D,d}-S_i^{\mathrm D}\ge A(b)-2k\varepsilon\) on \(G^\varepsilon\).
Both surpluses lie in \([0,m]\). Use the constants \(\ell_i,L_i,W_i\) from the proof of Theorem~\ref{thm:uniform}, evaluated at the present supply \(m\). By the utility comparison following \eqref{eq:derivatives}, the corresponding utility gain is at least \(\ell_i A(b)-2L_i k\varepsilon\) on \(G^\varepsilon\) and at least \(-W_i\) on its complement. Interim optimality gives
\(0\ge p_\varepsilon[\ell_i A(b)-2L_i k\varepsilon]-(1-p_\varepsilon)W_i\).
Since \(0\le M-c\le2\varepsilon\), we have \(\sum_{j=1}^k(b_j-b_{k+1})\le2k\varepsilon+A(b)\). Solving the preceding inequality gives
\begin{equation}\label{eq:th2-pooling}
\sum_{j=1}^k(b_j-b_{k+1})\le2k\varepsilon+\frac{2L_i k\varepsilon}{\ell_i}
+\frac{(1-p_\varepsilon)W_i}{p_\varepsilon\ell_i}=O(\varepsilon).
\end{equation}
The constants are independent of the equilibrium and of the own type within the reference region. The same argument applies to almost every type of bidder~1 in \(L^\varepsilon\cup H^\varepsilon\). In these regions, the first \(k\) values also exceed the \((k+1)\)st value by a fixed positive amount for small \(\varepsilon\).

\pstep{3}
Pass to a subsequence along which the joint distributions of all bids and the two active quantities capped at \(k+1\) converge weakly. Write \(\bar Q_i^\varepsilon=\min\{Q_i^\varepsilon,k+1\}\) for these capped quantities. Such a subsequence exists because bids lie in \([0,1]\) and quantities take finitely many values. Each active reference component has probability tending to one. The value caps and \eqref{eq:th2-pooling} imply that each active bidder's first \(k+1\) bids converge jointly in distribution to a common scalar \(C_i\in[0,y]\), while all remaining bids vanish in probability. The scalars \(C_1,C_2\) are independent because the own-type draws are independent at each \(\varepsilon\).

Retain the limiting capped quantities \(Q_i\) in the joint limit. If the scalars are positive and unequal, the higher bidder wins \(k+1\) units and the lower wins \(k\). At a positive tie, all \(2k+2\) active bids are separated from the vanishing bids. Thus one active bidder wins \(k+1\) units and the other wins \(k\), regardless of how the rankings approach the tie. If only one scalar is positive, that bidder wins \(k+1\) units. At a zero tie, retain only \(Q_1+Q_2\le m\). These restrictions follow from strict finite-auction rankings in neighborhoods of the corresponding limiting outcomes. Units beyond \(k+1\) contribute at most \(k\varepsilon\) to surplus on the own reference event. The type values differ from \(a_i\) by at most \(\varepsilon\), and \eqref{eq:th2-pooling} bounds the total difference between the first \(k+1\) bids and their common last coordinate. Consequently, on this event,
\(u_i(S_i^{\mathrm D,\varepsilon})=\varphi_i(b_{i,k+1}^\varepsilon,\bar Q_i^\varepsilon)+O(\varepsilon)\),
uniformly in the realized opponents and priority. For each possible quantity \(q\), \(\varphi_i(b,q)\) is continuous in \(b\). Since the quantities take finitely many values, this approximation and the joint convergence give the limiting utility \(\varphi_i(C_i,Q_i)\) from Lemma~\ref{lem:th2-pooled}.

Let \(\mathcal V_i^\varepsilon(t)\) be the supremum of interim expected utility against the equilibrium opponents. Bids above one can be clipped at one without lowering surplus, so it suffices to consider bid vectors in \(B_m\cap[0,1]^m\). For a fixed bid vector, the allocation does not depend on the own type. Hence \(|S_i^{\mathrm D}(t)-S_i^{\mathrm D}(t')|\le m\|t-t'\|_\infty\). All these surpluses lie in \([-m,m]\), where finite concavity makes \(u_i\) Lipschitz. Taking the supremum over bid vectors preserves the resulting common Lipschitz bound. Thus \(\mathcal V_i^\varepsilon(t)-\mathcal V_i^\varepsilon(a_i)=O(\varepsilon)\) uniformly on \(A_i^\varepsilon\). Pass to a further subsequence with \(\mathcal V_i^\varepsilon(a_i)\to v_i\).

For any bounded continuous function \(f\), interim optimality gives
\[
\E\big[(u_i(S_i^{\mathrm D,\varepsilon})-\mathcal V_i^\varepsilon(a_i))f(b_{i,k+1}^\varepsilon)\big]\longrightarrow0.
\]
Indeed, conditional on each reference type the difference of expected utilities is \(O(\varepsilon)\). Outside the reference component the integrand is uniformly bounded, and that component's complement has probability at most \(2\varepsilon\). Passing to the joint limit gives \(\E[(\varphi_i(C_i,Q_i)-v_i)f(C_i)]=0\). Since this holds for every such \(f\), it gives the conditional-payoff equality required by Lemma~\ref{lem:th2-pooled}.

Fix \(d\in(0,y)\) at which the rival's limiting distribution \(F_j\) is continuous. A reference type can bid \(d\) on the first \(k+1\) units and zero on the rest. In the limit, this deviation wins \(k+1\) units when \(C_j<d\) and \(k\) units when \(C_j>d\). The corresponding surpluses are \(w_i+y-(k+1)d\) and \(w_i-kd\). Compare this deviation at the reference type \(a_i\) with \(\mathcal V_i^\varepsilon(a_i)\). Boundedness and the absence of an atom at \(d\) give, upon taking limits,
\(v_i\ge v_i^-(d)+F_j(d)[v_i^+(d)-v_i^-(d)]\).
For any \(b\in[0,B]\), take continuity points \(d\downarrow b\). Right continuity of \(F_j\) gives \eqref{th2:pooled-obedience}, including at atoms. Lemma~\ref{lem:th2-pooled} now identifies the two limiting distributions. In particular, bidder~2's limiting distribution function is \eqref{eq:th2-f2}. Every convergent subsequence has this same bid limit, so the bid convergence holds along the entire equilibrium sequence.

\pstep{4}
Take any sequence of types of bidder~1 from \(L^\varepsilon\), or from \(H^\varepsilon\), at which interim optimality holds. Their limit is \(((X/k)^k,Y,0^k)\), with \(X=X_L\) or \(X_H\). Write \(U_{\mathrm{eq}}^\varepsilon\) for their interim equilibrium utility. Pass to a subsequence along which their \((k+1)\)st bids converge to \(b\). By \eqref{eq:th2-pooling}, all first \(k+1\) bids have that same limit.

If \(b>0\), the rival's limiting distribution function is continuous at \(b\). The two relevant quantities are therefore \(k\) and \(k+1\), with probabilities \(1-F_2(b)\) and \(F_2(b)\). The corresponding surplus limits are \(X-kb\) and \(X+Y-(k+1)b\), so the interim utility converges to \(\Phi(X,Y;b)\). If \(b=0\), the rival's scalar is positive almost surely. At most \(k\) positive-value units can then be won in the limit, so the limiting superior of interim utility is at most \(u(X)=\Phi(X,Y;0)\).

For every fixed \(d\in(0,B]\), let \(U_d^\varepsilon\) be the interim utility from bidding \(d\) on the first \(k+1\) units and zero on the rest. This deviation yields limiting utility \(\Phi(X,Y;d)\). Optimality gives \(U_{\mathrm{eq}}^\varepsilon\ge U_d^\varepsilon\). In either case,
\[
\Phi(X,Y;b)
\ge\limsup_{\varepsilon\downarrow0}U_{\mathrm{eq}}^\varepsilon
\ge\lim_{\varepsilon\downarrow0}U_d^\varepsilon
=\Phi(X,Y;d).
\]
Letting \(d\downarrow0\) includes the zero bid. Since \(\Phi\) strictly decreases above \(B\), every accumulation point \(b\) maximizes \(\Phi(X,Y;\cdot)\) on \([0,1]\).

If equilibria monotone for bidder~1 existed for arbitrarily small \(\varepsilon\), we could take the sequence above to consist of such equilibria. The strict separation in \eqref{eq:th2-separation} implies that, for all sufficiently small \(\varepsilon\) in the sequence, \(\sigma_{1,k+1}^\varepsilon(t)>q_+\) for almost every \(t\in L^\varepsilon\), whereas \(\sigma_{1,k+1}^\varepsilon(t)<q_-\) for almost every \(t\in H^\varepsilon\).
Otherwise, choose best-responding types from a violating set along a subsequence. Compactness would give an accumulation point outside the required maximizing set, contrary to the preceding argument. Fix a sufficiently small positive \(\varepsilon\) in the sequence and choose one type from each region in bidder~1's full-measure monotonicity set. The high type exceeds the low type coordinatewise but bids strictly less on unit \(k+1\). This is a contradiction in that finite auction. Consequently, for all sufficiently small positive \(\varepsilon\), the constructed distributions exclude a monotone pure equilibrium.

\par
Case 2. Suppose \(u_1(s)=A-Be^{-rs}\) for \(s\ge0\), with \(B,r>0\).

\pstep{1}
We construct the type distributions. With the normalization \(u_2(0)=0\), let \(\lambda=u'_{2,+}(0)>0\). Choose \(x_0>0\) such that \(0.99\lambda s\le u_2(s)\le\lambda s\) for \(0\le s\le x_0\).\footnote{The one-sided derivative bounds used in \eqref{eq:derivatives} give \(0<\lambda<\infty\). Concavity implies \(u_2(s)/s\le\lambda\) for \(s>0\), and this ratio converges to \(\lambda\) as \(s\downarrow0\).}
Set \(k=200\lceil500/r\rceil\) and \(R=rk\ge10^5\). Fix an integer \(m\ge\max\{100k^2,1.33\cdot10^{-4}k^2/x_0\}\) and put \(M=m-2k\). Set \(c=10^{-4}k/m\), \(a=c/2\), \(h=3c/4\), and \(a_+=4c/5\). Let \(d=2k/(M+1)\), \(d_+=d+0.001k/m\), and \(\tau=10^{-6}kc/m\). Finally, put \(\ell=0.1\) and \(\delta=e^{-0.7R}\).
These choices imply \(kc\le10^{-6}\), \(m\tau\le10^{-12}\), \(1.33kc\le x_0\), and \(0<\tau<a<h<a_+<c<d<d_+<1\).

Bidder~1's main type regions are \(L\) and \(H\). In \(L\), the first \(M\), next \(k\), and last \(k\) values lie in \([0.8,0.8001]\), \([0.35,0.3501]\), and \([0.349,0.3491]\), respectively. In \(H\), the corresponding intervals are \([0.9,0.9001]\), \([0.8,0.8001]\), and \([0.359,0.3591]\). Every high type strictly exceeds every low type coordinatewise. Bidder~2's main type regions are \(C\) and \(D\). In \(C\), the first \(k\), next \(k\), and remaining \(M\) values lie in \([0.9c,0.91c]\), \([0.4c,0.41c]\), and \([0,\tau]\), respectively. In \(D\), the first \(2k\) values lie in \([1-1/(100k),1]\) and the remaining \(M\) lie in \([0,\tau]\). Each region consists of the ordered vectors in the indicated intervals. Set \(E=T_m\cap[0,\tau]^m\). All five regions have positive volume.

Let \(\Delta=u_2(2k+10^{-12})-u_2(2k-1)>0\). Choose \(\rho>0\) so that \(2\rho<10^{-6}\), \(2\rho<e^{-r(m+2k)}\), and \(2\rho u_2(2k+1)<0.01\Delta\). The basic type distributions are as follows.
\[
\begin{aligned}
\mu_1&=(1-\rho)[\ell U_L+(1-\ell)U_H]+\rho U_E,\\
\mu_2&=(1-\rho)[(1-\delta)U_C+\delta U_D]+\rho U_E,\qquad
\mu_i=U_E\quad(i\ge3).
\end{aligned}
\]
Draw types independently from \(\widehat\mu_i=(1-\eta)\mu_i+\eta U_{T_m}\). For either bidder \(i\in\{1,2\}\), the event that the other bidder in \(\{1,2\}\) is drawn from one of her two main regions and all extra bidders are drawn from their basic components has probability \(W=(1-\rho)(1-\eta)^{n-1}\). Call this the basic opponent event. Conditional on it, the two main regions of the other bidder retain their relative weights specified above. Write \(\omega=1-W\). Because \(\omega\to\rho\) as \(\eta\downarrow0\), we can choose \(\eta>0\) so that \(\omega<\min\{10^{-6},e^{-r(m+2k)},0.01\Delta/u_2(2k+1)\}\). The resulting densities satisfy the bounds in Section~\ref{sec:model}. The mixture components are used to compute probabilities. The deviations depend only on the deviating bidder's own type.

Suppose an equilibrium \(\sigma=(\sigma_i)\) exists in which \(\sigma_1\) is monotone. By Lemma~\ref{lem:disc-caps}, the bids satisfy the no-overbidding bounds almost everywhere. We may impose the bounds everywhere by changing bids on own-type null sets. For \(F\in\{L,H\}\) and \(G\in\{C,D\}\), let \(\Prb_{F,G}\) and \(\E_{F,G}\) use independent draws \(t_1\sim U_F\) and \(t_2\sim U_G\). Each extra bidder draws independently from \(U_E\). The priority is as in Section~\ref{sec:model}. A conditional expectation given \(t_2\) holds bidder~2's type fixed and averages over the other draws and priority. Each own-type component is absolutely continuous with respect to the corresponding actual marginal.

\pstep{2}
For a type \(t\in L\cup H\) of bidder~1 and a bid vector \(b\), let \(S_1^{\mathrm D}[b]\) denote her surplus when she submits \(b\) against the equilibrium opponents. Write \(X(t)=V_M(t)\) and define
\(J_t(b)=e^{rX(t)}\E_{t_{-1},\pi}\!\left[e^{-rS_1^{\mathrm D}[b]}\right]\),
where \(t\) is fixed and the expectation averages over the actual opponent distributions and priority. All the bids of bidder~1 considered below respect the value caps. Their surpluses are nonnegative for every realization, so optimality implies \(J_t(\sigma_1(t))\le J_t(b)\) for each such deviation at almost every \(t\).

Bidder~1 can bid \(c\) on every unit. Against \(C\) on the basic opponent event, all opposing bids are below \(c\). She wins all \(m\) units, pays \(mc\), and obtains surplus at least \(X(t)+0.699k-mc\). Against \(D\), at most \(2k\) opposing bids can reach \(c\). She therefore wins at least \(M\) units. The first \(M\) give surplus \(X(t)-Mc\) and any further units contribute nonnegative surplus. Outside the basic opponent event she obtains nonnegative surplus. Using \(X(t)\le m\) and \(W\le1\) gives
\begin{equation}\label{eq:cara-cap-loss}
J_t(\sigma_1(t))
\le e^{-r(0.699k-mc)}+\delta e^{rMc}+\omega e^{rm}
<3e^{-0.6989R}=:K.
\end{equation}
The first term equals \(e^{-0.6989R}\) since \(mc=0.0001k\). The second is at most \(e^{-0.6999R}\), since \(Mc\le mc\). The third is less than \(e^{-2R}\) by the bound on \(\omega\).

If \(\sigma_{1,M+1}(t)\ge d\), the first \(M+1\) bids win against \(C\) on the basic opponent event, since all opposing bids are below \(d\). These units alone cost at least \((M+1)d=2k\). The value of all \(m\) units is at most \(X(t)+1.1592k\). Thus her original surplus is at most \(X(t)-0.8408k\) on that event. This yields \(J_t(\sigma_1(t))\ge W(1-\delta)e^{0.8408R}>K\). Consequently \(\sigma_{1,M+1}(t)<d\) for almost every type in \(L\cup H\).

\pstep{3}
A type in \(D\) can bid \(d_+\) on her first \(2k\) units and zero on the rest. The bids satisfy the value caps. By Step~2, at most \(M\) opposing bids can reach \(d_+\) on the basic opponent event. Hence the first \(2k\) bids win. Those units have total value at least \(2k-0.02\). Their payment is \(2kd_+=\tfrac{4k^2}{m-2k+1}+\tfrac{0.002k^2}{m}<0.05\). The deviation gives surplus at least \(2k-0.07\) on the basic opponent event and nonnegative surplus otherwise.

For a given type \(t_2\in D\), let \(P\) be the equilibrium probability of winning at least \(2k\) units under the actual opponent distributions. With fewer than \(2k\) units, the original surplus is at most \(2k-1\). Otherwise it is at most \(2k+10^{-12}\), since the remaining units have total value at most \(m\tau\). The expected original utility is therefore at most \(u_2(2k-1)+P\Delta\). Interim optimality and the deviating surplus bound imply
\[
P\Delta\ge Wu_2(2k-0.07)-u_2(2k-1)
\ge \frac{0.93}{1+10^{-12}}\Delta-\omega u_2(2k-0.07).
\]
For the last inequality, concavity bounds \(u_2(2k-0.07)\) below by the chord joining its values at \(2k-1\) and \(2k+10^{-12}\). By construction,
\(\omega u_2(2k-0.07)<0.01\Delta\). Dividing by \(\Delta>0\) gives \(P>0.9\).

Let \(P_H=\Prb_{H,D}(Q_2\ge2k\mid t_2)\). Decomposing the original probability over the basic opponent event and its complement gives \(P\le W[\ell+(1-\ell)P_H]+\omega\). If \(P_H\le 0.8\), the right side is at most \(0.82W+\omega<0.9\). Thus \(P_H>0.8\) for almost every \(t_2\in D\). Integrating over \(U_D\) gives \(\Prb_{H,D}(Q_2\ge2k)>0.8\).

\pstep{4}
Suppose \(\sigma_{1,m-k+1}(t)\ge h\) for almost every \(t\in H\). On the event \(Q_2\ge2k\) against \(D\), bidder~1 wins at most \(M\) units. Each winning bid is at least \(h\) because \(M<m-k+1\). Every one of the first \(M\) marginal values exceeds \(h\). Therefore, even when \(Q_1<M\),
\[
S_1^{\mathrm D}\le\sum_{j=1}^{Q_1}(t_{1j}-h)
\le\sum_{j=1}^{M}(t_{1j}-h)=X(t)-Mh.
\]
The basic event with opponents in \(D\) has probability \(W\delta\) for every own type. On the additional event \(Q_2\ge2k\), the normalized expression \(e^{rX(t)}e^{-rS_1^{\mathrm D}}\) is at least \(e^{rMh}\), independently of \(t\). Averaging over \(U_H\) therefore permits the use of the joint probability bound from Step~3 and gives
\[
\E_{t\sim U_H}[J_t(\sigma_1(t))]
\ge W\delta e^{rMh}\Prb_{H,D}(Q_2\ge2k)
>0.8W\delta e^{rMh}.
\]

Instead a high type can bid \(a\) on her first \(m-k\) units and zero on the rest. Against \(C\) on the basic opponent event, at most \(k\) opposing bids can reach \(a\), so the first \(m-k\) bids win. Those units are worth at least \(X(t)+0.8k\) and cost \((m-k)a\). Against \(D\), at most \(2k\) opposing bids can reach \(a\), so the first \(M\) units win. They give surplus \(X(t)-Ma\), and further units contribute nonnegative surplus. The deviation therefore gives an average \(J_t\) of at most
\(e^{-r(0.8k-(m-k)a)}+\delta e^{rMa}+\omega e^{rm}\).
Divide both bounds by \(\delta e^{rMa}\). The original lower bound exceeds \(9.7\), since \(M/m\ge 0.9999\) and \(0.8W e^{rM(h-a)} =0.8W e^{0.000025 R(M/m)} \ge 0.8W e^{2.49975}>9.7\). The deviating upper bound after division is less than \(1+e^{-(0.1-a)R}+e^{-1.3R}<2\). The first exponential follows from \((m-k)-M=k\) and \(\delta=e^{-0.7R}\). The second uses \(\omega e^{rm}<e^{-2R}\). These bounds contradict interim optimality averaged over \(U_H\).

\pstep{5}
Step~4 implies that a positive measure of high types bid strictly below \(h\) on unit \(m-k+1\). Choose such a type in the full-measure set where bidder~1's strategy is monotone. Every low type is below it coordinatewise. Hence \(\sigma_{1,m-k+1}(t)<h\) for almost every \(t\in L\).

\pstep{6}
A type in \(C\) can bid \(a_+\) on her first \(k\) units and zero on the rest. Against \(L\) on the basic opponent event, at most \(m-k\) opposing bids can reach \(a_+\), by Step~5 and \(a_+>h>\tau\). Thus the first \(k\) bids win. Those units have value at least \(0.9kc\) and cost \(ka_+=0.8kc\). The deviation gives surplus at least \(0.1kc\) on this event and nonnegative surplus elsewhere.

For the original bid of a fixed \(C\)-type, use the same extra bidders' bids and priority when comparing any low type and any high type in bidder~1's full-measure monotonicity set. Coordinatewise monotonicity of bidder~1's entire bid vector makes each of her bids weakly higher at the high type. This cannot improve the rank of any bid of the \(C\)-type, so she wins weakly fewer units. Accepted own bids form a prefix and each winning unit contributes nonnegative surplus. Thus her original utility against \(H\) is no greater than against \(L\) under this comparison.

Let \(V_C=V_m(t_2)\) for the fixed \(C\)-type. Then \(V_C\le 1.32kc+m\tau<1.33kc\le x_0\). Her equilibrium utility on the basic opponent event is at most \(\E_{L,C}[u_2(S_2^{\mathrm D})\mid t_2]\) by the comparison between \(H\) and \(L\). Her utility on the complement is at most \(u_2(V_C)\). Since the deviation yields at least \(u_2(0.1kc)\) with probability \(W\ell\), interim optimality gives
\(W\ell u_2(0.1kc) \le W\E_{L,C}[u_2(S_2^{\mathrm D})\mid t_2]+\omega u_2(V_C)\).
The local bounds on \(u_2\) give \(u_2(0.1kc)\ge 0.99\lambda(0.1kc)\), \(u_2(V_C)\le\lambda V_C\), and \(u_2(S_2^{\mathrm D})\le\lambda S_2^{\mathrm D}\). Dividing by \(W\lambda>0\), we obtain \(\E_{L,C}[S_2^{\mathrm D}\mid t_2]\ge 0.0099kc-\tfrac{\omega}{W}V_C>0.0098kc\). The last inequality follows from \(V_C<1.33kc\) and \(\omega<10^{-6}\).

Put \(j=k/200\). When \(Q_2<j\), her surplus is at most \(0.91cj=0.00455kc\), since each of the first \(k\) values is at most \(0.91c\). It is always at most \(1.33kc\). Thus
\[
0.0098kc
<\E_{L,C}[S_2^{\mathrm D}\mid t_2]
\le 0.00455kc+1.33kc\,\Prb_{L,C}(Q_2\ge j\mid t_2).
\]
Subtracting \(0.00455kc\) and dividing by \(1.33kc\) gives
\(\Prb_{L,C}(Q_2\ge j\mid t_2)>(0.0098-0.00455)/1.33>0.003\).
Integrating over \(U_C\) yields \(\Prb_{L,C}(Q_2\ge j)>0.003\).

\pstep{7}
When \(Q_2\ge j\), a low type of bidder~1 wins at most \(m-j\) units. The last \(j\) marginal values are each at least \(0.349\), while the full value of the last \(2k\) units is at most \(0.6992k\). Even at zero payments, her surplus is therefore at most \(X(t)+0.6992k-0.349j=X(t)+0.697455k\). The basic event with opponents in \(C\) has probability \(W(1-\delta)\) for every low type. On the additional event \(Q_2\ge j\), the normalized expression \(e^{rX(t)}e^{-rS_1^{\mathrm D}}\) is at least \(e^{-0.697455R}\), independently of \(t\). Averaging over \(U_L\) gives \(\E_{t\sim U_L}[J_t(\sigma_1(t))]\ge W(1-\delta)e^{-0.697455R}\Prb_{L,C}(Q_2\ge j)\). The bound \(\Prb_{L,C}(Q_2\ge j)>0.003\) from Step~6 then implies
\[
\E_{t\sim U_L}[J_t(\sigma_1(t))]>0.003W(1-\delta)e^{-0.697455R}>0.002e^{-0.697455R}.
\]
The last bound exceeds \(K\) in \eqref{eq:cara-cap-loss}, since \(e^{0.001445R}>1500\). But \eqref{eq:cara-cap-loss} bounds \(J_t(\sigma_1(t))\) by \(K\) for almost every low type. This contradiction excludes a pure equilibrium with monotone bids for bidder~1.

In each case, the construction uses a finite \(m\ge3\) and independent type distributions satisfying the density bounds in Section~\ref{sec:model}. It excludes a pure equilibrium with a monotone strategy for bidder~1 and hence excludes a monotone pure-strategy equilibrium.
\end{proof}

\subsection{Two-unit auctions with CARA bidders}\label{app:cara}\label{app:two}
This subsection considers two units and the three payment rules in Sections~\ref{sec:uniform} and~\ref{sec:discriminatory}. There are \(n\ge2\) bidders with independent types \(t_i=(x_i,y_i)\in T_2\). Each type distribution has a density, which need not have full support. We retain the bid space and the independent priority rule from Section~\ref{sec:model}. In place of \textup{(\ref{eq:utilities})}, fix \(r_i\ge0\). Up to positive affine transformations, take \(u_i(s)=1-e^{-r_i s}\) when \(r_i>0\) and \(u_i(s)=s\) when \(r_i=0\), for every \(s\in\R\). Thus the result also allows every bidder to be risk neutral.

\begin{proposition}\label{C:prop:existence}
Under highest-losing-bid uniform pricing, lowest-winning-bid uniform pricing, or discriminatory pricing, the two-unit auction has a monotone pure-strategy equilibrium with no overbidding. Optimality is tested against all ordered nonnegative bid vectors. Under highest-losing-bid pricing, the equilibrium can be chosen in the form \(\sigma_i(x,y)=(x,\beta_i(y))\), where \(\beta_i\) is nondecreasing and \(0\le\beta_i(y)\le y\).
\end{proposition}

\begin{proof}
We first normalize bids, then establish existence separately for highest-losing-bid pricing. For the other two rules, we construct equilibria on finite grids and pass to continuous bids.

\pstep{1}
Lemma~\ref{lem:caps} shows that, under either uniform price, capping each bid at its marginal value weakly increases surplus for every opposing bid profile and priority. Under highest-losing-bid pricing, subsequently raising the first bid to \(x\) also weakly increases surplus. Under discriminatory pricing, capping has the same property. An unchanged bid keeps its acceptance status by the rank observation in the proof of Theorem~\ref{thm:uniform}. A reduced bid that remains accepted costs less, while a unit lost after capping had negative marginal surplus.

Let \(G\subset[0,1]\) be a finite grid containing \(0\) and \(1\). Write \(\lfloor v\rfloor_G\) for the largest grid point at most \(v\), and \(\lceil v\rceil_G\) for the smallest grid point at least \(v\). Under discriminatory pricing, capping bids at \((\lfloor x\rfloor_G,\lfloor y\rfloor_G)\) weakly increases surplus. Every reduced grid bid previously exceeded its marginal value. Under uniform pricing, use the ceiling caps \((\lceil x\rceil_G,\lceil y\rceil_G)\). If a unit is lost after this reduction, its value is no greater than its new bid, which is no greater than the original price by the proof of Lemma~\ref{lem:caps}. The price of retained units weakly falls. Both grid normalizations preserve the ordering of bids.

\pstep{2}
Consider highest-losing-bid pricing. Suppose each opponent uses \((x_j,\beta_j(y_j))\), where \(0\le\beta_j(y_j)\le y_j\). Let \(X\ge Z\) be the two highest opposing bids. Then \(X=\max_{j\ne i}x_j\) is atomless and does not depend on the second-bid functions. Increasing those functions weakly raises \(Z\) for every opposing type profile.

For \(r=r_i>0\), define
\begin{equation*}
F_i(y,b)=-\E\left[
e^{r\max\{b,Z\}}\mathbf1_{\{X>b\}}
+e^{r(2X-y)}\mathbf1_{\{X\le b\}}\right].
\end{equation*}
For \(0\le b\le y\le x\), the expected utility of \((x,b)\) equals \(C_i(x)+e^{-rx}F_i(y,b)\), where \(C_i(x)=1+\E[(e^{r(Z-x)}-1)\mathbf1_{\{Z>x\}}]\). This term accounts for states in which neither bid wins and is independent of \(b\). For \(b'>b\), the difference between the statewise contributions to \(F_i(y,b')\) and \(F_i(y,b)\) is
\[
\begin{cases}
0,&X\le b,\\
e^{r\max\{b,Z\}}-e^{r(2X-y)},&b<X\le b',\\
e^{r\max\{b,Z\}}-e^{r\max\{b',Z\}},&X>b'.
\end{cases}
\]
Each expression is nondecreasing in \(y\) and \(Z\). Hence \(F_i\) has increasing differences between the bid and the second value. It also has increasing differences between the bid and the opponents' second-bid functions. Atomlessness of \(X\) gives continuity in \(b\) by dominated convergence. For \(r_i=0\), the corresponding objective is
\[
F_i^0(y,b)=-\E\left[
\max\{b,Z\}\mathbf1_{\{X>b\}}+(2X-y)\mathbf1_{\{X\le b\}}\right],
\]
whose differences have the same properties.

Choose the greatest maximizer over \([0,y]\). It is nondecreasing in \(y\) and in the opponents' second-bid functions. To verify the first assertion with the moving upper bound, suppose a lower second value has a larger selected bid. Both selected bids are then feasible at both values. Increasing differences and optimality imply that the larger bid is also optimal at the higher value, contrary to the selection of its greatest maximizer.

The nondecreasing functions \(\beta:[0,1]\to[0,1]\) satisfying \(\beta(y)\le y\) form a complete lattice under pointwise order. The product of these lattices over bidders is also complete. Tarski's fixed-point theorem therefore gives a fixed point of the greatest-best-reply maps. Every such function is measurable. Step~1 establishes optimality against all bid vectors, so this fixed point gives the claimed equilibrium.

\pstep{3}
Consider lowest-winning-bid pricing on a finite grid \(G\). Fix the opponents' strategies and include the random priority in all probabilities. Let \(E_1(a)\) be the event that a first bid \(a\) wins, and let \(E_2(b)\) be the event that a second bid \(b\) wins. The rank observation shows that these events depend only on their indicated coordinates. Write \(P_1(a)=\Prb(E_1(a))\) and \(P_2(b)=\Prb(E_2(b))\). Independence makes these probabilities independent of the bidder's type. Both acceptance probabilities are nondecreasing, and \(E_2(b)\subseteq E_1(a)\) when \(a\ge b\). With \(X\ge Z\) denoting the two highest opposing bids, the thresholds for these events are \(Z\) and \(X\), respectively.

When exactly one unit is won, its price is \(\min\{a,X\}\). When two units are won, their common price is \(b\). These formulas hold also at ties. For \(r=r_i>0\), multiplying expected utility by \(e^{rx}\) gives the equivalent objective
\[
H(x,y;a,b)=e^{rx}P_1(a)-\E[e^{r\min\{a,X\}}\mathbf1_{E_1(a)}]+\E[e^{rX}\mathbf1_{E_2(b)}]-e^{2rb-ry}P_2(b).
\]
This is a sum of a function of \((x,a)\) and a function of \((y,b)\). It is therefore modular in the bid vector. It has increasing differences in \((x,a)\), since \(P_1\) is nondecreasing, and in \((y,b)\), since \(e^{2rb}P_2(b)\) is nondecreasing. The risk-neutral objective is
\[
xP_1(a)-\E[\min\{a,X\}\mathbf1_{E_1(a)}]
+\E[X\mathbf1_{E_2(b)}]+(y-2b)P_2(b),
\]
with the same properties.

The ordered grid bids \(\{(a,b)\in G^2:a\ge b\}\) form a finite lattice. Supermodularity and increasing differences imply that the greatest best reply is nondecreasing in \((x,y)\). They also imply that best-reply sets are closed under coordinatewise maxima. Applying the ceiling caps from Step~1 preserves optimality and monotonicity of the selected best reply.

\pstep{4}
Consider discriminatory pricing on the same grid. Use the probabilities \(P_1,P_2\) from Step~3. For \(r=r_i>0\), expected utility is
\begin{equation}\label{C:eq:discriminatory-objective}
P_1(a)(1-e^{-r(x-a)})+e^{-r(x-a)}P_2(b)(1-e^{-r(y-b)}).
\end{equation}
Set \(h(b,y)=P_2(b)(1-e^{-r(y-b)})\). Its derivative with respect to \(y\) is \(r e^{-ry}e^{rb}P_2(b)\), which is nondecreasing in \(b\). Thus \(h\) has increasing differences. Define \(H_0(a,y)=\max\{h(b,y):b\in G,\ b\le a\}\). This function is nondecreasing in \(y\) and has increasing differences in \((a,y)\). Indeed, for \(a'>a\) and each newly feasible bid \(b\in(a,a']\cap G\), the difference \(h(b,y)-H_0(a,y)=\min_{z\in G,\ z\le a}\{h(b,y)-h(z,y)\}\) is nondecreasing in \(y\). Taking the maximum over these bids and zero proves the increasing-differences assertion.

After optimizing the second bid and multiplying \eqref{C:eq:discriminatory-objective} by \(e^{rx}\), the first-bid objective becomes \((e^{rx}-e^{ra})P_1(a)+e^{ra}H_0(a,y)\). It has increasing differences in \((a,x)\) and \((a,y)\). For the latter claim, \(H_0(a,y')-H_0(a,y)\) is nonnegative and nondecreasing in \(a\) whenever \(y'>y\). The greatest first-bid maximizer is consequently nondecreasing in both values. The greatest maximizer of \(h(b,y)\) over \(b\le a\) is nondecreasing in \(a\) and \(y\). These two selections give a monotone best reply.

Best-reply sets are also closed under coordinatewise maxima. The only case requiring verification consists of two optimal pairs \((a,b)\) and \((a',b')\) with \(a<a'\) and \(b>b'\). Both second bids are feasible with either first bid. Optimality of the two pairs gives both \(h(b,y)\ge h(b',y)\) and the reverse inequality. Thus \((a',b)\) is optimal. For \(r_i=0\), the objective is \((x-a)P_1(a)+(y-b)P_2(b)\), which is modular and has increasing differences. Finally, the floor caps from Step~1 preserve optimality and monotonicity of the selected best reply.

\pstep{5}
We now apply \citet[Theorems~4.1 and~4.3]{Reny2011} to each of the last two rules on \(G\). The action lattice is finite, so payoffs are continuous in actions. They are bounded and measurable. Each type distribution is atomless. The countable order condition also holds. A set of positive type probability has positive Lebesgue measure. Near an interior density point it contains two strictly coordinatewise ordered points with a rational point between them.

Steps~3 and~4 give a monotone best reply against every opposing strategy profile. They also show that the set of monotone best replies is closed under pointwise coordinatewise maxima. The set of all measurable strategies satisfying the relevant grid caps almost everywhere is closed under pointwise coordinatewise maxima. It is also closed under measurable pasting and almost-everywhere pointwise limits. This set contains the zero strategy and a monotone best reply. Reny's restricted-strategy result therefore gives a monotone equilibrium satisfying these caps. It remains an equilibrium against every ordered grid bid.

\pstep{6}
For lowest-winning-bid or discriminatory pricing, take these equilibria on \(G_N=\{0,1/N,\ldots,1\}\), with strategies \(\sigma_i^N\). The compactness result for monotone functions in \citet[Lemma~A.10]{Reny2011} gives a common subsequence converging almost surely, for every bidder, to a measurable monotone strategy \(\sigma_i\). Both grid caps imply \(\sigma_{ij}(t_i)\le t_{ij}\) in the limit. We retain the index \(N\) for this subsequence.

We must account for discontinuities of quantities at ties. For a realized limiting bid profile \(b=(\sigma_i(t_i))_{i=1}^n\), let \(\mathcal Q(b)\) be the finite set of quantity vectors obtained by ranking the bids and resolving boundary ties in every possible way. Earlier own units retain priority. Put \(Q_i^+(b)=\max_{q\in\mathcal Q(b)}q_i\). For discriminatory pricing, write \(S_i(k;b)=\sum_{j=1}^k(t_{ij}-b_{ij})\) for \(k\in\{0,1,2\}\). For lowest-winning-bid pricing, write \(p(b)=b_{(2)}\) and \(S_i(k;b)=V_k(t_i)-kp(b)\). The price \(p(b)\) is continuous in bids and independent of the resolution of ties. Because bids are capped, \(S_i(Q_i^+(b);b)\ge S_i(q_i;b)\) for every \(q\in\mathcal Q(b)\). Define \(W_i^+(t)=u_i(S_i(Q_i^+(b);b))\).

Fix \(\delta>0\). On grid \(G_N\), deviate to the upward rounding of \(\min\{\sigma_i(t_i)+\delta\mathbf1,\mathbf1\}\). Types have densities, so each limiting own bid is below one almost surely and is strictly raised by this deviation. For almost every type profile, the deviating bidder eventually wins at least \(Q_i^+(b)\) units. Under discriminatory pricing, each accepted own bid rises by at most \(\delta\) in the limit. Under lowest-winning-bid pricing, the limiting price rises by at most \(\delta\). Any additional accepted unit has original bid at least \(p(b)-\delta\), so its value is at least \(p(b)-\delta\). In either case, the limiting inferior of deviating surplus is at least \(S_i(Q_i^+(b);b)-4\delta\). Grid rounding errors vanish.

Let \(U_i^N\) denote ex ante equilibrium utility. Boundedness of utilities on the relevant surplus interval and equilibrium optimality give \(\liminf_N U_i^N\ge\E[u_i(S_i(Q_i^+(b);b)-4\delta)]\). Letting \(\delta\downarrow0\) yields
\begin{equation}\label{C:eq:favorable}
\liminf_N U_i^N\ge\E W_i^+.
\end{equation}

\pstep{7}
For each type profile, define \(D(t)=\min_{q\in\mathcal Q(b)}\sum_i[W_i^+(t)-u_i(S_i(q_i;b))]\). Each summand is nonnegative. Every strict comparison between limiting bids holds eventually along the sequence. Thus the grid quantity vector eventually belongs to \(\mathcal Q(b)\). For a fixed quantity, the corresponding payments converge. The reverse Fatou inequality for bounded utilities gives \(\limsup_N\sum_i U_i^N\le\sum_i\E W_i^+-\E D\). Combining this with \eqref{C:eq:favorable} gives \(D=0\) almost surely.

A boundary tie is competitive if bids from different bidders tie for fewer remaining units than those tied bids request. Independence and the type densities imply
\(\Prb(t_{ij}=\sigma_{h\ell}(t_h))=0\) whenever \(i\ne h\). There are finitely many such pairs. Hence, almost surely, every unit involved in a competitive boundary tie has marginal value strictly above its tied bid. This bid is the price of that unit under either payment rule. At any allocation in \(\mathcal Q(b)\), at least one bidder receives fewer units than \(Q_i^+(b)\). Her lost units have strictly positive marginal surplus. Strict increase of utility then makes the corresponding summand in \(D\) positive. Since \(\mathcal Q(b)\) is finite, a competitive boundary tie implies \(D>0\). This contradicts the preceding conclusion and rules out such ties, including ties at zero.

Without competitive boundary ties, each bidder's quantity is locally constant at the limiting bid profile. Payments therefore converge under the original priority rule. Dominated convergence gives \(U_i^N\to U_i(\sigma)\), where \(\sigma=(\sigma_1,\ldots,\sigma_n)\) is the limiting strategy profile and \(U_i(\sigma)\) is its ex ante utility.

\pstep{8}
Let \(d_i:T_2\to B_2\) be any measurable deviation. By Step~1, it suffices to consider \(0\le d_{ij}(t_i)\le t_{ij}\). Write \(b^d=(d_i(t_i),b_{-i})\) for the realized bid profile after this deviation. Raise only bidder~\(i\)\textquotesingle s vector \(d_i(t_i)\) by \(\delta\), cap each coordinate at one, and round upward to \(G_N\). The argument in Step~6 applies against the converging opposing bids. Equilibrium optimality, followed by \(N\to\infty\) and \(\delta\downarrow0\), gives
\[
U_i(\sigma)\ge
\E\left[u_i\bigl(S_i(Q_i^+(b^d);b^d)\bigr)\right]
\ge U_i(d_i,\sigma_{-i}).
\]
The second inequality holds because the capped deviation weakly benefits from receiving every unit available under a favorable resolution of ties.

These inequalities imply interim optimality almost everywhere. To make this implication explicit, upward rational approximations to any capped bid achieve a limiting utility at least as high as its actual utility. This gives a countable set of ordered rational bids sufficient to detect any profitable deviation. If interim optimality failed on a set of positive type probability, one such bid would be profitable on a measurable subset of positive probability. Using that bid on the subset gives a measurable deviation. Capping it at each type's marginal values weakly raises utility by Step~1, so the ex ante inequality gives a contradiction. Step~1 also covers bids above one. The limiting profile is therefore a monotone pure-strategy equilibrium against all ordered nonnegative bids.
\end{proof}

The proposition gives a two-unit existence result with the type order used in the nonexistence proofs. It allows heterogeneous CARA coefficients and densities without full support. The conclusion concerns a no-overbidding equilibrium and does not assert admissibility.


\clearpage
\section{Appendix B}\label{app:further-results}

\subsection{Additional uniform prices for \texorpdfstring{$\alpha\in(0,1)$}{0<alpha<1}}\label{app:interior-prices}
\begin{proof}[Proof of Remark~\ref{rem:uniform_prices}]
Use the type distributions constructed in Appendix~\ref{app:uniform-proof}, with the parameters fixed as in Step~6 there. We show that one sufficiently large finite index \(r\) excludes an admissible monotone equilibrium for every \(\alpha\in[0,1]\).

The capping argument for Lemma~\ref{lem:caps} applies to every price \(p_\alpha\). Capping weakly lowers both price order statistics. A unit lost after capping has value at most the original highest losing bid, which is at most \(p_\alpha\). The same opposing bid profile used in that lemma gives strict improvement when an own bid exceeds its marginal value. Thus it suffices to exclude equilibria with no overbidding in which bidder~1's last bid is nondecreasing on a full measure set. Call these candidate equilibria.

Fix \(\alpha\in[0,1]\) and a candidate equilibrium \(\sigma^r\) at an index where one exists. Write \(b_{ij}^r=\sigma_{ij}^r(t_i^r)\) for realized bids and \(\beta_r=b_{1m}^r\). Put \(\bar p^r=b_{(m)}^r\) and \(\underline p^r=b_{(m+1)}^r\), so \(p^r=(1-\alpha)\bar p^r+\alpha\underline p^r\). Use the outcome notation, basic-distribution expectations, and constants \(\ell_i,L_i,W_i,e_{ir}\) from Step~2 of Appendix~\ref{app:uniform-proof}. All deviations below satisfy the value caps, so original and deviating surpluses lie in \([0,m]\). The utility change under a deviation is \(H_i^{r,d}=u_i(S_i^{r,d})-u_i(S_i^r)\). Equation~\eqref{eq:derivatives} gives \(H_i^{r,d}\ge\ell_i g-L_i h\) whenever \(\Delta S_i^{r,d}\ge g-h\), with \(g,h\ge0\).

We apply \eqref{eq:obedience} with \(\nu_1=\mu_1^r\) for bidder~1. Her expected gain is \(\E^rH_1^{r,d}=(1-\delta)\E_C^rH_1^{r,d}+\delta\E_D^rH_1^{r,d}\). For bidder~2, we use \(\nu_2=U_{D_m(r)}\) or \(U_{C_m(r)}\), giving the expectation \(\E_D^r\) or \(\E_C^r\), respectively. In each case, \eqref{eq:obedience} bounds the indicated own-type average by \(e_{ir}\). Bidder~1's deviations below change her strategy only on \(L_m\cup H_m\).

\pstep{1}
Let \(F_r=\{Q_1^r=1,Q_2^r=m-1\}\) and \(P_{1r}^D=\Prb_D^r(Q_1^r=0)\). For \(t_1\in L_m\cup H_m\), let \(d_{1j}^{r,+h}(t_1)=\min\{\sigma_{1j}^r(t_1)+h,t_{1j}\}\). Each of the two price order statistics rises by at most \(h\). Hence \(p^{r,+h}\le p^r+h\), and the surplus loss is at most \(mh\), as in Step~4 of Appendix~\ref{app:uniform-proof}. With \(h=2\rho_r\), a type winning nothing in state \(D\) gains at least \(z/2\) for large \(r\). Let \(J_r\) indicate this event, including state \(D\). Then \(\Delta S_1^{r,+2\rho_r}\ge(z/2)J_r-2m\rho_r\), so \eqref{eq:derivatives} gives \(H_1^{r,+2\rho_r}\ge\ell_1(z/2)J_r-2L_1m\rho_r\). Since \(\E^rJ_r=\delta P_{1r}^D\), \eqref{eq:obedience} with \(\nu_1=\mu_1^r\) gives
\[
e_{1r}\ge\E^rH_1^{r,+2\rho_r}\ge\delta\ell_1(z/2)P_{1r}^D-2L_1m\rho_r.
\]
Rearranging yields
\begin{equation}\label{ext:q}
P_{1r}^D\le\frac{e_{1r}+2L_1m\rho_r}{\delta\ell_1z/2}=O(r^{-1}).
\end{equation}

Define the weighted gaps \(\widetilde X_r=(1-\alpha)\E_D^r[(\bar p^r-b_{12}^r)\1_{F_r}]\) and \(\widetilde Y_r=(1-\alpha)\E_D^r[(b_{11}^r-b_{12}^r)\1_{F_r}]\). Apply the first-bid reduction \(d_1^{r,-}\) from Step~4 of Appendix~\ref{app:uniform-proof}, replacing \(b_{11}^r\) by \(\min\{b_{11}^r,\max\{b_{12}^r,2\rho_r\}\}\). It preserves the acceptance of every bid under the basic distributions. The highest losing bid is therefore unchanged, and the price weakly falls. On \(F_r\), the old lowest winning bid is \(\bar p^r=\min\{b_{11}^r,b_{2,m-1}^r\}\ge b_{12}^r\). The new one is at most \(b_{12}^r+2\rho_r\). Thus \eqref{eq:derivatives} gives utility gain at least \(\ell_1(1-\alpha)(\bar p^r-b_{12}^r-2\rho_r)_+\) there. The gain is nonnegative elsewhere in either state. Equation~\eqref{eq:obedience} with \(\nu_1=\mu_1^r\) yields
\[
e_{1r}\ge(1-\delta)\E_C^rH_1^{r,-}+\delta\E_D^rH_1^{r,-}
\ge\delta\ell_1(1-\alpha)\E_D^r[(\bar p^r-b_{12}^r-2\rho_r)_+\1_{F_r}].
\]
On \(F_r\), \(\bar p^r-b_{12}^r\le(\bar p^r-b_{12}^r-2\rho_r)_++2\rho_r\). Multiply by \((1-\alpha)\1_{F_r}\), take \(\E_D^r\), and use the preceding bound to obtain
\begin{equation}\label{ext:X}
\widetilde X_r\le\frac{e_{1r}}{\delta\ell_1}+2(1-\alpha)\rho_r\le\frac{e_{1r}}{\delta\ell_1}+2\rho_r=O(r^{-1}).
\end{equation}

The four-event argument in Step~4 of Appendix~\ref{app:uniform-proof} now applies with \(\bar p^r\) in place of its price variable. Put \(h_r=r^{-2/3}\) and \(\theta_r=r^{-1/3}\). The events where the bid gap is at most \(h_r\), or where \(\bar p^r-b_{12}^r>h_r/2\), contribute at most \(h_r\) and \(2\widetilde X_r/h_r\) to \(\widetilde Y_r\). On the remaining profiles, \(\bar p^r<b_{11}^r\), so \(\bar p^r=b_{2,m-1}^r\). If \(t_{12}^r-\bar p^r\le\theta_r\), no overbidding places \(t_{12}^r\) in an interval of length \(\theta_r+h_r/2\) around \(b_{2,m-1}^r\). Independence and its density bound \(1/\varepsilon\) bound the probability of this third event by \((\theta_r+h_r/2)/\varepsilon\). Let \(E_r\) be the fourth event in \(F_r\), on which \(b_{11}^r-b_{12}^r>h_r\), \(\bar p^r-b_{12}^r\le h_r/2\), and \(t_{12}^r-\bar p^r>\theta_r\). The capped increase by \(h_r\) wins a second unit there. Since \(p^r\le\bar p^r\) and \(p^{r,+h_r}\le p^r+h_r\), its surplus gain is at least \(\theta_r-mh_r>0\) for large \(r\). Equation~\eqref{eq:derivatives} gives utility gain at least \(\ell_1(\theta_r-mh_r)\) on this event in state \(D\), and utility gain at least \(-L_1mh_r\) elsewhere. Equation~\eqref{eq:obedience} with \(\nu_1=\mu_1^r\) therefore gives
\[
e_{1r}\ge\E^rH_1^{r,+h_r}
\ge\delta\ell_1(\theta_r-mh_r)\Prb_D^r(E_r)-L_1mh_r.
\]
Thus \(\Prb_D^r(E_r)\le(e_{1r}+L_1mh_r)/[\delta\ell_1(\theta_r-mh_r)]\). The weighted bid gap is at most one, so this also bounds its contribution to \(\widetilde Y_r\). Combining the four contributions gives
\begin{equation}\label{ext:Y}
\widetilde Y_r\le h_r+\frac{2\widetilde X_r}{h_r}+\frac{\theta_r+h_r/2}{\varepsilon}
+\frac{e_{1r}+L_1mh_r}{\delta\ell_1(\theta_r-mh_r)}=O(r^{-1/3}).
\end{equation}
The constants in these bounds do not depend on \(\alpha\).

\pstep{2}
Set \(R_*=\tfrac12(0.90/m+0.90/(m-1))\) and \(\kappa=0.90-(m-1)R_*=0.45/m>0\). Use bidder~2's deviation \eqref{eq:expensivestrategy}, raising her first \(m-1\) bids to at least \(R_*\) on \(D_m(r)\). If \(Q_2^r\le m-2\), she obtains \(m-1\) units at a price no greater than \(R_*\). The surplus gain is at least \(V_{m-1}(t_2^r)-(m-1)R_*-V_{m-2}(t_2^r)=t_{2,m-1}^r-(m-1)R_*\ge\kappa\). On \(F_r\), the acceptance of every bid is unchanged, so \(\underline p^r\) is unchanged. The lowest winning bid rises by at most \(b_{11}^r-b_{12}^r\). Her surplus loss is consequently at most \((1-\alpha)(m-1)(b_{11}^r-b_{12}^r)\). If she already wins all units, both price order statistics are unchanged. The remaining event is \(R_r=\{Q_2^r=m-1,Q_1^r=0\}\), where the utility loss is at most \(W_2\). Equation~\eqref{eq:derivatives} converts the surplus gain on \(\{Q_2^r\le m-2\}\) into utility gain at least \(\ell_2\kappa\). It bounds the utility loss on \(F_r\) by \(L_2(1-\alpha)(m-1)(b_{11}^r-b_{12}^r)\). Set \(P_{2r}^D=\Prb_D^r(Q_2^r\le m-2)\). Since \(\Prb_D^r(R_r)\le P_{1r}^D\), \eqref{eq:obedience} with \(\nu_2=U_{D_m(r)}\) gives
\[
e_{2r}\ge\E_D^rH_2^{r,D}\ge\ell_2\kappa P_{2r}^D-L_2(m-1)\widetilde Y_r-W_2P_{1r}^D.
\]
There is no factor \(\delta\) here because bidder 2's own type is integrated only over her normalized \(D\) component. As \(Q_1^r\ge2\) implies \(Q_2^r\le m-2\), rearranging gives
\begin{equation}\label{ext:a}
K_r^D\le P_{2r}^D\le\frac{e_{2r}+L_2(m-1)\widetilde Y_r+W_2P_{1r}^D}{\ell_2\kappa}=O(r^{-1/3}).
\end{equation}

Next use bidder~1's flat-bid deviation from Step~5 of Appendix~\ref{app:uniform-proof}. On \(L_m\cup H_m\), set \(d_{1j}^{r,\mathrm f}(t_1)=\min\{\sigma_{1m}^r(t_1)+2\rho_r,t_{1j}\}\). It wins at least \(m-1\) units in state \(C\) and at least one in state \(D\), at a price at most \(\beta_r+2\rho_r\). In state \(C\), its quantity is no smaller than before. If she originally wins all units, replacing all bids by \(\beta_r\) preserves acceptance under the common bidder priority. Raising these bids subject to their value caps also preserves acceptance. Each new bid is at most its old level plus \(2\rho_r\), so both price order statistics increase by at most \(2\rho_r\). Her surplus loss in state \(C\) is therefore at most \(2m\rho_r\).

In state \(D\), zero original units give a nonnegative gain. With one original unit, \(\beta_r\) is rejected, so \(p^r\ge\beta_r\). The surplus gain is at least \((p^r-\beta_r)-2\rho_r\). Apply \eqref{eq:derivatives} with \(g=p^r-\beta_r\ge0\) and \(h=2\rho_r\) to obtain utility gain at least \(\ell_1(p^r-\beta_r)-2L_1\rho_r\). With at least two original units, the utility loss is at most \(W_1\). The statewise expected gains therefore satisfy
\[
\E_C^rH_1^{r,\mathrm f}\ge-2L_1m\rho_r,
\qquad \E_D^rH_1^{r,\mathrm f}\ge\ell_1T_r-2L_1\rho_r\Prb_D^r(Q_1^r=1)-W_1K_r^D,
\]
where \(T_r=\E_D^r[(p^r-\beta_r)\1_{\{Q_1^r=1\}}]\). Equation~\eqref{eq:obedience} with \(\nu_1=\mu_1^r\) averages both rival states, giving
\[
\begin{aligned}
e_{1r}&\ge(1-\delta)\E_C^rH_1^{r,\mathrm f}+\delta\E_D^rH_1^{r,\mathrm f}\\
&\ge\delta\ell_1T_r-2L_1\rho_r[(1-\delta)m+\delta\Prb_D^r(Q_1^r=1)]-\delta W_1K_r^D\\
&\ge\delta\ell_1T_r-2L_1m\rho_r-\delta W_1P_{2r}^D.
\end{aligned}
\]
The last inequality uses \((1-\delta)m+\delta\Prb_D^r(Q_1^r=1)\le m\) and \(K_r^D\le P_{2r}^D\). Hence
\begin{equation}\label{ext:T}
T_r
\le\frac{e_{1r}+2L_1m\rho_r+\delta W_1P_{2r}^D}{\delta\ell_1}=O(r^{-1/3}).
\end{equation}

Set \(\widetilde Z_r=(1-\alpha)\E^r(b_{11}^r-\beta_r)=(1-\alpha)\E_D^r(b_{11}^r-\beta_r)\), using independence. On \(F_r\), the rejected last bid gives \(\underline p^r\ge\beta_r\). Hence \(p^r-\beta_r\ge (1-\alpha)(\bar p^r-\beta_r)\), which implies \((1-\alpha)(b_{11}^r-\beta_r)\le (1-\alpha)(b_{11}^r-b_{12}^r)+(p^r-\beta_r)\). Also \(\Prb_D^r(F_r^c)\le P_{2r}^D+P_{1r}^D\), and the weighted bid gap is at most one. It follows that
\begin{equation}\label{ext:Z}
0\le \widetilde Z_r\le \widetilde Y_r+T_r+P_{2r}^D+P_{1r}^D=O(r^{-1/3}).
\end{equation}

\pstep{3}
Use bidder~2's deviation \eqref{eq:cheapstrategy}, raising only her first bid to \(t_{21}\) on \(C_m(r)\). Write \(g_r=b_{11}^r-\beta_r\) and \(A_r^C(a)=\{Q_1^r=m,\beta_r\le a\}\), where \(0<a<c\). If her first bid already wins, the acceptance of every bid and \(\underline p^r\) are unchanged. The old lowest winning bid is at least \(\beta_r\), while the new one is at most \(\max\{b_{11}^r,\rho_r\}\). Her surplus loss is at most \((1-\alpha)m(g_r+\rho_r)\). If her first bid loses, her original surplus is zero and the new surplus is nonnegative. On \(A_r^C(a)\), she wins exactly one unit after deviating. Bidder 1's last bid becomes the highest losing bid, so the new price is at most \((1-\alpha)b_{11}^r+\alpha\beta_r\). Her new surplus is at least \(c-\beta_r-(1-\alpha)g_r\ge c-a-(1-\alpha)g_r\). Combining the cases gives \(\Delta S_2^{r,C}\ge(c-a)\1_{A_r^C(a)}-(1-\alpha)m(g_r+\rho_r)\). Apply \eqref{eq:derivatives} with \(g=(c-a)\1_{A_r^C(a)}\) and \(h=(1-\alpha)m(g_r+\rho_r)\) to obtain
\[
H_2^{r,C}\ge\ell_2(c-a)\1_{A_r^C(a)}-L_2(1-\alpha)m(g_r+\rho_r).
\] Equation~\eqref{eq:obedience} with \(\nu_2=U_{C_m(r)}\) gives
\[
\begin{aligned}
e_{2r}\ge\E_C^rH_2^{r,C}
&\ge\ell_2(c-a)K_r^C(a)-L_2(1-\alpha)m(\E_C^rg_r+\rho_r)\\
&=\ell_2(c-a)K_r^C(a)-L_2m(\widetilde Z_r+(1-\alpha)\rho_r).
\end{aligned}
\]
The equality uses \((1-\alpha)\E_C^rg_r=\widetilde Z_r\), since bidder 1's type distribution is independent of bidder 2's component. Again there is no factor \(1-\delta\), because bidder 2's own type is integrated over the normalized \(C\) component. Rearranging yields
\begin{equation}\label{ext:KC}
K_r^C(a)\le\frac{e_{2r}+L_2m(\widetilde Z_r+(1-\alpha)\rho_r)}{\ell_2(c-a)}\longrightarrow0.
\end{equation}

We have established the two bounds in \eqref{eq:statebounds}, uniformly over \(\alpha\in[0,1]\) and all candidate equilibria. Step~6 of Appendix~\ref{app:uniform-proof} now gives a contradiction. Its flat-bid deviations have price at most the flat bid for every \(\alpha\). When the original last bid is rejected, \(p^r\ge\beta_r\), so the original-surplus bounds used there also remain valid. The positive utility margins in \eqref{eq:margins} are independent of \(\alpha\). Thus the comparisons \eqref{eq:lowfinal} and \eqref{eq:highfinal} hold with the same fixed parameters. One sufficiently large finite \(r\) excludes candidate equilibria for all \(\alpha\in[0,1]\). The initial capping argument then proves the claim for admissible monotone equilibria.
\end{proof}


\subsection{All-pay auctions}\label{app:allpay}\label{cc:sec:construction}
Use the model in Section~\ref{sec:model}. Under all-pay pricing, bidder \(i\) pays every own bid, including rejected bids. Her surplus is \(S_i^{\mathrm{AP}}=V_{Q_i}(t_i)-\sum_{j=1}^m b_{ij}\), and expected utility is \(\E[u_i(S_i^{\mathrm{AP}})]\). All ordered nonnegative bids are feasible. A bid vector satisfying the marginal-value caps can still give negative surplus because rejected bids also cost money.

\begin{theorem}\label{thm:allpay}
For every \(n\ge2\), \(m\ge3\), and utility profile satisfying \textup{(\ref{eq:utilities})}, there exist independent type distributions on \(T_m\) with densities satisfying \(0<\underline f_i\le f_i(t_i)\le\overline f_i<\infty\) almost everywhere such that the continuous-bid all-pay auction has no pure equilibrium whose third bid \(\sigma_{13}(t_1)\) is coordinatewise nondecreasing on a full-measure set. In particular, there is no monotone pure equilibrium. The conclusion does not impose admissibility or a no-overbidding restriction. It also holds when the feasible bids are restricted to \([0,1]\).
\end{theorem}

We first establish bid caps. We then construct an auxiliary contest with unique equilibrium distributions. Its second bidder's distribution gives two ordered types of bidder~1 strictly reversed third bids. Finally, we approximate the contest by full-dimensional type distributions with full support.

\begin{lemma}\label{ap:lem:caps}
Every all-pay best response satisfies \(b_{ij}\le t_{ij}\) for every \(j\). This holds for continuous bids and for any finite bid grid containing zero, whenever monetary utility is strictly increasing.
\end{lemma}
\begin{proof}
Fix an own type and suppose \(b_{ij}>t_{ij}\). Delete bid \(j\), shift all later bids one position forward, and append zero. The new vector is feasible in either model. Fix the opponents' bids and the priority. The relative rankings of the retained bids do not change, so quantity falls by at most one. If quantity falls from \(Q\) to \(Q-1\), the deleted bid was accepted and \(Q\ge j\). The lost value is therefore \(t_{iQ}\le t_{ij}\). Payment falls by exactly \(b_{ij}\), so surplus rises by at least \(b_{ij}-t_{ij}>0\). If quantity is unchanged, surplus rises by \(b_{ij}>0\). Strict increase of utility makes this deviation profitable.
\end{proof}

Equilibrium bids consequently lie in \([0,1]^m\). All deviations used below lie in that set as well, so their surpluses lie in \([-m,m]\). Finite concavity gives continuity of each utility on this compact interval. The scalar construction uses concavity only at positive surpluses.

Fix \(n\ge2\) and \(m\ge3\). Bidders~1 and~2 are the active bidders. Their positive bid blocks have lengths \(2\) and \(m-1\). All other bidders' types will concentrate near zero. Bidder~1 will also have a small positive probability of a comparison type. A further uniform component gives every bidder full support. Throughout this subsection write \(u=u_1\). The utility assumptions remain those in \textup{(\ref{eq:utilities})}, including strict concavity of \(u\).

\phantomsection\label{cc:subsec:parameters}
Set \(s=m-1\), \(A=H=\tfrac34\), \(v_*=\tfrac1{4s}\), and \(w_2=(m-2)H\). Let \(\kappa=u(m+1)-u(m)>0\) and \(M=4[u(1/2)-u(1/4)]>0\). For \(1/2\le x<y\le m\), the secant-slope comparison for concave functions gives \(\kappa(y-x)\le u(y)-u(x)\le M(y-x)\). Define \(D(B)=u(A+B)-u(A)\) and \(\rho=2M/D(v_*)\). Choose
\begin{equation}\label{cc:eq:B-choice}
0<B_L<B_H<\min\{v_*/4,\kappa/(2\rho M)\}.
\end{equation}
Let \(I_L=u(A+B_L+C)-u(A+B_L)\) and \(I_H=u(A+B_H+C)-u(A+B_H)\). By continuity, we can choose \(C\in(0,B_L)\) sufficiently small that \(2I_L<D(B_L)\). Strict concavity gives \(I_L>I_H>0\), since a fixed positive increment in surplus is worth strictly less at a higher initial surplus. There is therefore a unique \(v_\ell\in(0,B_L)\) such that
\(D(v_\ell)=I_L+I_H=:d_0\).
In particular,
\begin{equation}\label{cc:eq:key-increments}
2I_H<d_0<2I_L,\qquad d_0<D(B_L)<D(B_H).
\end{equation}
These finite-increment comparisons are the strict curvature inequalities needed below. No derivative of \(u\) is assumed.

For \(v>0\) and \(z\ge0\), write \(\Delta_v(z)=u(A+v-2z)-u(A-2z)\).
Initially choose \(0<a<\min\{C,v_\ell/2\}\). For \(0\le z\le a\), define \(F_\ell(z)=[u(A)-u(A-2z)]/\Delta_{v_\ell}(z)\). The restriction \(2a<v_\ell\) gives \(0<F_\ell(z)<1\) for \(0<z\le a\). Put \(K=u(A-2a)+F_\ell(a)\Delta_{v_*}(a)\), \(L=[A+v_*-u^{-1}(K)]/2\), and \(v_2=sL\). The inverse is well defined. Indeed, \(K-u(A)=F_\ell(a)[u(A+v_*-2a)-u(A+v_\ell-2a)]>0\), whereas \(K<u(A+v_*-2a)\). Consequently, \(a<L<v_*/2,\qquad 0<v_2<sv_*/2=1/8<H\).
As \(a\downarrow0\), we have \(F_\ell(a)\to0\), \(K\to u(A)\), \(L\to v_*/2\), and \(v_2\to1/8\). Reduce \(a\), if necessary, so that
\(B_H<L\).
All further restrictions on \(a\) will leave \(B_L,B_H,C,v_\ell\) fixed. Since \(v_*\le1/8\), every \(z\in[0,L]\) satisfies \(A-2z\ge A-v_*\ge5/8\) and \(w_2-sz\ge w_2-v_2>5/8\). The corresponding surpluses after winning the contested unit are below \(m\). Thus all arguments in the following formulas lie in \([1/2,m]\).

Define
\begin{equation}\label{cc:eq:G2}
G_2(z)=
\begin{cases}
F_\ell(z),&0\le z\le a,\\
\dfrac{K-u(A-2z)}{\Delta_{v_*}(z)},&a\le z\le L,
\end{cases}
\end{equation}
and, for \(0\le z\le L\),
\begin{equation}\label{cc:eq:G1}
G_1(z)=\frac{u_2(w_2)-u_2(w_2-sz)}{u_2(w_2+v_2-sz)-u_2(w_2-sz)}.
\end{equation}
Extend both functions by zero below zero and by one above \(L\). Let \(p_\ell=G_1(a)\in(0,1)\).

\begin{lemma}\label{cc:lem:CDFs}
The functions \(G_1,G_2\) are continuous distribution functions, strictly increasing on \([0,L]\), with value zero at zero and one at \(L\). Moreover, \(G_2\) is \(\rho\)-Lipschitz on \([a,\infty)\).
\end{lemma}
\begin{proof}
The two expressions for \(G_2\) agree at \(a\). The definition of \(L\) gives \(G_2(L)=1\). On the first interval, \(u(A)\) lies strictly between the two state utilities whenever \(0<z\le a\). On the second interval, \(u(A-2z)<K<u(A+v_*-2z)\) whenever \(a\le z<L\). The ratios therefore belong to \((0,1)\) away from the endpoints. Each ratio strictly increases. Indeed, increasing the bid strictly lowers both state utilities, so maintaining the prescribed expected utility requires a strictly higher winning probability. The same argument applies to \(G_1\), using \(sL=v_2\). Continuity follows from continuity of utility and positivity of the denominators.

For \(z\in[a,L]\), put \(N(z)=K-u(A-2z)\) and \(D_*(z)=\Delta_{v_*}(z)\). Both are positive. Concavity makes \(D_*\) nondecreasing and gives \(D_*(z)\ge D(v_*)\). Thus, for \(a\le z<w\le L\),
\[
0\le G_2(w)-G_2(z)
\le\frac{N(w)-N(z)}{D_*(w)}
\le\frac{2M(w-z)}{D(v_*)}=\rho(w-z).
\]
The constant extension beyond \(L\) preserves the bound.
\end{proof}

\phantomsection\label{cc:subsec:scalar}
Consider an auxiliary contest with two bidders. Bidder~1 always receives baseline value \(A\). Her value for one contested unit is \(v_\ell\) with probability \(p_\ell\), and \(v_*\) with probability \(1-p_\ell\). Bidder~2 always receives baseline value \(w_2\) and has contested-unit value \(v_2\). Their nonnegative scalar bids are \(x\) and \(z\). Their payments are \(2x\) and \(sz\), respectively. The higher bid wins the contested unit. A fair random priority resolves equal bids. Baseline values are received even at a zero bid. This convention belongs to the auxiliary contest. The approximation argument below will establish the required baseline allocations in the auction limit.

The corresponding auction types are \(t_1^\ell=(A,v_\ell,0,\ldots,0)\), \(t_1^*=(A,v_*,0,\ldots,0)\), and \(t_2^0=(\underbrace{H,\ldots,H}_{m-2},v_2,0)\). The bid vectors are \(d_1(x)=(x,x,0,\ldots,0)\) and \(d_2(z)=(\underbrace{z,\ldots,z}_{s},0)\).
For positive unequal scalar bids, bidder~1 receives one unit and wins a second exactly when \(x>z\). Bidder~2 receives \(m-2\) units and wins one more exactly when \(z>x\). Their surpluses are then those of the auxiliary contest.

\begin{proposition}\label{cc:prop:unique-scalar}
The auxiliary contest has a mixed equilibrium with bidder~2's bid distribution \(G_2\) and bidder~1's aggregate bid distribution \(G_1\). Bidder~1's conditional distribution is \(G_1(z)/p_\ell\) on \([0,a]\) at the low type and \((G_1(z)-p_\ell)/(1-p_\ell)\) on \([a,L]\) at the high type. These conditional distributions and bidder~2's distribution are the unique mixed equilibrium distributions. The utilities of the low type, the high type, and bidder~2 are, respectively,
\(u(A)\), \(K\), and \(u_2(w_2)\).
\end{proposition}
\begin{proof}
\pstep{1}
We verify the proposed equilibrium. Against \(G_2\), the expected utility of bidder~1 with contested value \(v\) from scalar bid \(z\) is
\(U_v(z)=u(A-2z)+G_2(z)\Delta_v(z)\).
It equals \(u(A)\) for the low type on \([0,a]\) and \(K\) for the high type on \([a,L]\). On \([0,L]\), we have \(U_{v_*}(z)-U_{v_\ell}(z)=G_2(z)J(z)\), where \(J(z)=u(A+v_*-2z)-u(A+v_\ell-2z)\). Concavity makes \(J\) nondecreasing, and \(J\) is positive. Since \(G_2\) strictly increases, their product strictly increases. The high type therefore strictly prefers her support to bids below \(a\). The low type strictly prefers her support to bids above \(a\). Bids above \(L\) win surely and cost strictly more than \(L\). Equation~\eqref{cc:eq:G1} makes bidder~2 indifferent on \([0,L]\), with utility \(u_2(w_2)\). Her bids above \(L\) are also worse. The stated conditional distributions aggregate to \(G_1\), proving existence.

\pstep{2}
Let \(F_1,F_2\) be the aggregate bid distributions in any mixed equilibrium. A payment above the contested value gives surplus strictly below the baseline even after winning. A zero bid gives at least baseline utility. Thus a type of bidder~1 with value \(v\) never bids above \(v/2\), and bidder~2 never bids above \(v_2/s=L\). All equilibrium bids lie in a compact interval.

There is no shared atom, including at zero. At any such atom, a type that assigns it positive probability can bid slightly more. The deviation converts a tie against the opponent's atom into a certain win. The limiting utility gain is positive because the contested value is positive, whereas the payment increase tends to zero.

There is no positive atom held by only one bidder either. Suppose bidder~\(i\) has an atom at \(z>0\). The opponent has no atom at \(z\). Every opponent type strictly prefers a bid just above \(z\) to all bids in a sufficiently short interval directly below \(z\). To see this uniformly, the gain from winning against the atom is bounded below by its mass times a strictly positive utility increment. The minimum of this increment on the relevant compact surplus interval is positive. The utility cost of the additional payment tends to zero by uniform continuity. There are only finitely many types, so one interval \((z-\delta,z)\), with \(0<\delta<z\), works for all of them. The opponent has no mass in \((z-\delta,z]\). A type that bids \(z\) with positive probability can then shade to \(z-\delta/2\), keep her winning probability, and pay strictly less. This is a contradiction.

Neither bidder can concentrate at zero. Against such a bidder, every positive bid can be lowered while remaining positive. A zero bid is worse than a sufficiently small positive bid because it does not win the contested unit surely. The opponent would therefore have no best response.

The positive supports coincide. If one bidder had a positive support point outside the other's support, it would lie in an open interval with constant winning probability. Shading within that interval strictly lowers payment. This contradicts optimality at the support point. Expected utility is continuous at every positive bid because there are no positive atoms. Moreover, each point in an aggregate support belongs to the support of some type, since the type set is finite.

The common positive support has lower endpoint zero and no gap. A positive lower endpoint could be shaded without changing the probability of beating the opponent's possible zero atom. At the upper endpoint of a gap, shading into the gap also preserves winning probability because the opponent has no positive atom. Hence the closure of the common positive support is \([0,T]\) for some \(T>0\). Both CDFs are continuous and strictly increasing on \((0,T]\). They may still have atoms at zero.

\pstep{3}
At a positive scalar bid \(z\), the difference between the high and low types' payoffs is
\(F_2(z)J(z)\).
It strictly increases on \((0,T]\). If an optimal high-type bid were below an optimal low-type bid, adding their optimality inequalities would contradict this strict increase. The bids in the common support satisfy \(z\le L<v_*/2\). Hence \(A-2z\ge5/8\) and \(w_2-sz>5/8\), so the concavity comparisons use arguments in \([1/2,m]\).

Every positive low-type bid therefore precedes every positive high-type bid. If bidder~1 has an atom at zero, bidder~2 has none. The same comparison then includes zero because \(F_2(0)=0\). Write \(p_i=F_i(0)\). Whenever \(p_1<p_\ell\), there is a unique cutoff \(c\in(0,T)\) with \(F_1(c)=p_\ell\). The low type's support ends at \(c\), and the high type's support starts there. Both payoff equations hold at \(c\) by continuity.

\pstep{4}
Suppose first that \(p_1=p_2=0\). Bidder~2's support accumulates at zero, so her equilibrium utility is \(u_2(w_2)\). For \(0<z\le T\), indifference gives \(F_1(z)=[u_2(w_2)-u_2(w_2-sz)]/[u_2(w_2+v_2-sz)-u_2(w_2-sz)]\).
At \(T\), the equality \(F_1(T)=1\) implies \(sT=v_2\), so \(T=L\) and \(F_1=G_1\). The cutoff is \(a\), since \(G_1(a)=p_\ell\). The low type's support accumulates at zero. Her utility is therefore \(u(A)\), which forces \(F_2=F_\ell\) on \([0,a]\). The high type's utility at \(a\) is \(K\). Her indifference then forces the second expression in \eqref{cc:eq:G2} on \([a,L]\). Thus \(F_2=G_2\).

\pstep{5}
Suppose \(p_2>0\) and \(p_1=0\). Bidder~2's zero bid gives exactly \(u_2(w_2)\). Her indifference again gives \(F_1=G_1\), \(T=L\), and cutoff \(a\). The high type's payoff at \(L\) is \(u(A+v_*-2L)=K\). Her indifference at \(a\) gives \(F_2(a)=F_\ell(a)\), so the low type's payoff there is \(u(A)\). But the low type's positive support accumulates at zero, where her limiting payoff is \(u(A)+p_2[u(A+v_\ell)-u(A)]>u(A)\). This contradicts indifference.

\pstep{6}
Suppose \(p_1>0\) and \(p_2=0\). Taking bidder~2's positive bids down to zero gives equilibrium utility \(U_2^{\mathrm{eq}}=u_2(w_2)+p_1[u_2(w_2+v_2)-u_2(w_2)]>u_2(w_2)\). For \(0<z\le T\), her indifference implies
\begin{equation}\label{cc:eq:F1-zeroatom}
F_1(z)=\frac{U_2^{\mathrm{eq}}-u_2(w_2-sz)}{u_2(w_2+v_2-sz)-u_2(w_2-sz)}>G_1(z).
\end{equation}
At the upper endpoint, \(U_2^{\mathrm{eq}}=u_2(w_2+v_2-sT)\), so \(T<L\).

If \(p_1<p_\ell\), the cutoff satisfies \(c<a\) by \eqref{cc:eq:F1-zeroatom}. Sorting implies that only the low type assigns positive probability to zero. Her utility is \(u(A)\), and indifference gives \(F_2(z)=F_\ell(z)\) on \([0,c]\). The high type's utility at the cutoff is
\(K(c)=u(A)+F_\ell(c)J(c)\).
This expression strictly increases for \(0<c\le a\). Hence \(K(c)<K(a)=K\). Her payoff at \(T<L\), however, is \(u(A+v_*-2T)>u(A+v_*-2L)=K\), a contradiction.

If \(p_1\ge p_\ell\), all low types bid zero by sorting. The high type either has a zero atom or has positive support accumulating at zero. Since \(p_2=0\), her utility is \(u(A)\) in either case. She has some positive bids because bidder~1 is not concentrated at zero. Her payoff at the upper endpoint is again strictly above \(K>u(A)\), a contradiction.

Shared zero atoms were excluded in Step~2. Thus both distributions are atomless at zero, and Step~4 determines them uniquely. Sorting and the fixed type probabilities also determine the two conditional distributions of bidder~1.
\end{proof}

\subsubsection{Comparison types}\label{cc:subsec:probes}

Let bidder~2 submit $(Z,\ldots,Z,0)$, with $s=m-1$ copies of $Z$ and $Z\sim G_2$.
Consider bidder~1's two comparison types $t^L=(A,B_L,C,0,\ldots,0)$ and $t^H=(A,B_H,C,0,\ldots,0)$.

\begin{proposition}\label{cc:prop:probe-reversal}
For all sufficiently small $a>0$, every best response of $t^L$ has third bid greater than $3a/4$, whereas every best response of $t^H$ has third bid less than $a/4$.
Each type also has a fixed deviation with positive second bid and expected utility strictly greater than $u(A)$.
\end{proposition}

\begin{proof}
\pstep{1}
Repeated bid deletion as in Lemma~\ref{ap:lem:caps} improves any action violating a marginal-value cap and produces a capped vector after at most $m$ deletions.
It therefore suffices to consider capped actions, which have zero bids after the third coordinate at either comparison type.
A bid vector with zero second bid wins at most one unit against the opponent's $m-1$ positive bids, so its utility is at most $u(A)$.
If the second bid is positive, the first unit is certain. Lowering the first bid to the second preserves the allocation and saves its payment difference.
Indeed, the second unit is obtained exactly when the second bid exceeds $Z$, and the third unit exactly when the third bid exceeds $Z$. Ties have probability zero.
Thus every best response with positive second bid has the form $(x,x,y,0,\ldots,0)$, where $0\le y\le\min\{C,x\}$ and $0<x\le B$, with $B=B_L$ or $B_H$.
Its expected utility is
\begin{align}\label{cc:eq:Phi}
\Phi_B(x,y)
={}&u(A-2x-y)+G_2(x)\bigl[u(A+B-2x-y)-u(A-2x-y)\bigr]\notag\\
&+G_2(y)\bigl[u(A+B+C-2x-y)-u(A+B-2x-y)\bigr].
\end{align}
Extend this function continuously to $(0,0)$ by setting $\Phi_B(0,0)=u(A)$.
At that point it bounds the actual payoff from above. We will obtain feasible positive-second-bid actions with strictly higher payoffs.

\pstep{2}
Every feasible $(x,y)$ satisfies $x\le B_H<v_*/4$ and $y\le C<B_H$.
Hence $A-2x-y>A-3v_*/4>1/2$, and all surplus arguments in \eqref{cc:eq:Phi} belong to $[1/2,m]$.
Suppose $x\ge a$ and $x+h\le B$, with $h>0$. Raising $x$ by $h$ increases payment by $2h$.
At fixed allocation probabilities, the resulting utility loss is at least $2\kappa h$.
The probability of obtaining the second unit rises by at most $\rho h$, and the utility gain from that unit is at most $MB$.
Consequently, $\Phi_B(x+h,y)-\Phi_B(x,y)\le(\rho MB-2\kappa)h<0$.
Likewise, when $y\ge a$ and $y+h\le\min\{C,x\}$, we have $\Phi_B(x,y+h)-\Phi_B(x,y)\le(\rho MC-\kappa)h<0$.
These comparisons first charge the additional payment at the old allocation probabilities, then add the extra winning probability at the new payment.
The strict signs follow from \eqref{cc:eq:B-choice} and $C<B_L$.
If $y>a$, lowering $y$ to $a$ therefore improves utility. Once $y\le a$, lowering any $x>a$ to $a$ also improves utility.
Thus every maximizer lies in
\begin{equation}\label{cc:eq:small-triangle}
0\le y\le x\le a.
\end{equation}
This comparison covers every feasible larger bid.

\pstep{3}
Let $\lambda_A=u'_-(A)$, which is finite and satisfies $\lambda_A\ge\kappa>0$, and put $c_*=2\lambda_A/d_0$.
Uniformly over $0\le X\le1$,
\begin{equation*}
\frac{G_2(aX)}a=\frac{u(A)-u(A-2aX)}{a\Delta_{v_\ell}(aX)}\longrightarrow c_*X.
\end{equation*}
The definition of the left derivative gives a uniform $o(a)$ remainder for increments in $[0,2a]$.
The denominator converges uniformly to $d_0>0$.
Applying the same argument to the payment $a(2X+Y)$ in \eqref{cc:eq:Phi}, and continuity to the remaining utility increments, gives $[\Phi_B(aX,aY)-u(A)]/a\to\alpha_BX+\beta_BY$ uniformly on $0\le Y\le X\le1$.
Here $\alpha_B=c_*D(B)-2\lambda_A$ and $\beta_B=c_*[u(A+B+C)-u(A+B)]-\lambda_A$.
By \eqref{cc:eq:key-increments}, $\alpha_{B_L}>0$, $\alpha_{B_H}>0$, $\beta_{B_L}>0$, and $\beta_{B_H}<0$.
The limiting objectives therefore have unique maximizers $(1,1)$ and $(1,0)$ for the low and high types, respectively.
Restricting the low objective to $Y\le3/4$ lowers its maximum by at least $\beta_{B_L}/4$.
Restricting the high objective to $Y\ge1/4$ lowers its maximum by at least $-\beta_{B_H}/4$.
Uniform convergence preserves both strict gaps for sufficiently small $a$.

The limiting gains at $(1,1)$ and $(1,0)$ are $\alpha_{B_L}+\beta_{B_L}>0$ and $\alpha_{B_H}>0$.
For sufficiently small $a$, the actual deviations $(a,a,a,0,\ldots,0)$ and $(a,a,0,\ldots,0)$ therefore give their respective types utility above $u(A)$.
The continuous function on the compact triangle in \eqref{cc:eq:small-triangle} attains its maximum.
Every maximizer has positive second bid and is an actual best response, by Steps~1 and 2.
The strict gaps give the claimed third-bid inequalities.
\end{proof}

\subsubsection{Concentrated distributions and full support}\label{pt:section}

Fix $a$ satisfying the preceding proposition and retain all the resulting parameters.
Write $\mathcal B_m=\{b\in[0,1]^m:b_1\ge\cdots\ge b_m\}$ and set $r_1=2$, $r_2=m-1$.
Then $r_1+r_2=m+1$. By Lemma~\ref{ap:lem:caps}, equilibrium actions belong to $\mathcal B_m$.
The reference types are $\bar t_{1\ell}=(A,v_\ell,0,\ldots,0)$, $\bar t_{1h}=(A,v_*,0,\ldots,0)$, and $\bar t_2=(H,\ldots,H,v_2,0)$, where $\bar t_2$ has $m-2$ copies of $H$.
Bidder~1's reference types have probabilities $p_\ell$ and $1-p_\ell$.
The scalar contest has baseline values $w_1=A$, $w_2=(m-2)H$ and payments $r_i z_i$.
Its equilibrium bid distributions are unique, including the two conditional distributions for bidder~1.
Bidder~2's distribution $G_2$ has no atoms, including at zero.

\begin{lemma}\label{pt:relaxed}
For $Z\sim G_2$ and $(t,b)\in T_m\times\mathcal B_m$, define
\begin{equation}\label{pt:Psi}
\Psi(t,b)=\E\left[u_1\left(V_{1+\sum_{j=2}^m\mathbf1_{\{b_j>Z\}}}(t)-\sum_{j=1}^m b_j\right)\right].
\end{equation}
The function $\Psi$ is continuous.
Against bidder~2's reference bid block and zero bids by all other opponents, it is the actual payoff whenever $b_1>0$.
It bounds the actual payoff from above at the all-zero action.
If $b_2=0$, then
\begin{equation}\label{pt:zero-second}
\Psi(t,b)=u_1(t_1-b_1)\le u_1(t_1).
\end{equation}
There are positive-volume compact boxes $\mathcal L,\mathcal H\subset\operatorname{int}T_m$, constants $c_L>c_H>0$, and $g>0$ with the following properties.
Every type in $\mathcal H$ strictly exceeds every type in $\mathcal L$ coordinatewise, and
\begin{align}
b\in\arg\max_{a\in\mathcal B_m}\Psi(t,a),\quad t\in\mathcal L
&\ \Longrightarrow\ b_3>c_L,\label{pt:ideal-low}\\
b\in\arg\max_{a\in\mathcal B_m}\Psi(t,a),\quad t\in\mathcal H
&\ \Longrightarrow\ b_3<c_H.\label{pt:ideal-high}
\end{align}
Both inequalities hold with a positive uniform margin.
For each $R\in\{\mathcal L,\mathcal H\}$ there is a fixed action $a^R$ with $a^R_2>0$ such that, for every $t\in R$,
\begin{equation}\label{pt:benchmark}
\Psi(t,a^R)\ge u_1(t_1)+g.
\end{equation}
\end{lemma}

\begin{proof}
If $(t^k,b^k)\to(t,b)$, the allocations inside \eqref{pt:Psi} eventually agree whenever $Z$ differs from all coordinates of $b$.
This event has probability one because $G_2$ is atomless.
Bounded convergence gives continuity.
There are $m-1$ strictly positive opposing bids almost surely.
If $b_1>0$, the first unit is certain, and unit $j\ge2$ is obtained exactly when $b_j>Z$, except at null tie events.
At the all-zero action, at most one unit can be received.
This proves the payoff assertions and \eqref{pt:zero-second}.

At $t^L,t^H$, Proposition~\ref{cc:prop:probe-reversal} identifies all maximizers of $\Psi$.
Indeed, their values exceed $u(A)$, while zero-second-bid actions give at most $u(A)$.
Compactness and continuity preserve the separated third bids in relative neighborhoods of these types.
The two fixed deviations from that proposition also preserve a uniform positive gain over $u_1(t_1)$ after these neighborhoods are reduced.

Choose interior types in the neighborhoods by giving coordinates $4,\ldots,m$ small, strictly decreasing positive values.
Increase the first, third, and all later coordinates of the high type slightly relative to the low type.
Its second coordinate is already strictly larger.
Sufficiently small closed boxes around these two types have the required coordinate order and lie in $\operatorname{int}T_m$.
Compactness of the maximizer sets permits common strict margins and the stated $c_L,c_H,g$.
\end{proof}

For each sufficiently small $\varepsilon>0$, choose positive-volume boxes
$N_{1\ell}^\varepsilon,N_{1h}^\varepsilon,N_2^\varepsilon$ within distance $\varepsilon$ of their respective reference types in the sup norm.
On bidder~$i$'s reference boxes, require coordinates after $r_i$ to be at most $\varepsilon$.
For $j\ge3$, choose a positive-volume box $N_j^\varepsilon\subset\operatorname{int}T_m$ with every coordinate at most $\varepsilon$.
For $0<\eta<1/2$, define independent type distributions by
\begin{equation*}
\mu_1^{\varepsilon,\eta}=(1-\eta)\bigl[p_\ell U_{N_{1\ell}^\varepsilon}+(1-p_\ell)U_{N_{1h}^\varepsilon}\bigr]+\frac\eta2(U_{\mathcal L}+U_{\mathcal H}),
\end{equation*}
$\mu_2^{\varepsilon,\eta}=U_{N_2^\varepsilon}$, and $\mu_j^{\varepsilon,\eta}=U_{N_j^\varepsilon}$ for $3\le j\le n$.
Let $g_i^{\varepsilon,\eta}$ be their bounded densities.
For $0\le\xi<1/2$, set $\widehat\mu_i^{\varepsilon,\eta,\xi}=(1-\xi)\mu_i^{\varepsilon,\eta}+\xi U_{T_m}$ independently across bidders. Its density is
\begin{equation}\label{fs:ap:density}
\widehat f_i^{\varepsilon,\eta,\xi}(t)=(1-\xi)g_i^{\varepsilon,\eta}(t)+\xi m!.
\end{equation}
When $\xi>0$, these densities satisfy, almost everywhere on $T_m$,
\[
0<\xi m!\le\widehat f_i^{\varepsilon,\eta,\xi}(t)\le(1-\xi)\|g_i^{\varepsilon,\eta}\|_\infty+\xi m!<\infty.
\]
The parameter $\eta$ gives positive probability to both comparison boxes, while $\xi$ gives each bidder full support.

\begin{proposition}\label{pt:joint}
There exists $\tau>0$ such that, for $0<\varepsilon,\eta<\tau$ and $0\le\xi<\tau$, every pure equilibrium under \eqref{fs:ap:density} satisfies, almost everywhere on the indicated boxes,
\begin{equation}\label{pt:actual-reversal}
\sigma_{13}(t)>c_L\quad(t\in\mathcal L),\qquad \sigma_{13}(t)<c_H\quad(t\in\mathcal H).
\end{equation}
Consequently, no such equilibrium has a third bid for bidder~1 that is coordinatewise nondecreasing on a full-measure set.
Any fixed sufficiently small positive triple $(\varepsilon,\eta,\xi)$ therefore supplies the densities in Theorem~\ref{thm:allpay}.
\end{proposition}

\begin{proof}
\pstep{1}
Consider any sequence $\varepsilon_k,\eta_k,\xi_k\to0$ with $\varepsilon_k,\eta_k>0$ and $\xi_k\ge0$ along which pure equilibria exist. Write $\sigma^k$ for a selected equilibrium. At a realized type profile, put $b_i=\sigma_i^k(t_i)$ and use $b_{ij}$ for its coordinates, suppressing $k$ in these realized bids.
Call an own type regular when the best-response inequalities and Lemma~\ref{ap:lem:caps} hold there.
Regular types have full measure under each positive-weight own mixture component.
Mixture components may be represented by independent labels. These labels are used only to compute probabilities and are not observed by bidders.
A reference draw means a draw from an original shrinking box, excluding the comparison and uniform components.

Let $\mathcal R_{-i}^k$ be the event that all opponents are reference draws.
Its probability is independent of the own type, with $P_{1,k}=(1-\xi_k)^{n-1}$ and $P_{i,k}=(1-\eta_k)(1-\xi_k)^{n-1}$ for $i\ge2$.
Put $e_k=\eta_k+(n-1)\xi_k$, so $1-P_{i,k}\le e_k\to0$.
On $[-m,m]$, define
\begin{align}
\omega_i(d)&=\max\{|u_i(x)-u_i(y)|:x,y\in[-m,m],\ |x-y|\le d\},\label{pt:modulus}\\
\underline\omega_i(d)&=\min_{-m\le x\le m-d}[u_i(x+d)-u_i(x)],\qquad0<d\le m.\label{pt:increment-minimum}
\end{align}
Then $\omega_i(d)\to0$ as $d\downarrow0$, and $\underline\omega_i(d)>0$ for every fixed $0<d\le m$.
Write $\mathcal W_i=u_i(m)-u_i(-m)$.
All actions considered below lie in $\mathcal B_m$, so their surpluses belong to $[-m,m]$.
The same bound holds when an allocation gain is evaluated before charging an additional payment.

Interim optimality, applied before integration over any selected own component, gives
\begin{equation*}
\E[\mathcal U_i(d_i,b_{-i})-\mathcal U_i(b_i,b_{-i})\mid\mathcal R_{-i}^k]
\le\mathcal W_i\frac{1-P_{i,k}}{P_{i,k}}\longrightarrow0,
\end{equation*}
where the expectation also averages over that own component and priority, and $\mathcal U_i=u_i(S_i^{\mathrm{AP}})$.
The deviation may depend on the own type, but not on a mixture label.
No division by the probability of the own component is needed. Each component is absolutely continuous with respect to the actual own distribution.
At every regular type,
\(0\le b_{ij}\le t_{ij}\).
Hence all reference bids of the additional bidders, and all active reference bids after coordinate $r_i$, are at most $\varepsilon_k$.

\pstep{2}
For active bidder $i\in\{1,2\}$ at any regular type, put $c_i=\max\{b_{i,r_i},2\varepsilon_k\}$ and replace $b_{ij}$ by $\min\{b_{ij},c_i\}$ for $j<r_i$.
This leaves the bid vector ordered.
On $\mathcal R_{-i}^k$, every reduced bid remains above $\varepsilon_k$ and wins.
There are at most $r_i-2$ earlier own bids and $r_{3-i}$ opposing bids ahead of it, with $(r_i-2)+r_{3-i}=m-1$.
Unchanged own bids retain their acceptance status, so the allocation is unchanged.
Let $D_i$ be the payment saving.
If $D_i\ge d$ for a fixed $0<d\le m$, the gain on this event is at least $\underline\omega_i(d)$, while the gain elsewhere is at least $-\mathcal W_i$.
Thus optimality requires $(1-e_k)\underline\omega_i(d)\le e_k\mathcal W_i$.
For every fixed $0<d\le m$, this fails for all large $k$.
The bound is uniform over regular own types, so there is a deterministic $\zeta_k\to0$ such that $D_i\le\zeta_k$ for both active bidders.
Consequently,
\begin{equation}\label{pt:pool}
0\le b_{ij}-b_{i,r_i}\le2\varepsilon_k+\zeta_k=o(1),\qquad j<r_i.
\end{equation}

\pstep{3}
At an active reference type, raise the first $r_i-1$ bids to at least $2\varepsilon_k$.
This costs at most $2(r_i-1)\varepsilon_k$, never lowers quantity, and secures at least $r_i-1$ units on $\mathcal R_{-i}^k$.
The values of these units on the reference boxes have a common positive lower bound $v_0$ for all large $k$.
If fewer than $r_i-1$ units were originally won, the allocation gain on $\mathcal R_{-i}^k$ is at least $v_0$.
Before the additional payment, the utility gain is at least $\underline\omega_i(v_0)$ on that event and is nonnegative elsewhere.
Charging the payment costs at most $\omega_i(2(r_i-1)\varepsilon_k)$ in utility.
Optimality therefore implies
\begin{equation}\label{pt:base-units}
\Prb\{Q_i<r_i-1,\mathcal R_{-i}^k\mid t_i\}
\le\frac{\omega_i(2(r_i-1)\varepsilon_k)}{\underline\omega_i(v_0)}\longrightarrow0.
\end{equation}
This holds uniformly over regular reference types.

\pstep{4}
Pass to a subsequence along which the scalar bid distributions $b_{i,r_i}$ converge under each of the three active reference components.
Let $F_1$ be bidder~1's limiting mixture with weights $p_\ell,1-p_\ell$, and let $F_2$ be bidder~2's limit.
We show that these distributions have no common atom, including at zero.

Suppose $z$ were a common atom. For small $h>0$, let $E_i$ denote the event that bidder~$i$ is a reference draw and $|b_{i,r_i}-z|\le h$.
At any own type satisfying this bid inequality, consider raising its first $r_i$ bids to at least $z+2h$.
This deviation depends only on the own type and its prescribed bid. Its additional payment is at most $3r_i h$.
The reference marginal-value caps put every limiting scalar bid strictly below one, so the deviation lies in $\mathcal B_m$ for sufficiently small $h$.

On $E_1\cap E_2$, for large $k$, \eqref{pt:pool} puts every active opposing reference bid below $z+2h$.
If all additional bidders are reference draws, their bids are also below this threshold.
Each of the two unilateral deviations would then secure its bidder $r_i$ units.
The original allocation cannot give both bidders these quantities because $r_1+r_2=m+1$.
Thus at least one deviation raises value by a fixed amount $\beta>0$, where $\beta$ is a common lower bound on the active reference marginal prizes.
Before the additional payments, the sum of the utility gains is at least $\min\{\underline\omega_1(\beta),\underline\omega_2(\beta)\}$ on this event.
At every other profile both allocation gains remain nonnegative.

Integrate the own-type best-response inequalities over $E_i$ and add them.
Restricting this integration to a mixture component does not make the deviation depend on its unobserved label.
Independence gives
\begin{equation}\label{pt:no-common-atoms}
\min\{\underline\omega_1(\beta),\underline\omega_2(\beta)\}\Prb(E_1)\Prb(E_2)(1-\xi_k)^{n-2}
\le\sum_{i=1}^2\omega_i(3r_i h)\Prb(E_i).
\end{equation}
Choose $h$ with continuity endpoints for both limiting distributions, first let $k\to\infty$, then let $h\downarrow0$.
The right side tends to zero. A common atom would leave a strictly positive left side, a contradiction.

\pstep{5}
By \eqref{pt:pool}, active reference bid vectors converge to blocks with first $r_i$ coordinates equal and remaining coordinates zero.
Their independent scalar limits $Z_1,Z_2$ are unequal almost surely by Step~4.
The higher block wins its marginal prize. The lower block receives its $r_i-1$ base units, apart from events whose probabilities vanish by \eqref{pt:base-units}.
Values and payments after the reference block contribute at most $m\varepsilon_k$ each.
The event of a nonreference opponent has vanishing probability and contributes an error bounded by $\mathcal W_i$ times that probability.
It follows that, under each active reference component, equilibrium expected utility converges to
\begin{equation}\label{pt:limit-payoff}
\E\left[u_i\left(w_i+B_i\mathbf1_{\{Z_i>Z_{3-i}\}}-r_iZ_i\right)\right],
\end{equation}
where $B_i=v_\ell,v_*$ or $v_2$, according to that component.
This conclusion also covers a zero scalar bid, since the two limiting scalars are never both zero with positive probability.

Fix $0<x\le1$ at a continuity point of the opposing limiting distribution.
The deviation with first $r_i$ bids equal to $x$ and all remaining bids zero secures the base units against reference opponents for large $k$.
Its expected utility converges to the scalar-contest payoff from $x$.
Integrate optimality separately over each own reference component and pass to the limit.
The payoff in \eqref{pt:limit-payoff} is therefore at least the payoff from each such fixed deviation.
Approaching any desired bid from above by positive continuity points includes zero and opposing atoms.
The resulting bound awards a tie to the deviator, so it is at least the payoff under the prescribed tie priority.
The endpoint one is above all opposing limiting bids. Bids above one are unprofitable, since $r_ix>B_i$ and even winning gives less than the baseline payoff available at zero.

Thus the limiting mixed action of each reference type yields at least the payoff of every scalar deviation.
Its average payoff cannot exceed the supremum of these payoffs, so it assigns probability one to best responses.
The limiting distributions form a mixed equilibrium of the scalar contest.
Its uniqueness identifies them as $G_1,G_2$, including bidder~1's two conditional distributions.
Every convergent subsequence has this same limit.
In particular, bidder~2's full bid distribution converges to $(Z,\ldots,Z,0)$ with $Z\sim G_2$, and all additional bidders' bid distributions converge to zero.

\pstep{6}
Take any sequence of regular comparison types $t^k$ in one compact box $R\in\{\mathcal L,\mathcal H\}$ and their equilibrium actions $b^k=\sigma_1^k(t^k)$.
Pass to a subsequence with $t^k\to t\in R$ and $b^k\to b$.
If $b_2>0$, expected equilibrium utility converges to $\Psi(t,b)$.
The first unit is secure against reference opponents for large $k$, and the remaining allocation thresholds are continuous except when $Z$ equals a bid coordinate.
Those ties have probability zero.
If $b_2=0$, then
\begin{equation}\label{pt:probe-limsup}
\limsup_k\E\left[u_1\left(V_{Q_1}(t^k)-\sum_j b_j^k\right)\right]
\le u_1(t_1-b_1)=\Psi(t,b).
\end{equation}
Indeed, fix $v>0$. When the opposing reference scalar exceeds $v$ and $b_2^k<v$, at most one unit can be obtained.
The limiting probability of the remaining reference event is at most $G_2(v)$, which tends to zero as $v\downarrow0$.
Nonreference opponents have vanishing probability, and utility is uniformly bounded.

For any fixed action $d$ with $d_2>0$, its deviation payoff converges to $\Psi(t,d)$.
Optimality and the preceding upper bound imply $\Psi(t,b)\ge\Psi(t,d)$ for every such $d$.
These actions are dense in $\mathcal B_m$. Continuity therefore makes $b$ a maximizer of $\Psi(t,\cdot)$.
The benchmark \eqref{pt:benchmark} also excludes $b_2=0$.
The uniform strict margins in \eqref{pt:ideal-low}--\eqref{pt:ideal-high} now imply the required third-bid inequalities along every sequence under consideration.

To obtain a single threshold $\tau$, suppose none existed.
For each integer $k$, choose $0<\varepsilon_k,\eta_k<1/k$, $0\le\xi_k<1/k$, an equilibrium, and a regular comparison type violating \eqref{pt:actual-reversal}.
After passing to one comparison box and a convergent subsequence, Step~6 would give a limiting maximizer violating its uniform strict margin.
This contradiction proves the joint threshold assertion.

Both comparison boxes have positive probability for every fixed $\eta>0$ and $\xi<1$.
If bidder~1's third bid were nondecreasing on a full-measure set, choose a regular type from each box in that set.
The high type strictly exceeds the low type in every coordinate, but its third bid is strictly smaller by \eqref{pt:actual-reversal} and $c_H<c_L$.
This is the required contradiction.
\end{proof}

\begin{proof}[Completion of Theorem~\ref{thm:allpay}]
Fix the reference parameters and the comparison boxes above.
Choose any strictly positive $\varepsilon,\eta,\xi$ below the threshold of Proposition~\ref{pt:joint}.
The independent densities in \eqref{fs:ap:density} have positive lower bounds and finite upper bounds on $T_m$.
The proposition excludes a pure equilibrium with a coordinatewise nondecreasing third bid for bidder~1, without an admissibility assumption.
All conclusions concern this fixed finite auction.
The same strict reversal also holds for the concentrated distributions with $\xi=0$.
\end{proof}

\subsection{Finite bid grids}\label{app:grid}
Let $G\subset[0,1]$ be a finite grid containing $0$ and $1$. Each bidder
submits an ordered vector in $G^m$. Types remain independent and have
full-dimensional densities. Write $h(G)$ for the largest gap between consecutive grid points. Let $\lceil v\rceil_G$ be the smallest grid point at least $v$, and let $\lfloor v\rfloor_G$ be the largest grid point at most $v$.
Under uniform pricing, call an equilibrium \emph{nontrivial} if
$\sigma_{ik}(t_i)\le\lceil t_{ik}\rceil_G$ almost everywhere. This condition
allows overbidding by at most $h(G)$. We impose no such restriction in
the discriminatory and all-pay results. Full support is required only
where it is expressly stated below.

\subsubsection{Uniform-price auctions}\label{sec:grid-uniform}
The first result discretizes the distributions used in the proof of
Theorem~\ref{thm:uniform}. The second uses a separate construction with
seven bid levels.

\begin{proposition}\label{ug:prop:common}
Let $\mu_i^r$ be the basic type distributions in Step~1 of Appendix~\ref{app:uniform-proof}, with the parameters fixed as in Step~6 there. Use $\mu_i^r$ as the actual distributions, so the uniform-background
weight is zero. Put $N_r=\lceil r^2/\varepsilon\rceil$ and
$G_r^{\mathrm{pair}}=\{0,1/N_r,\ldots,1\}$.
For every sufficiently large integer $r$, neither highest-losing-bid nor
lowest-winning-bid pricing on $G_r^{\mathrm{pair}}$ admits a nontrivial pure equilibrium
in which bidder~1's last bid is coordinatewise nondecreasing. The same
type distributions exclude no-overbidding pure equilibria with this
monotonicity property when bids are continuous.
\end{proposition}

\begin{proof}
Write $h_r=1/N_r\le\varepsilon/r^2$ and
$\bar\rho_r=\rho_r+h_r$, where $\rho_r=\varepsilon/r$.
A nontrivial bid satisfies $b_k\le t_k+h_r$. Every winning unit therefore
has marginal surplus at least $-h_r$, so total surplus lies in
$[-mh_r,m]$. Since utility is finite, increasing and concave on $\R$,
there are constants $0<\ell_i\le L_i<\infty$ bounding its secant slopes
on $[-m,m]$. Consequently a monetary gain bounded below by $g-h$, with
$g,h\ge0$, gives a utility gain of at least $\ell_i g-L_i h$.

For an ordered continuous deviation $d_i$, use the ordered grid deviation
$D_r(t_i,d_i)_k=\lceil\min\{d_{ik},t_{ik}\}\rceil_{G_r^{\mathrm{pair}}}$.
Relative to the value-capped deviation, every coordinate rises by at
most $h_r$. Quantity cannot fall, and the price rises by at most $h_r$.
Each additional unit has marginal surplus at least $-h_r$.
By Lemma~\ref{lem:caps}, capping gives nonnegative surplus and weakly
improves upon $d_i$. Rounding lowers the capped surplus by at most $mh_r$,
so the grid deviation gives at least
$\max\{S_i^{\mathrm{UP}}(d_i,b_{-i};t_i),0\}-mh_r$ statewise. Here $S_i^{\mathrm{UP}}$ is the model's uniform-price surplus at the indicated bid profile and type. The utility cost of
rounding is at most $L_i mh_r$.

We apply the finite deviation estimates in Steps~3--6 of Appendix~\ref{app:uniform-proof}. Interim optimality, integrated over any of the
own-type components, gives an expected deviation gain of at most zero.
All opponent coordinates that were bounded by $\rho_r$ are now bounded
by $\bar\rho_r$. The bound on bidder~1's second bid becomes
$y_H+\varepsilon+h_r<R_*$ for large $r$, where $R_*$ is the fixed threshold
in Step~3 of the proof in Appendix~\ref{app:uniform-proof}.

For completeness, the required changes to the deviations are as follows.
Raise the expensive rival's first $m-1$ bids to at least
$\lceil R_*\rceil_{G_r^{\mathrm{pair}}}$, and raise the cheap rival's first bid to
$\lceil t_{21}\rceil_{G_r^{\mathrm{pair}}}$. Replace every capped uniform increase and
flat bid by $D_r$. For the compression deviation, replace bidder~1's
first bid by
$\min\{b_{11},\max\{b_{12},\lceil2\bar\rho_r\rceil_{G_r^{\mathrm{pair}}}\}\}$.
This compression preserves quantity and weakly lowers the price.
The other deviations retain the quantity guarantees in that proof.
Their surplus bounds acquire errors of at most a constant times $mh_r$.
In particular, the expensive rival's fixed gain $\kappa=0.45/m$ can be
replaced by $\kappa/2$ for all large $r$.

The only estimate that also uses the value cap directly is the
small-surplus event in Step~4 of Appendix~\ref{app:uniform-proof}. On that event,
$b_{12}\le t_{12}+h_r$ enlarges the interval containing $t_{12}$ by at
most $h_r$. Its probability is therefore bounded by
$(\theta_r+\zeta_r/2+h_r)/\varepsilon$, where
$\zeta_r=r^{-2/3}$ and $\theta_r=r^{-1/3}$ are the bid increase and surplus
threshold used there. These changes preserve the bounds on the first
bid spread and on the exceptional quantities. With the notation of
that proof, for either pricing rule we obtain
$K_r^D=\Prb_D^r(Q_1^r\ge2)\to0$ and
$K_r^C(a)=\Prb_C^r(Q_1^r=m,\ b_{1m}^r\le a)\to0$ for every fixed
$a\in(0,c)$.
Indeed, the new errors are $O(\rho_r+h_r)$ before division by the
thresholds, and $h_r=O(r^{-2})$ is smaller than those in the continuous
proof. The utility bounds remain valid on $[-m,m]$ even when ceiling
bids produce a small negative surplus.

Finally, round up the two fixed flat deviations in Step~6 of Appendix~\ref{app:uniform-proof}. For large
$r$, they remain below every marginal value in the low and high type
regions. Their utility guarantees differ by at most $L_1mh_r$ from the
ones in that step. The fixed strict payoff gaps therefore give the
same contradiction. A positive measure of low types must bid above
$a$ on the last unit, monotonicity implies that almost every high type
does so, and the high types then have a profitable deviation. The
bounds hold uniformly over candidate equilibria at each $r$. Taking
$r$ above the thresholds for both pricing rules proves the grid claim.
For continuous bids the same proof applies with background weight
zero, so its integrated optimality error is zero.
\end{proof}

The proposition concerns the paired sequence $(\mu^r,G_r^{\mathrm{pair}})$. It does
not assert nonexistence on every finer grid after one $r$ is fixed.
The next result uses different type regions and allows every price
$p_\alpha=(1-\alpha)b_{(m)}+\alpha b_{(m+1)}$.

\begin{proposition}\label{ug:prop}
For every $n\ge2$, $m\ge3$, and utility profile satisfying
\textup{(\ref{eq:utilities})}, there are independent full-dimensional type
densities bounded above and a grid
$G=\{0,s,2s,c,c+s,X,1\}$, where
$0<s<2s<c<c+s<X<1$, such that, simultaneously for every
$\alpha\in[0,1]$, no nontrivial pure equilibrium has a coordinatewise
nondecreasing last bid for bidder~1. In particular, no admissible
monotone pure equilibrium exists.
\end{proposition}

\begin{proof}
\pstep{1} We first verify the ceiling restriction for admissible bids.
For type $t$, let $v_k=\lceil t_k\rceil_G$ and replace $b_k$ by
$\widehat b_k=\min\{b_k,v_k\}$. The new vector is ordered. Its price
weakly falls. If own unit $k$ is lost, the opposing bid that displaces
it is at least $v_k\ge t_k$ and was originally rejected. Thus the old
price was at least $t_k$, and the lost unit had nonpositive surplus.
Each retained unit costs weakly less. This proves statewise weak
improvement for every $\alpha$.

If $b_j>v_j$, let one rival bid $b_j$ on every unit and let all other
rivals bid zero. Whenever this bidder precedes that rival in priority,
unit $j$ originally wins at price $b_j$ and loses after clipping.
Its original surplus is strictly negative. Clipping therefore strictly
improves surplus on a positive-probability priority event and never
reduces it. Strict increase of utility proves weak dominance. Every
admissible grid bid satisfies the ceiling restriction.

\pstep{2} We choose the parameters and distributions. On $[-m,m]$, fix
positive lower and finite upper secant-slope bounds $\ell_i,L_i$ for
$u_i$. Set $h=3/4$ and choose $X>0$ so small that
\begin{equation}\label{ug:eq:slack}
X<\min\left\{\frac1{2m},\frac{h}{4(m-1)}\right\},\qquad
L_2(m-1)X<\frac{\ell_2 h}{4}.
\end{equation}
Then $\kappa:=h-(m-1)X>3h/4$.
Put $x_*=X/2$ and $z_*=x_*/4$. Choose two initial decreasing value
vectors with common first coordinate $x_*$, second coordinates $x_*/2$
and $3x_*/4$, and common later coordinates
$z_*+(m-k)x_*/(12m)$ for $3\le k\le m$.
Their first-$m-1$ totals satisfy $v_L^*<v_H^*$, and their last values
are $z_*$. Let $D_L(0)=u_1(v_L^*+z_*)-u_1(v_L^*)$ and
$D_H(0)=u_1(v_H^*+z_*)-u_1(v_H^*)$. Strict concavity gives
$D_L(0)>D_H(0)>0$.
Choose $c\in(0,z_*/m)$ so small that
$D_L(c):=u_1(v_L^*+z_*-mc)-u_1(v_L^*)>D_H(0)$.
Set $C_*=u_1(x_*)-u_1(x_*-c)>0$ and
$\delta=[D_L(c)+D_H(0)]/[D_L(c)+D_H(0)+2C_*]$.
Define $F_0(t)=(1-\delta)u_1(V_{m-1}(t))+\delta u_1(t_1)$,
$F_+(t)=(1-\delta)u_1(V_m(t)-mc)+\delta u_1(t_1-c)$, and
$G_c(t)=(1-\delta)u_1(V_m(t))+\delta u_1(t_1-c)$.
The initial low vector satisfies $F_+>F_0$, and the initial high vector
satisfies $F_0>G_c$. Add a sufficiently small positive constant to
every coordinate of the high vector. Both strict inequalities persist,
and the high vector strictly exceeds the low vector coordinatewise.
Choose small closed rectangles $\mathcal L,\mathcal H\subset\operatorname{int}T_m$
around them. Every high type strictly exceeds every low type, and
throughout the rectangles $t_1<X$ and $t_m>mc$. For some $\lambda>0$,
the respective inequalities on $\mathcal L$ and $\mathcal H$ are
\begin{equation}\label{ug:eq:gaps}
F_+(t)-F_0(t)\ge\lambda,\qquad F_0(t)-G_c(t)\ge\lambda.
\end{equation}
These rectangles are separate from the type regions in the proof of
Theorem~\ref{thm:uniform}.

For sufficiently large integers $r$, put $s_r=c/r$,
$\epsilon_r=c/r^2$, and
$G_r=\{0,s_r,2s_r,c,c+s_r,X,1\}$. Bidder~1's distribution is
$\tfrac12U_{\mathcal L}+\tfrac12U_{\mathcal H}$. Bidder~2's distribution
is $(1-\delta)U_{C_r}+\delta U_{D_r}$. Here $C_r$ is the set of
types in $T_m$ satisfying $c-\epsilon_r<t_1<c$ and
$0<t_m<\cdots<t_2<\epsilon_r$. The set $D_r$ is defined by
$h-\epsilon_r<t_{m-1}<\cdots<t_1<h+\epsilon_r$ and
$0<t_m<\epsilon_r$.
Every remaining bidder is uniform on
$\{0<t_m<\cdots<t_1<\epsilon_r\}$. These independent distributions
have full-dimensional densities bounded above. Call bidder~2's two
components states $C$ and $D$.

\pstep{3} Fix $\alpha$ and suppose that candidate equilibria exist along
a sequence $r\to\infty$. Retain each bid's index among the seven ordered
grid levels, in addition to the types, numerical bids, quantities and
priority. Write $b_{ik}^r=\sigma_{ik}^r(t_i)$ for realized bids. Take a joint weakly convergent subsequence. On a Skorokhod
representation, the finite indices, quantities and priority are
eventually constant on each path. Allocation depends on these indices
and priority, not on the distances between grid levels. Distinct levels
that approach the same number therefore retain their original ranking.
The ceiling restriction gives $b_{1k}^r\le X$. In $C_r$, it gives
$b_{21}^r\le c$ and $b_{2k}^r\le s_r$ for $k\ge2$; in $D_r$, it
gives $b_{2m}^r\le s_r$. Every bid from $i\ge3$ is at most $s_r$.
Write $b=(b_1,\ldots,b_m)$ for bidder~1's limiting bid vector.
Its coordinates belong to $\{0,c,X\}$. Bidder~2's limiting types are
$(c,0,\ldots,0)$ in $C$ and $(h,\ldots,h,0)$ in $D$.
Independence across bidders and from priority is preserved. Draw one
bidder~1 input independently from each rectangle. The last bid at the
high input is at least the last bid at the low input in every game;
this order is closed and hence also holds in the limit.

All original and deviating surpluses below lie in $[-m,m]$.
Original utilities converge, because quantities and grid indices have
been retained. For each deviation, its limiting lower payoff bound
passes to expectations by bounded Fatou. Interim optimality, integrated
over any own-type component, bounds its expected gain by zero.
A vanishing surplus loss has vanishing utility cost, while a fixed
positive surplus gain has a fixed positive utility gain. The argument
also applies when the integrated optimality inequalities have errors
tending to zero.

\pstep{4} Put $v_*:=\min_{\mathcal L\cup\mathcal H}t_m>mc$.
Raise every bid of bidder~1 to at least $2s_r$. Quantity cannot fall,
and the price rises by at most $2s_r$. If quantity increases, each
additional winning bid is $2s_r$, so the new price is at most $2s_r$.
The surplus loss is at most $2ms_r$, and the gain when quantity
increases is at least $v_*-2ms_r$.
The deviation guarantees $m-1$ units in $C$ and one in $D$.
It guarantees two in $D$ if bidder~2's $(m-1)$st bid is at most $s_r$.
Optimality therefore implies these quantity bounds almost surely in
the limit.

We will prove
\begin{equation}\label{ug:eq:restrictions}
\begin{gathered}
\text{in }D,\quad Q_2=m-1,\quad Q_1=1,\quad p_\alpha\ge b_m,\\
\text{in }C,\quad Q_1\le m-1\quad\text{on }\{b_m<c\}.
\end{gathered}
\end{equation}
For $\alpha<1$ we will also show that all coordinates of $b$ are equal.
Write $\rho^r=b_{2,m-1}^r$ in state $D$, with numerical limit $\rho$.
Let $F=\{Q_2=m-1\}$ and $H=\{Q_2\le m-2\}$ in that state.
The quantity bound above makes these events exhaustive and implies
$Q_1=1$ on $F$.

\pstep{5} Suppose $\theta:=1-\alpha>0$. Compress bidder~1's first bid
to $b_1^{r\prime}=\min\{b_1^r,\max\{2s_r,b_2^r\}\}$, leaving
her other bids unchanged.
Quantity is preserved on every profile. If at least two bids originally
win, the unchanged second bid retains its rank among opposing bids,
and the new first bid remains at least as high. If exactly one bid wins
and changes, its new level is at least $2s_r$; at most $m-1$ opposing
bids exceed $s_r$, so it still wins. Lowering an accepted bid without
changing quantity weakly lowers the price.

On $F$, the highest rejected bid tends to $b_2$, and the compressed
price is at most $b_2^r+2s_r$. Optimality implies
\begin{equation}\label{ug:eq:partial-price}
p_\alpha=b_2\quad\text{on }F.
\end{equation}
The least accepted bid on $F$ is $\min\{b_1,\rho\}$ in the limit.
Since $\theta>0$, we also obtain
\(\min\{b_1,\rho\}=b_2\quad\text{on }F\).

\pstep{6} Let $B=F\cap\{b_1>b_2\}$. On this event,
$\rho=b_2\in\{0,c\}$. We claim that
\begin{equation}\label{ug:eq:priority-bound}
\Prb(B\mid D)\le\Prb(H\mid D).
\end{equation}
The retained grid indices allow the priority comparison to be made
before levels merge.

First suppose $b_2=\rho=0$. The cases $\rho^r\le s_r$ cannot occur
on $F$, by Step~4. If $\rho^r=2s_r$ and $b_2^r\le s_r$, bidder~1
originally wins one unit on the input event $b_1>b_2$.
Whenever she precedes bidder~2 in priority, raising her low bids to
$2s_r$ wins a second unit. The relative priority has probability $1/2$
independently of the inputs. This deviation has a strictly positive
limiting gain on that priority event and only a vanishing loss
elsewhere. Optimality excludes the entire input event. The remaining
zero-limit case has equal levels $\rho^r=b_2^r=2s_r$.

Next suppose $b_2=\rho=c$. The levels $\rho^r=c$ and $b_2^r=c+s_r$
cannot give $F$, since bidder~1 then wins at least two units.
For the opposite levels, $\rho^r=c+s_r$ and $b_2^r=c$, raise all of
bidder~1's bids equal to $c$ to $c+s_r$. This preserves order and
raises price by at most $s_r$. The surplus loss is at most $ms_r$.
On the input event under consideration, bidder~1 originally wins one
unit, but wins a second after the deviation whenever she precedes
bidder~2. The limiting gain is at least $v_*-c>0$ there. Optimality
again excludes that input event. Thus the only remaining cases have
equal levels $\rho^r=b_2^r=c$ or $\rho^r=b_2^r=c+s_r$.

In each remaining case the common level is $2s_r$, $c$, or $c+s_r$.
Bidder~1's first bid is higher, and every residual opposing bid is
lower. Conditional on these inputs, bidder~2 preceding bidder~1 gives
$Q_2=m-1$. Reversing their relative priority gives bidder~1 at least
two units, hence $Q_2\le m-2$. These orders are equally likely.
Summing over the disjoint input events proves
\eqref{ug:eq:priority-bound}.

\pstep{7} Raise bidder~2's first $m-1$ bids in state $D$ to $1$, retaining
her last bid. The deviation wins at least $m-1$ units. If it wins exactly
$m-1$, the price is at most $X$. If it wins all $m$, her last bid forces
the price to be at most $s_r$, and her last value tends to zero.
On $H$, the limiting surplus gain is therefore at least
$h-(m-1)X=\kappa$.
On $F$, only accepted bids rise, so quantity is unchanged.
The new price tends to $\theta b_1+\alpha b_2$; the original price is
$b_2$ by \eqref{ug:eq:partial-price}. The limiting surplus change is
$-(m-1)\theta(b_1-b_2)$. It is zero outside $B$ and at least
$-(m-1)X$ on $B$.
Optimality gives
$\ell_2\kappa\Prb(H\mid D)\le L_2(m-1)X\Prb(B\mid D)$.
By \eqref{ug:eq:priority-bound},
$[\ell_2\kappa-L_2(m-1)X]\Prb(H\mid D)\le0$.
The coefficient is positive by \eqref{ug:eq:slack}.
Thus $H$ and $B$ have probability zero. In state $D$, bidder~2 wins
$m-1$ units, bidder~1 wins one, and $p_\alpha=b_1=b_2$.
Independence and $\delta>0$ imply $b_1=b_2$ under bidder~1's entire
limiting distribution.

\pstep{8} Replace bidder~1's vector by the flat grid bid
$\max\{2s_r,b_m^r\}$. If $b_m^r=X$, the vector was already flat.
Otherwise the new bid is at most $c+s_r<v_*$, so every additional unit
has positive surplus. In state $C$, the original limiting quantity is
$m-1$ or $m$. With $m-1$ units the original price is at least $b_m^r$,
whereas the deviation guarantees that many units at price at most
$b_m^r+2s_r$. With $m$ units, flattening at the last bid preserves all
units, since opposing bids tied there still lose under the same bidder
priority. The price rises by at most $2s_r$. The limiting gain in $C$
is therefore nonnegative.

In $D$, the original quantity is one at price $b_1$. The deviation
guarantees one unit at price at most $b_m^r+2s_r$, with profitable
additional units if any. Its limiting gain is at least $b_1-b_m$.
Optimality and independence imply
$0\ge\delta\E[u_1(t_1-b_m)-u_1(t_1-b_1)]$.
The integrand is nonnegative and strictly positive when $b_1>b_m$.
Thus
\begin{equation}\label{ug:eq:pooling}
b=(\beta,\ldots,\beta),\qquad \beta\in\{0,c,X\}.
\end{equation}
This establishes the state-$D$ restrictions in
\eqref{ug:eq:restrictions} when $\alpha<1$.

\pstep{9} Raise the cheap rival's first bid to $c+s_r$, retaining her
other bids. On $\{\beta=0\}$ this wins a unit at a price tending to zero.
If she already won a unit, her limiting utility is unchanged;
otherwise her limiting gain is $u_2(c)-u_2(0)>0$.
On $\{\beta=c\}$, both her original and deviating surpluses tend to zero.
On $\{\beta=X\}$, she loses before and after the deviation.
Optimality forces her to win a unit whenever $\beta=0$.
By \eqref{ug:eq:pooling}, this is exactly $\{b_m<c\}$, which proves
the state-$C$ restriction for $\alpha<1$.

\pstep{10} Consider the remaining endpoint $\alpha=1$.
Step~4 again bounds the expensive rival's quantity by $m-1$.
Raise her first $m-1$ bids to $1$. If she already wins $m-1$ units,
only accepted bids rise, leaving quantity and the highest rejected bid
unchanged. Otherwise the limiting gain is at least $\kappa$, as in
Step~7. Optimality gives $Q_2=m-1$ and $Q_1=1$ in $D$.
Bidder~1's last bid is rejected, so $p_1\ge b_m$.

For the cheap rival, raise the first bid to $c+s_r$. If it already
wins, quantity and the highest rejected bid are unchanged. Otherwise
the deviation adds at most one unit at price at most $c+s_r$.
Her first value is at least $c-\epsilon_r$, so the surplus loss is at
most $s_r+\epsilon_r$. On the limiting event
$\{C,Q_1=m,b_m<c\}$, the rival originally wins none. Her new first bid
wins, displacing bidder~1's last bid, which becomes the highest rejected
bid. Her limiting utility gain is $u_2(c-b_m)-u_2(0)>0$.
Elsewhere the lower limiting gain is nonnegative. Optimality excludes
this event and completes \eqref{ug:eq:restrictions}.

\pstep{11} The flat grid bids $c+s_r$ and $2s_r$ give the respective
limiting conditional utility guarantees $F_+(t)$ and $F_0(t)$.
The first wins all units in $C$ and at least one in $D$.
The second wins at least $m-1$ in $C$ and one in $D$.
Prices do not exceed the flat bids, and every additional unit is
profitable because $v_*>mc$.

Suppose a positive measure of low inputs has $b_m<c$.
Choose a continuity threshold $a<c$ such that the low component assigns
positive probability to $b_m<a$. At the original low types with
$b_m^r<a$, use the first flat deviation. By
\eqref{ug:eq:restrictions}, their limiting original expected utilities
are at most $F_0(t)$, whereas the deviation gives $F_+(t)$.
The positive gap in \eqref{ug:eq:gaps} contradicts optimality.
Hence $b_m\ge c$ almost surely on the low rectangle. The order between
the low and high components, retained in Step~3, implies the same on
the high rectangle. There, \eqref{ug:eq:restrictions} bounds original
expected utility by $G_c(t)$. This bound grants all units for free in $C$ and uses
one unit at price at least $c$ in $D$. The second flat deviation gives
$F_0(t)$, contradicting the other gap in \eqref{ug:eq:gaps}.
This excludes any bad sequence for a fixed $\alpha$, including
sequences with vanishing integrated optimality errors.

Finally, allocations do not depend on $\alpha$, and changing it to
$\alpha'$ changes any surplus by at most $m|\alpha-\alpha'|$.
The expected utility difference is at most
$L_i m|\alpha-\alpha'|$. If arbitrarily large indices admitted a
candidate equilibrium for some price parameter, take a subsequence
with $r\to\infty$ and $\alpha_r\to\bar\alpha$.
At $\bar\alpha$, its integrated optimality errors are at most
$2L_i m|\alpha_r-\bar\alpha|\to0$. The argument above, including
Step~10 if $\bar\alpha=1$, gives the same contradiction.
Thus one sufficiently large $r$ supplies the distributions and grid
for all $\alpha\in[0,1]$. Step~1 gives the admissibility conclusion.
\end{proof}

The seven-level construction has large gaps. The result does not assert
that it survives arbitrary additional bid levels. Its parameters depend
on the fixed utilities, and its type regions differ from those in
Proposition~\ref{ug:prop:common}.

\subsubsection{Discriminatory auctions}\label{dg:section}

We first transfer a fixed continuous-bid counterexample to all sufficiently
fine grids. The type distributions do not change with the grid. The main
issue is the possible appearance of ties in a limit of equilibria.

\begin{lemma}\label{dg:closure}
Fix $n,m$, independent type distributions with Lebesgue densities on $T_m$,
and continuous, strictly increasing utilities on $\R$. Suppose that
$\Prb(t_{i1}<a)>0$ for every $a>0$ and every bidder $i$.
If grids $G_\ell$ satisfy $h(G_\ell)\to0$ and each corresponding
discriminatory auction has a monotone pure equilibrium, then the
continuous-bid auction has a monotone pure equilibrium. The conclusion
allows unrestricted nonnegative bids. Concavity is not needed.
\end{lemma}

\begin{proof}
\pstep{1}
Let $\sigma^\ell$ be the grid equilibria and write $h_\ell=h(G_\ell)$.
The argument of Lemma~\ref{lem:disc-caps} also applies on a grid. Replace
each overbid by the largest grid point at most its marginal value. An
unchanged bid retains its acceptance, a reduced winning bid costs less,
and any unit lost after the reduction had negative surplus. Thus a winning
overbid yields a strict improvement. Independence and the positive
probability of arbitrarily low first values then exclude all overbids,
as in that lemma. Hence $0\le \sigma^\ell_{ik}(t_i)\le t_{ik}$ for every
$k$ at almost every $t_i$.

Let $U_i(\sigma)$ denote ex ante expected utility. Given any measurable ordered
deviation $a_i:T_m\to[0,1]^m$, cap it coordinatewise at $t_i$ to obtain
$c_i$, then round up to $d_i^\ell=\lceil c_i\rceil_{G_\ell}$.
Capping weakly raises discriminatory surplus. Rounding up cannot reduce
quantity. Each retained unit costs at most $h_\ell$ more, while each
additional unit has surplus at least $-h_\ell$. Hence, state by state,
\[
S_i^{\mathrm D}(d_i^\ell,\sigma_{-i}^\ell;t_i)
\ge S_i^{\mathrm D}(a_i,\sigma_{-i}^\ell;t_i)-mh_\ell.
\]
All these surpluses lie in $[-m,m]$. Let $\omega_i$ be a modulus of
continuity of $u_i$ on this interval. Grid optimality implies
\begin{equation}\label{dg:approx-obedience}
U_i(\sigma^\ell)\ge U_i(a_i,\sigma_{-i}^\ell)-\omega_i(mh_\ell).
\end{equation}
The error tends to zero uniformly over measurable deviations.

\pstep{2}
We extract almost-everywhere limits of the monotone bids.
On a common full-probability set $E_i$, all $\sigma_i^\ell$ are monotone and
value capped. Extend each coordinate to $T_m$ by
$F_{ik}^\ell(t)=\sup\{\sigma_{ik}^\ell(s):s\in E_i,\ s\le t\}$,
with the supremum of the empty set equal to zero. These extensions agree
with the original bids on $E_i$, remain ordered across coordinates, and
satisfy $0\le F_{ik}^\ell(t)\le t_k$.
Bounded coordinatewise monotone functions are continuous outside a
Lebesgue-null set in the interior of $T_m$. For example, their rational
superlevel boundaries meet a line parallel to $(1,\ldots,1)$ in at most
one point, so Fubini's theorem gives this assertion.
Diagonal extraction on a countable dense set, followed by monotone
squeezing at continuity points, gives a subsequence with
$\sigma_i^\ell(t_i)\to \sigma_i(t_i)$ almost everywhere. Absolute continuity of
the type distributions is sufficient here; full support is unnecessary.
The limit is measurable, monotone, ordered, and value capped.

\pstep{3}
We exclude competitive ties in the limit. For the realized limiting bid profile $b(t)=(\sigma_i(t_i))_{i=1}^n$, let $\mathcal Q(b)$ contain all quantity allocations obtained by selecting
the $m$ highest bids, resolving boundary ties in any way that selects a
prefix of each bidder's bids. Put $Q_i^+(b)=\max_{q\in\mathcal Q(b)}q_i$,
so $Q_i^+$ awards bidder $i$ every relevant boundary tie. Write
$S_i^{\mathrm D}(k;b)=\sum_{j=1}^{k}(t_{ij}-b_{ij})$ for $k\in\{0,\ldots,m\}$ and
$W_i^+(t)=u_i(S_i^{\mathrm D}(Q_i^+(b(t));b(t)))$.
For fixed $\delta>0$, the deviation
$\min\{\sigma_i(t_i)+\delta\mathbf1,\mathbf1\}$ eventually wins at least
$Q_i^+(b(t))$ units against $\sigma_{-i}^\ell$, almost surely. Indeed, every
own bid is strictly below one almost surely, while opposing bids
converge. Its surplus is eventually at least
$S_i^{\mathrm D}(Q_i^+(b(t));b(t))-m\delta$, since the limiting bids are value capped.
Apply \eqref{dg:approx-obedience}, Fatou's lemma for bounded payoffs,
and then let $\delta\downarrow0$. This gives
\begin{equation}\label{dg:favorable-limit}
\liminf_\ell U_i(\sigma^\ell)\ge\E W_i^+.
\end{equation}

Consider the nonnegative quantity
\[
D(t)=\min_{q\in\mathcal Q(b(t))}
\sum_i\bigl[W_i^+(t)-u_i(S_i^{\mathrm D}(q_i;b(t)))\bigr].
\]
Every strict comparison of limiting bids eventually holds in the
sequence. The realized allocation therefore eventually belongs to
$\mathcal Q(b(t))$, for every priority realization. Replacing payments
at $\sigma_i^\ell(t_i)$ by payments at $\sigma_i(t_i)$ changes utility by a term tending
to zero, uniformly over the finitely many quantities. Reverse Fatou gives
$\limsup_\ell\sum_i U_i(\sigma^\ell)\le\sum_i\E W_i^+-\E D$.
Together with \eqref{dg:favorable-limit}, this implies $D=0$ almost surely.

Independence and atomless marginal values imply
$\Prb(t_{ik}=\sigma_{jr}(t_j))=0$ whenever $i\ne j$.
At a cross-bidder tie, each tied unit therefore has strictly positive
surplus almost surely, by the value caps. If $\mathcal Q(b)$ contains
more than one quantity allocation, every allocation gives some bidder
fewer than $Q_i^+(b)$ units. That bidder loses a strictly positive surplus
at the boundary. Strict increase of utility then makes $D>0$.
Thus competitive ties have probability zero, including ties at zero.
Quantities and payoffs converge, so $U_i(\sigma^\ell)\to U_i(\sigma)$.

\pstep{4}
Cap any measurable deviation $a_i$ at its marginal values.
For this capped deviation, repeat the favorable-tie argument with
$\min\{a_i+\delta\mathbf1,\mathbf1\}$ in
\eqref{dg:approx-obedience}. Payoff convergence and
$\delta\downarrow0$ imply that $U_i(\sigma)$ is at least the payoff from
$a_i$ with every boundary tie awarded to bidder $i$. This payoff is
at least the payoff under the stated priority rule, since every added
unit has nonnegative surplus. Hence no measurable deviation is profitable.
These inequalities give interim optimality almost everywhere. To see
this directly, upward rational approximations to a capped bid lose at
most $m$ times the approximation error in surplus. Thus a countable
family of ordered rational bids detects any profitable interim
deviation; a profitable deviation on a positive-measure set would give
a profitable measurable deviation. Finally, capping also weakly improves
bids above one. The limit is therefore an equilibrium with unrestricted
nonnegative bids.
\end{proof}

\begin{proposition}\label{dg:transfer}
Fix a continuous-bid discriminatory auction satisfying the assumptions
of Lemma~\ref{dg:closure}. If it has no monotone pure equilibrium,
there is $h_0>0$ such that every grid $G$ with $h(G)<h_0$ also has no
monotone pure equilibrium. In particular, the same utilities, supply,
and type distributions give nonexistence on
$G_N=\{0,1/N,\ldots,1\}$ for every sufficiently large integer $N$.
\end{proposition}

\begin{proof}
Otherwise, choose a grid with mesh below $1/\ell$ admitting a monotone
pure equilibrium for each positive integer $\ell$.
Lemma~\ref{dg:closure} gives a monotone pure equilibrium in the fixed
continuous-bid auction, a contradiction.
\end{proof}

Apply this proposition to any fixed counterexample from
Theorem~\ref{thm:discriminatory}, including its full-support type
distributions. It yields nonexistence on every sufficiently fine grid.
The threshold may depend on that auction and is not numerical.
The argument requires monotonicity of every bidder. In the exponential
case, the following explicit grid requires only bidder~1 to be monotone.

\begin{corollary}\label{dg:sixpoint}
Suppose the utilities satisfy \textup{(\ref{eq:utilities})} and
$u_1(s)=A-Be^{-rs}$ on $[0,\infty)$, with $B,r>0$.
Take the parameters and full-support type distributions from Case~2
of Appendix~\ref{app:discriminatory-proof}. In particular,
$k=200\lceil500/r\rceil$, the integer $m$ satisfies the bounds specified
there, and $M=m-2k$. Put $c=k/(10000m)$ and
$d_+=2k/(M+1)+k/(1000m)$, and define
\begin{equation}\label{dg:sixlevels}
G_* =\{0,c/2,4c/5,c,d_+,1\}.
\end{equation}
Every finite grid $G$ containing $G_*$ has no discriminatory pure
equilibrium in which bidder~1 is monotone. The same type distributions
work for every such grid.
\end{corollary}

\begin{proof}
The parameter bounds in Case~2 give six distinct levels in $[0,1]$.
The grid version of Lemma~\ref{lem:disc-caps}, proved in Step~1 of
Lemma~\ref{dg:closure}, excludes all equilibrium overbids.
The four deviations used in Steps~2--7 of that case are
\[
(c^m),\qquad (d_+^{2k},0^M),\qquad
((c/2)^{m-k},0^k),\qquad ((4c/5)^k,0^{m-k}),
\]
where superscripts denote repetitions. They remain feasible on $G$.
The levels $3c/4$ and $2k/(M+1)$ serve only as thresholds.
All surplus and probability bounds in those steps are therefore unchanged,
and give the same contradiction.
\end{proof}

The integer $N_*=100000m(M+1)$ is a common denominator for
\eqref{dg:sixlevels}. Multiplication by $N_*$ sends the four interior
levels, in order, to
$5k(M+1)$, $8k(M+1)$, $10k(M+1)$, and
$200000km+100k(M+1)$.
Consequently the corollary applies to
$\{0,1/(LN_*),\ldots,1\}$ for every positive integer $L$.
This explicit inclusion argument covers integer-multiple refinements.
Proposition~\ref{dg:transfer} separately covers every sufficiently large
denominator, with an unspecified threshold.

\subsubsection{All-pay auctions}\label{ag:subsec}

The all-pay construction also survives every sufficiently fine grid
with fixed type distributions. We then give a separate three-point
construction in which priority ties produce a reversal of the last bid.

\begin{proposition}\label{ag:prop:fine}
Fix the utilities, $n\ge2$, and $m\ge3$ as in
Theorem~\ref{thm:allpay}. The parameters in \eqref{fs:ap:density}
can be chosen with $\varepsilon,\eta,\xi>0$ so that, for some $h_0>0$,
every grid $G$ with $h(G)<h_0$ has no all-pay pure equilibrium whose
third bid for bidder~1 is coordinatewise nondecreasing. These same
full-support densities also give the continuous-bid conclusion of
Theorem~\ref{thm:allpay}. The statement also holds for a suitable fixed
choice of the concentrated densities, with $\xi=0$.
\end{proposition}

\begin{proof}
We extend Proposition~\ref{pt:joint} to a vanishing grid mesh.
Suppose $\varepsilon_\ell,\eta_\ell,\xi_\ell,h_\ell\to0$, with
$\varepsilon_\ell,\eta_\ell,h_\ell>0$ and $\xi_\ell\ge0$, and select
pure equilibria on grids $G_\ell$ of mesh at most $h_\ell$. The cap lemma is valid on every grid. All the utility bounds
and opponent-event probabilities in that proposition remain valid.
We check the deviations used to identify the limiting bids.

\pstep{1}
Use the grid clipping height
$\max\{b_{i,r_i},\lceil2\varepsilon_\ell\rceil_{G_\ell}\}$.
The rank count still leaves allocation
unchanged whenever all opponents come from their concentrated components.
The saved payment tends uniformly to zero by
\eqref{pt:modulus}--\eqref{pt:increment-minimum}. Thus
$0\le b_{ij}-b_{i,r_i}\le2\varepsilon_\ell+h_\ell+\zeta_\ell$ for
$j<r_i$, where $\zeta_\ell\to0$ uniformly over the relevant types.
To secure the first $r_i-1$ units, raise those bids to at least
$\lceil2\varepsilon_\ell\rceil_{G_\ell}$. The additional payment is
at most $(r_i-1)(2\varepsilon_\ell+h_\ell)$, so the probability of
losing any of these units again tends to zero.

\pstep{2}
At a proposed common limiting atom $z$, on the own event
$|b_{i,r_i}-z|\le v$, raise the first $r_i$ bids to at least
$\lceil z+2v\rceil_{G_\ell}$. For each fixed small $v>0$, the payment
increase is at most $r_i(3v+h_\ell)$ for all sufficiently large $\ell$.
The allocation gain in the common-atom argument is unchanged.
Letting $\ell\to\infty$ and then $v\downarrow0$ yields the
contradiction in \eqref{pt:no-common-atoms}.
These deviations lie below one for small $v$, because the limiting
scalar bids are bounded by marginal values strictly below one.

For a fixed positive scalar deviation $x<1$ at a continuity point of
the opponent's limiting distribution, use a block of height
$\lceil x\rceil_{G_\ell}$. Its payment error is at most $r_i h_\ell$.
The expected payoff therefore converges to the same scalar payoff.
Approximation from above covers zero and atoms, as in
Proposition~\ref{pt:joint}. The limiting distributions consequently
satisfy all scalar best-response inequalities. Uniqueness identifies
bidder~2's limit as the atomless distribution $G_2$.

\pstep{3}
Consider any convergent sequence of regular comparison types
and their equilibrium bids. The payoff upper bound
\eqref{pt:probe-limsup} is unchanged, including when the limiting second
bid is zero. For each fixed deviation $a\in\mathcal B_m$ with $a_2>0$, round each coordinate of $a$ upward to $G_\ell$. This preserves order and changes total
payment by at most $mh_\ell$. The first unit is guaranteed against
concentrated opponents for large $\ell$, and the remaining allocation
thresholds meet the atomless limiting bid with probability zero.
Its expected payoff converges to $\Psi(t,a)$.
Equilibrium optimality therefore implies
$\Psi(t,b)\ge\Psi(t,a)$ for every such $a$.
Continuity of $\Psi$ and density of these deviations give the inequality
for all $a\in\mathcal B_m$. Every limiting comparison bid must thus maximize $\Psi$,
so the strict third-bid bounds in \eqref{pt:actual-reversal} persist.

This sequential argument gives a common threshold for all sufficiently
small $\varepsilon,\eta,\xi,h$, allowing $\xi=0$.
Otherwise a sequence of such parameters, equilibria, and regular types
violating a strict bound would yield a limiting maximizer contradicting
that bound. Fix positive $\varepsilon,\eta,\xi$ below this threshold
and the threshold in Proposition~\ref{pt:joint}. Then choose $h_0$
below the common threshold. The resulting fixed densities give the
third-bid reversal on every grid with mesh below $h_0$ and in the
continuous auction. Their positive lower and finite upper bounds
follow from \eqref{fs:ap:density}. The same reasoning with $\xi=0$
proves the last assertion.
\end{proof}

\begin{proposition}\label{ag:prop}
Fix $n\ge2$, $m\ge3$, and utilities satisfying \textup{(\ref{eq:utilities})}.
There are $c\in(0,1)$ and independent type distributions with bounded
densities on positive-volume subsets of $T_m$ for which the all-pay
auction on $\{0,c,1\}$ has no pure equilibrium whose last bid for
bidder~1 is coordinatewise nondecreasing. No admissibility or
no-overbidding restriction is imposed.
\end{proposition}

\begin{proof}
\pstep{1}
We construct two ordered comparison types.
By Lemma~\ref{ap:lem:caps}, every best response satisfies $b_k\le t_k$.
Thus a type with $t_1<c$ bids zero, while a type with $t_1<1$ never
bids one. Fix $0<w<B_L<B_H<A<1$ and consider
$t^j(c)=(A,B_j,w^{m-3},C(c))$ for $j=L,H$.
The repeated block is empty when $m=3$.
Write $u=u_1$, $S_j=A+B_j+(m-3)w$, and $x_j(c)=S_j-(m-1)c$.
For small $c>0$, put $H(c)=u(A-(m-1)c)-u(A-mc)>0$ and choose
$d(c)\in(0,w/2)$ to satisfy
\begin{equation}\label{ag:eq:tuning}
\frac12\sum_{j\in\{L,H\}}[u(x_j(c)+d(c))-u(x_j(c))]=H(c).
\end{equation}
The left side is continuous and strictly increasing in $d$, starts
at zero, and has a positive limit at $d=w/2$ as $c\downarrow0$.
Since $H(c)\to0$, a unique solution exists for all small $c$ and
$d(c)\to0$. Set $C(c)=c+d(c)$. Then $c<C(c)<w$ for small $c$,
so both types belong to $T_m$.

Suppose bidder~2 submits $(c^{m-1},0)$ and every other opponent bids
zero. The only possible best responses of either comparison type are
$a^k=(c^k,0^{m-k})$, $0\le k\le m$.
Action $a^1$ wins one unit. For $k\ge2$, action $a^k$ wins $k$ units
if bidder~1 has priority over bidder~2 and one unit otherwise.
Writing $V_k(t)=\sum_{j=1}^k t_j$, its expected utility for $k\ge2$ is
$F_t(k)=\tfrac12u(V_k(t)-kc)+\tfrac12u(A-kc)$.
Also, $F_t(1)=u(A-c)<u(A)$ and
$F_t(0)=u(A)/n+(1-1/n)u(0)\le u(A)$.
The last expression accounts for the single remaining unit at a zero bid.

Let $J_j(c)=u(x_j(c)+d(c))-u(x_j(c))$.
Strict concavity gives $J_L(c)>J_H(c)$, while
\eqref{ag:eq:tuning} gives $(J_L(c)+J_H(c))/2=H(c)$.
For $j=L,H$,
\[
F_{t^j(c)}(m)-F_{t^j(c)}(m-1)=\tfrac12[J_j(c)-H(c)].
\]
This difference is positive at the low type and negative at the high type.
For small $c$, every action with $k<m-1$ is strictly worse than
$a^{m-1}$. Indeed, the latter payoff tends to
$[u(S_j)+u(A)]/2>u(A)$, and each $2\le k<m-1$ has a strictly
smaller limit because all the first $m-1$ marginal values are positive.
There are finitely many actions. The unique best responses are therefore
$a^m$ at the low type and $a^{m-1}$ at the high type.

\pstep{2}
We give bidder~2 a strict incentive to submit $a^{m-1}$.
Fix $v_R\in(0,1)$ and consider $t^R(c)=(v_R^{m-1},c/4)$, taking $c<v_R$.
Value caps leave only actions $a^k$, $0\le k\le m-1$.
Against zero bids, bidder~2 wins all units if she has the highest
priority and otherwise wins exactly $k$. With $R(c)=(m-1)v_R+c/4$,
her expected utility is
$W_c(k)=u_2(R(c)-kc)/n+(1-1/n)u_2(kv_R-kc)$.
For $k<m-1$, the difference $W_c(m-1)-W_c(k)$ tends to
$(1-1/n)[u_2((m-1)v_R)-u_2(kv_R)]>0$ as $c\downarrow0$.
Fix $c$ sufficiently small for these and the preceding strict comparisons.

\pstep{3}
Replace the types by positive-volume compact boxes in
$\operatorname{int}T_m$. A box $\mathcal R$ near $t^R(c)$ can have
first value below one, last value below $c/2$, and first $m-1$ values
above $c$. By continuity, $a^{m-1}$ remains the unique best response against
zero bids, with a uniform payoff gap $\delta_R>0$ over every other
value-capped grid action.
Likewise choose boxes $\mathcal L,\mathcal H$ near the comparison types
on which $a^m,a^{m-1}$, respectively, remain unique best responses
against $a^{m-1}$ and zero bids, with a uniform gap $\delta_P>0$.
The centers can first be perturbed into the interior, separating
repeated coordinates and slightly raising the previously equal
coordinates of the high type. Strict payoff inequalities survive.
For every $h\in\mathcal H$ and $\ell\in\mathcal L$, the boxes can
therefore satisfy
\begin{equation}\label{ag:eq:box-order}
h_k>\ell_k\quad(k=1,\ldots,m).
\end{equation}
For bidder~1 and every bidder $j\ge3$, choose a positive-volume compact
box $\mathcal D_j\subset\operatorname{int}T_m$ with $t_1<c/2$.
These types bid zero.

\pstep{4}
Put $\Omega_2=u_2(m)-u_2(-m)$ and choose
$0<\eta<\min\{1/2,\delta_R/[2(\delta_R+\Omega_2)]\}$.
Use the independent distributions
$\mu_1=(1-\eta)U_{\mathcal D_1}+\tfrac\eta2(U_{\mathcal L}+U_{\mathcal H})$,
$\mu_2=U_{\mathcal R}$, and $\mu_j=U_{\mathcal D_j}$ for $j\ge3$.
Their densities are bounded and both comparison boxes have positive
probability. For a type in $\mathcal R$, all opponents bid zero with
probability $1-\eta$. All grid-action surpluses lie in $[-m,m]$.
Thus the expected advantage of $a^{m-1}$ over any other value-capped action
is at least
\((1-\eta)\delta_R-\eta\Omega_2>\delta_R/2>0\).
Every equilibrium therefore assigns $a^{m-1}$ to almost every type in
$\mathcal R$. Against these bids, almost every type in $\mathcal L$
chooses $a^m$, while almost every type in $\mathcal H$ chooses
$a^{m-1}$. Their last bids are $c$ and zero, respectively, contradicting
\eqref{ag:eq:box-order}. Positive probability of both boxes makes the
contradiction unaffected by changes on null sets.
\end{proof}

This three-point construction uses a separate choice of parameters and
densities. It does not assert nonexistence after arbitrary bid levels
are added to the grid.



\newpage

\begin{thebibliography}{99}
\addcontentsline{toc}{section}{References}

\bibitem[Agence France Tr\'{e}sor(n.d.)]{AFT}
Agence France Tr\'{e}sor (n.d.). ``Issuance Techniques.''  2026.

\bibitem[Armantier and Sba\"{i}(2006)]{ArmantierSbai2006}
Armantier, Olivier, and Erwann Sba\"{i} (2006). ``Estimation and Comparison of Treasury Auction Formats when Bidders Are Asymmetric.'' \emph{Journal of Applied Econometrics} 21(6), 745--779.


\bibitem[Ausubel et~al.(2014)]{AusubelEtAl2014}
Ausubel, Lawrence M., Peter Cramton, Marek Pycia, Marzena Rostek, and Marek Weretka (2014). ``Demand Reduction and Inefficiency in Multi-Unit Auctions.'' \emph{Review of Economic Studies} 81(4), 1366--1400.

\bibitem[Back and Zender(1993)]{BackZender1993}
Back, Kerry, and Jaime F. Zender (1993). ``Auctions of Divisible Goods: On the Rationale for the Treasury Experiment.'' \emph{Review of Financial Studies} 6(4), 733--764.

\bibitem[Gentry, Li, and Lu(2015)]{GentryLiLu2015}
Gentry, Matthew, Tong Li, and Jingfeng Lu (2015). ``Existence of Monotone Equilibrium in First Price Auctions with Private Risk Aversion and Private Initial Wealth.'' \emph{Games and Economic Behavior} 94, 214--221.

\bibitem[German Finance Agency(n.d.)]{GermanFinanceAgency}
German Finance Agency (n.d.). ``Auction Process.''  2026.

\bibitem[H\"{a}fner(2023)]{Hafner2023}
H\"{a}fner, Samuel (2023). ``Risk Aversion in Share Auctions: Estimating Import Rents from TRQs in Switzerland.'' \emph{Quantitative Economics} 14(2), 419--470.

\bibitem[Jackson and Swinkels(2005)]{JacksonSwinkels2005}
Jackson, Matthew O., and Jeroen M. Swinkels (2005). ``Existence of Equilibrium in Single and Double Private Value Auctions.'' \emph{Econometrica} 73(1), 93--139.

\bibitem[Lizzeri and Persico(2000)]{LizzeriPersico2000}
Lizzeri, Alessandro, and Nicola Persico (2000). ``Uniqueness and Existence of Equilibrium in Auctions with a Reserve Price.'' \emph{Games and Economic Behavior} 30(1), 83--114.

\bibitem[Markakis and Telelis(2012)]{MarkakisTelelis2012}
Markakis, Evangelos, and Orestis Telelis (2012). ``Uniform Price Auctions: Equilibria and Efficiency.'' In \emph{Algorithmic Game Theory}, Lecture Notes in Computer Science 7615, 227--238. Springer.

\bibitem[Maskin and Riley(1984)]{MaskinRiley1984}
Maskin, Eric, and John Riley (1984). ``Optimal Auctions with Risk Averse Buyers.'' \emph{Econometrica} 52(6), 1473--1518.

\bibitem[Maskin and Riley(2000)]{MaskinRiley2000}
Maskin, Eric, and John Riley (2000). ``Equilibrium in Sealed High Bid Auctions.'' \emph{Review of Economic Studies} 67(3), 439--454.

\bibitem[McAdams(2003)]{McAdams2003}
McAdams, David (2003). ``Isotone Equilibrium in Games of Incomplete Information.'' \emph{Econometrica} 71(4), 1191--1214.

\bibitem[McAdams(2006)]{McAdams2006}
McAdams, David (2006). ``Monotone Equilibrium in Multi-Unit Auctions.'' \emph{Review of Economic Studies} 73(4), 1039--1056.

\bibitem[McAdams(2007a)]{McAdams2007a}
McAdams, David (2007a). ``Uniqueness in Symmetric First-Price Auctions with Affiliation.'' \emph{Journal of Economic Theory} 136(1), 144--166.

\bibitem[McAdams(2007b)]{McAdams2007b}
McAdams, David (2007b). ``On the Failure of Monotonicity in Uniform-Price Auctions.'' \emph{Journal of Economic Theory} 137(1), 729--732.


\bibitem[Olszewski, Reny, and Siegel(2026)]{OlszewskiRenySiegel2026}
Olszewski, Wojciech, Philip J. Reny, and Ron Siegel (2026). ``Equilibrium Existence in First-Price Auctions With Private Values.'' \emph{Econometrica} 94(1), 193--224.

\bibitem[Prokopovych and Yannelis(2019)]{PY2019}
Prokopovych, Pavlo, and Nicholas C. Yannelis (2019). ``On Monotone Approximate and Exact Equilibria of an Asymmetric First-Price Auction with Affiliated Private Information.'' \emph{Journal of Economic Theory} 184, 104925.

\bibitem[Pycia and Woodward(2026)]{PyciaWoodward2026}
Pycia, Marek, and Kyle Woodward (2026). ``A Case for Pay-as-Bid Auctions.'' \emph{Journal of Political Economy} 134(2), 795--845.

\bibitem[Reny(1999)]{Reny1999}
Reny, Philip J. (1999). ``On the Existence of Pure and Mixed Strategy Nash Equilibria in Discontinuous Games.'' \emph{Econometrica} 67(5), 1029--1056.

\bibitem[Reny(2011)]{Reny2011}
Reny, Philip J. (2011). ``On the Existence of Monotone Pure-Strategy Equilibria in Bayesian Games.'' \emph{Econometrica} 79(2), 499--553.

\bibitem[Reny and Zamir(2004)]{RenyZamir2004}
Reny, Philip J., and Shmuel Zamir (2004). ``On the Existence of Pure Strategy Monotone Equilibria in Asymmetric First-Price Auctions.'' \emph{Econometrica} 72(4), 1105--1125.

\bibitem[U.K. Debt Management Office(n.d.)]{UKDMO}
U.K. Debt Management Office (n.d.). ``Glossary.'' Entry ``Auction (gilt).''  2026.

\bibitem[U.S. Treasury(n.d.)]{USTreasury}
U.S. Department of the Treasury (n.d.). ``About Auctions.'' TreasuryDirect.  2026.

\bibitem[Wilson(1979)]{Wilson1979}
Wilson, Robert (1979). ``Auctions of Shares.'' \emph{Quarterly Journal of Economics} 93(4), 675--689.

\end{thebibliography}
\end{document}